\documentclass[aps,prd,twocolumn,showpacs,nofootinbib,superscriptaddress]{revtex4-2}

\usepackage[title,titletoc]{appendix}

\usepackage{graphicx}
\usepackage{dcolumn}
\usepackage{bm}
\usepackage[colorlinks, allcolors=black]{hyperref}
\usepackage[usenames, dvipsnames]{color}
\usepackage{subfig}
\usepackage{lipsum}
\usepackage{titlesec}
\titlespacing\section{0pt}{12pt plus 4pt minus 4pt}{1pt plus 20pt minus 2pt}
\usepackage{xcolor}
\usepackage{braket}
\usepackage{tabularx}
\usepackage{comment}
\usepackage{afterpage}
\usepackage{placeins}
\usepackage{booktabs}
\usepackage{multirow}
\usepackage{array}
\usepackage{setspace}
\graphicspath{{Figs/}}
\usepackage{hhline}
\usepackage{xfrac}
\usepackage{mathtools}
\usepackage{listings}
\usepackage[normalem]{ulem}
\usepackage{titlesec}
\usepackage{amsfonts}
\usepackage[version=4]{mhchem}

\usepackage{epstopdf}
\catcode`@11
\def\seceqaa{\@addtoreset{equation}{section}
\def\theequation{A\arabic{equation}}}
\def\seceqbb{\@addtoreset{equation}{section}
\def\theequation{B\arabic{equation}}}
\def\seceqcc{\@addtoreset{equation}{section}
\def\theequation{C\arabic{equation}}}
\def\seceqdd{\@addtoreset{equation}{section}
\def\theequation{D\arabic{equation}}}
\def\seceqee{\@addtoreset{equation}{section}
\def\theequation{E\arabic{equation}}}
\def\seceqff{\@addtoreset{equation}{section}
\def\theequation{F\arabic{equation}}}
\def\seceqgg{\@addtoreset{equation}{section}
\def\theequation{G\arabic{equation}}}
\def\seceqhh{\@addtoreset{equation}{section}
\def\theequation{H\arabic{equation}}}
\def\nn{\nonumber}
\catcode`@11

\newcommand{\zd}[1]{{ #1}}

\begin{document}
\title{Chiral Magnetic Conductivity in the Tight-Binding Model of Dirac Semimetals}
\author{Mustafa Bohra}
\altaffiliation{These authors contributed equally to this work.}
\affiliation{Department of Physics, Ariel University, Ariel-40700, Israel}

\author{Yuexiang Zhang}
\altaffiliation{These authors contributed equally to this work.}
\affiliation{Department of Physics, Ariel University, Ariel-40700, Israel}

\author{Mikhail Zubkov}
\email{mikhailzu@ariel.ac.il}
\affiliation{Department of Physics, Ariel University, Ariel-40700, Israel}

\date{\today}

\begin{abstract}

\noindent We consider the typical tight – binding model of Dirac semimetal in the presence of both magnetic and electric fields. The electric conductivity reveals dependence on magnetic field. We calculate this dependence in the limit of strong magnetic field, when the given model is described effectively by the one – dimensional SSH model because the dynamics in the plane orthogonal to magnetic field is reduced to that of the lowest Landau level (LLL). Considering the small temperature limit we take into account dissipation due to scattering on impurities. The corresponding dissipation rate is calculated explicitly.  The obtained results confirm that the source of the magnetoconductivity in this system is chiral magnetic effect. 
\end{abstract}

\maketitle


\section{Introduction}
\vspace{0.3cm}
Weyl and Dirac semimetals possess negative magnetoresistance - their electrical resistance 
decrease when an external magnetic field is applied. Since its experimental observation, this behavior has attracted 
significant interest from both theory and experiment. It is generally understood to be a 
signature of the chiral magnetic effect (CME), which describes how an imbalance between 
left- and right-handed charge carriers drives an electric current along the direction of 
an applied magnetic field. Despite broad acceptance of this link, the theoretical 
arguments underpinning it are based on a number of assumptions that, while physically 
reasonable, have not been derived from first principles. The CME originally was considered as a representative of the family of non - dissipative transport effects (chiral separation effect \cite{zubkov2023effect}, chiral vortical effect \cite{abramchuk2018anatomy}, quantum Hall effect \cite{zhang2019hall,selch2025non}, etc). These effects reveal deep relation between relativitic quantum field theory and  fermionic superfluids \cite{volovik2013nambu}, and the condensed matter systems in general \cite{volovik2015scalar,katsnelson2013euler}. Topology plays important role in description of these effects \cite{zubkov2012momentum,zubkov2017topology}. This application of topology is common for condensed matter and high energy physics \cite{volovik2017standard}. In the latter case the other application of topology is worth to be mentioned, namely, the physics of topological defects in quark matter   \cite{bakker1999central,bakker2005standard,zubkov2018momentum}.

The commonly cited heuristic picture of chiral magnetic effect \cite{CMEZrTe5, Kaushik2017} proceeds as follows.
Parallel electric and magnetic fields trigger the chiral anomaly, a quantum mechanical 
effect of pumping the  electron-hole pairs of opposite chirality out of the 
Dirac sea. The result is a net imbalance between right- and left-handed fermions, 
quantified by the chiral charge density:
\begin{equation}
    \rho_{5} = \rho_{R} - \rho_{L} = N_{f}\frac{\vec{E}\cdot\vec{B}}{2\pi^{2}}\tau_{5},
\end{equation}
where $N_{f}$ is the number of Dirac fermion species and $\tau_{5}$ is the timescale over 
which this imbalance relaxes. Throughout this paper we work in natural units 
$\hbar = c = |e| = k_{B} = 1$. {The electron charge is $q=-|e|$. Thus the kinetic momentum is $\boldsymbol{\Pi}=\mathbf p+|e|\mathbf A$; after setting $|e|=1$ we write $\boldsymbol{\Pi}=\mathbf p+\mathbf A$. Factors of $|e|$ are restored explicitly in the final SI formulas.}

One then introduces the chiral chemical potential $\mu_{5}$, defined as half the difference 
between the chemical potentials of the two chiralities, to characterize this imbalance. By 
analogy with the standard relation between charge density and chemical potential in 
equilibrium, one writes, in the weak-field regime:
\begin{equation}
    \rho_{5} \approx \frac{\mu_{5}}{3v_{F}^{3}}\Big(T^{2} + \frac{\mu^{2}}{\pi^{2}}\Big),
\end{equation}
with $v_{F}$ the Fermi velocity, $T$ the temperature in energy units, and $\mu$ the 
ordinary chemical potential. When the magnetic field is strong enough such that only the lowest Landau 
level (LLL) is relevant \cite{Kaushik2017}, the relation reads:
\begin{equation}
    \rho_{5} \approx N_{f}\frac{\mu_{5}}{2\pi^{2}v_{F}}|\vec{B}|.
\end{equation}

The final ingredient is the CME current formula itself,
\begin{equation}
    \vec{j} = N_{f}\frac{1}{2\pi^{2}}\mu_{5}\vec{B},
\end{equation}
which tells us that a nonzero $\mu_{5}$ in a magnetic field generates a current flowing 
along the field axis. Putting these pieces together yields the longitudinal 
magnetoconductivity:
\begin{equation}
    \sigma_{zz} = N_{f}\frac{v_{F}|\vec{B}|}{2\pi^{2}}\tau_{5}.
\end{equation}
The same result is recovered via the Kubo formula in \cite{Gorbar2014, Lu2015, Li2023}, 
and it predicts that conductivity along the field direction grows linearly with $|\vec{B}|$ 
--- which is precisely what negative magnetoresistance means: the stronger the field, the 
better the material conducts.

Multiple theoretical frameworks have been developed to connect negative magnetoresistance in Weyl and Dirac semimetals, as well as vector currents in quantum chromodynamics (QCD), with the chiral magnetic effect ~\cite{Kaushik2017, Fukushima2008, Stephanov2012, Burkov2014, Son2013, Gorbar2016, Gao2012, Lin2020, Hattori2017, Huang2018} including lattice simulations ~\cite{Valgushev2015}. Chiral kinetic theory has been extensively utilized to reproduce transport phenomena ~\cite{Stephanov2012, Gao2012, Son2012, Sekine2017, Sekine2021} and to examine quantum dynamics ~\cite{Lin2020, Hattori2017}. Furthermore, quasiclassical~\cite{Burkov2014, Son2013, Gorbar2016} and semiclassical models have been introduced to describe this effect. All of these approaches, despite their physical motivation and internal consistency, are based on a shared set of assumptions that remain hypotheses rather than rigorously derived results from microscopic, first-principles analysis. An analogous manifestation of the chiral magnetic effect has been identified in superfluid helium-3 in its A phase (³He-A), where low-energy quasiparticles are electrically neutral. In this context, the effect originates from an emergent gauge field \cite{volovik2017chiral}, demonstrating that chiral magnetic responses are not exclusive to charged relativistic fermions but may also occur in systems with effective relativistic excitations and emergent gauge fields.

 Defining a chiral chemical potential in a system affected by the chiral anomaly is problematic because chiral charge is not conserved, which prevents the establishment of a true equilibrium state with finite \( \mu_{5} \). The anomaly leads to continuous creation chiral charge. Its annihilation occurs in parallel due to the interactions, which dirves the system out of  equilibrium. Rigorous analyses demonstrate that in thermodynamic equilibrium, the current of chiral magnetic effect (CME)  described by Eq.~(4) vanishes identically ~\cite{Valgushev2015, Buividovich2015, Buividovich2014, Buividovich2014a, Zubkov2016, Zubkov2016a, Vazifeh2013,Yamamoto2015, Banerjee2021}. The CME arises only in nonequilibrium scenarios ~\cite{Banerjee2022}, such as when \( \mu_{5} \) is time-dependent or when the system is subject to dissipation. This observation indicates that the standard heuristic derivation, which treats \( \mu_{5} \) as an equilibrium parameter, is not formally consistent.

The conceptual shortcomings identified previously require a more robust theoretical framework. The Keldysh-Schwinger formalism, also known as the non-equilibrium Green’s function technique, provides such an approach. Developed to describe systems outside of equilibrium, this formalism facilitates the calculation of observable quantities directly from microscopic dynamics, thus eliminating reliance on ill-defined constructs such as the chiral chemical potential. Dissipative effects, which are essential for  stabilizing chiral charge dynamics, emerge naturally from the imaginary components of self-energies associated with scattering processes.

The present study adopts  approach that is not based on continuum effective theory. Instead  it employs a complete tight-binding model of a Dirac semimetal defined on a three-dimensional lattice. This model retains the full band structure without any linearization approximation. The lattice regularization is incorporated from the outset, which offers several important advantages.

First, the relaxation of chirality, and thus the finiteness of \( \tau_{5} \), arises intrinsically from the lattice structure. In continuum theories, chiral symmetry breaking must be introduced explicitly, such as through a Dirac mass term or is to be implied as the effect of ultraviolet regularization. In contrast, in crystalline systems, the lattice itself breaks chiral symmetry because the right- and left-handed branches of the spectrum are connected across the Brillouin-zone boundary. The Nielsen–Ninomiya mechanism thereby provides a natural channel for chirality-violating transitions, ensuring that the numbers of right- and left-handed fermions are not conserved independently, even in the absence of an explicit mass term. Within the tight-binding framework, this mechanism is incorporated at the microscopic level.

Second, the tight-binding approach enables a controlled low-energy expansion, explicitly revealing the structure of the effective Hamiltonian under a strong magnetic field. Utilizing the symmetric gauge and projecting onto the lowest Landau level, the system exhibits significant dimensional reduction. The three-dimensional problem reduces to an effective one-dimensional theory, and the resulting Hamiltonian precisely matches the Su-Schrieffer-Heeger (SSH) model ~\cite{PhysRevLett.42.1698, BOHRA2026116276, Su1980, bernevig2013topological, Lima2026, McCann2026}, a well-established one-dimensional topological insulator originally formulated for polyacetylene. {This is an exact algebraic correspondence for the LLL-projected Hamiltonian after the controlled transverse continuum expansion; it is not an exact identity for the unprojected three-dimensional lattice model.} The algebraic structure governing conducting polymers also characterizes Dirac semimetals in the quantum limit of strong magnetic fields.

Third, impurity scattering calculations are particularly transparent within the lattice representation. Disorder is modeled as random delta-function impurities characterized by specific concentrations and scattering strengths. The impurity self-energy is evaluated explicitly using the eigenstates of the SSH Hamiltonian. 

This study examines the zero-temperature limit, which significantly simplifies the analysis because phonon-mediated scattering processes vanish as \(T \to 0\). In this regime, elastic scattering on impurities remains the sole source of dissipation, persisting even at absolute zero. Omitting phonons allows for the isolation of the essential physics of impurity-induced relaxation and enables full analytical control over the calculations. This approach is sufficient to demonstrate the validity of the CME-based framework and to derive explicit expressions for the magnetoconductivity.

The paper is organized as follows. In Section II we derive the effective Hamiltonian by projecting the full $(3+1)$-dimensional lattice Dirac system onto the Lowest Landau Level, showing explicitly how the problem reduces to a one-dimensional SSH-type model along the magnetic field direction. Section III introduces the Keldysh formalism adapted to the tight-binding lattice description of Dirac fermions, and constructs the relevant Green's functions including the lesser component. Section IV sets up the impurity scattering problem within the Dyson series framework on the lattice, and Section V evaluates the resulting self-energy integral analytically in the LLL regime in the absence of the electric field, obtaining an explicit expression for the dissipation rate. Section VI calculates the correction to the self-energy induced by the electric field and performs the exact resummation of the impurity ladder (the vertex correction); the details of these calculations are collected in Appendices \ref{app:source} and \ref{app:kernel}. Section VII uses the Wigner-Keldysh formalism to derive the axial charge density and the electric current, including the vertex contribution, and establishes a direct relation between the two. Section VIII concludes with a summary of the main results and a discussion of possible extensions, including higher Landau levels and finite-temperature regimes.
\section{Lowest Landau Level Effective Hamiltonian}\label{Hamiltonian}

The effective Hamiltonian of a three-dimensional lattice Dirac system under a uniform magnetic field $\mathbf{B}=B\hat{z}$ and a parallel electric field $\mathbf{E}=E\hat{z}$ is obtained by projecting the full Hamiltonian onto the Lowest Landau Level (LLL). This approach isolates the low-energy sector relevant in strong magnetic fields, where transverse motion is quantized and suppressed, resulting in dynamics that are predominantly longitudinal along the field direction.

We begin from the lattice-regularized Dirac Hamiltonian
\begin{equation}
H(\mathbf{p})=v_F\sum_{i=1}^{3}\gamma_{0}\gamma_{i}\sin p_{i}
+v_F\gamma_{0}\Bigl(m+\sum_{i=1}^{3}(1-\cos p_{i})\Bigr),
\end{equation}
Here, $\gamma_\mu$ satisfy the Clifford algebra $\{\gamma_\mu,\gamma_\nu\}=2g_{\mu\nu}$ with the metric signature $(+,-,-,-)$. 

{{\it We set the lattice spacing $a=1$ in this section. The dimensionless variables are $p_i a$, $m a$, $B a^2$, $\mu a/(\hbar v_F)$ and $\epsilon a/(\hbar v_F)$. These replacements must be used when $v_F$ and $\hbar$ are restored.}}

{ The parameter $m$ introduces the finite mass of the fermions via the substitution $\sum_i (1 - \cos p_i) \to m + \sum_i (1 - \cos p_i)$ of the conventional model of gapless Dirac semimetal; in the original units it corresponds to $(1-\cos(p a)) \to m a + 1 - \cos(pa)$, so that $m$ has the dimension of inverse length and the physical Dirac mass equals $\Delta = \hbar v_F m$. The massless model  \cite{Abramchuk_2026} corresponds to $m = 0$.} 

 The sine terms correspond to the lattice analogue of linear Dirac dispersion, whereas the cosine terms regularize the theory at large momenta and introduce an effective mass structure. The lattice spacing is set to unity, so $p_i\in[-\pi,\pi]$ spans the Brillouin zone.

Electromagnetic fields are introduced via {the electron minimal coupling},
\begin{equation}
p_i\rightarrow \Pi_i=p_i+A_i,
\end{equation}
with gauge choice~\cite{sakurai2017modern, Shankar1994}:
\begin{equation}
A_x=-\frac{B}{2}y,\qquad A_y=\frac{B}{2}x,\qquad {A_z=-E_3t,\qquad A_0=0}.
\end{equation}
{The symmetric gauge represents the magnetic field but preserves ordinary translational invariance only along $z$; transverse translations are magnetic translations. The temporal gauge $A_z=-E_3t$ generates the uniform field $E_z=E_3$.} Consequently, the kinetic momenta are
\begin{equation}
\Pi_x = -i\partial_x - \frac{B}{2}y,\quad \Pi_y = -i\partial_y + \frac{B}{2}x,\quad 
{\Pi_z=p_z-E_3t}.
\end{equation}
The electric field manifests as the time-dependent electron kinetic momentum {$\theta=p_z-E_3t$}, in agreement with $\dot p_z=-|e|E_3$.

Substituting $\Pi_i$ into the Hamiltonian gives
\begin{equation}
\begin{aligned}
H=&v_F\gamma_0\gamma_1\sin\Pi_x+v_F\gamma_0\gamma_2\sin\Pi_y+v_F\gamma_0\gamma_3\sin\Pi_z \\
&+v_F\gamma_0\Bigl(m+\sum_{i=1}^{3}(1-\cos\Pi_i)\Bigr).
\end{aligned}
\end{equation}
Because the magnetic field acts only in the $x$–$y$ plane, we decompose
\begin{equation}
H=H_\perp+H_\parallel,
\end{equation}
where
\begin{equation}
\begin{aligned}
H_\perp=&v_F\gamma_0\gamma_1\sin\Pi_x+v_F\gamma_0\gamma_2\sin\Pi_y\\
&+v_F\gamma_0[(1-\cos\Pi_x)+(1-\cos\Pi_y)],
\end{aligned}
\end{equation}
\begin{equation}
H_\parallel=v_F\gamma_0\gamma_3\sin\Pi_z+v_F\gamma_0(m+1-\cos\Pi_z).
\end{equation}
This separation arises because the magnetic field quantizes transverse motion into discrete Landau levels, whereas longitudinal motion remains continuous. The details of this are in appendix \ref{Hamiltonianlattice}.

Low-energy states near the band minima or Weyl nodes can be described by expanding the transverse trigonometric functions for small $\Pi_{x,y}$,
\begin{equation}
\sin\Pi_i\approx\Pi_i,\qquad 
1-\cos\Pi_i\approx \frac{\Pi_i^2}{2},\qquad (i=x,y).
\end{equation}

The condition for the small-angle expansion is $|\Pi_{x,y}| \ll 1$, which requires the magnetic length to be much larger than the lattice spacing. In natural units:

\begin{align}
   \sin \Pi_i \approx \Pi_i, \qquad 1 - \cos \Pi_i \approx \frac{\Pi_i^2}{2}
\end{align}

These approximations are valid when:

\begin{align}
    &|\Pi_{x,y}| \ll 1 \implies 1 \gg k_ia+\frac{eBa^2}{\hbar}\\& \implies \frac{1}{\sqrt{|B|}} \gg a \implies |B| \ll \frac{1}{a^2}
\end{align}

In SI units, restoring $\hbar$ and $e$:

\begin{align}
   |B| \ll \frac{\hbar}{ea^2} \approx 10^3~{\rm T}
\end{align}

(the numerical estimate corresponds to $a \approx 8$~\AA; for a typical $a \sim 3$~\AA\ the bound is $\sim 10^{4}$~T)

This is the limit condition: the cyclotron radius must greatly exceed the lattice spacing, so that the continuum Landau level approximation remains valid and lattice discretization effects are negligible. 

The perpendicular Hamiltonian becomes
\begin{equation}
H_\perp\approx{v_F\gamma_0\gamma_1\Pi_x+v_F\gamma_0\gamma_2\Pi_y
+\frac{v_F\gamma_0}{2}(\Pi_x^2+\Pi_y^2)}.
\end{equation}
The essential ingredient of Landau quantization is the non-commutativity
\begin{equation}
[\Pi_x,\Pi_y]={-iB},
\end{equation}
which follows directly from the gauge structure. This algebra is identical to that of a harmonic oscillator. Introducing ladder operators
\begin{align}
&a=\frac{1}{\sqrt{2|B|}}\bigl(\Pi_x-i\,\mathrm{sign}(B)\Pi_y\bigr),\nn\\& 
a^\dagger=\frac{1}{\sqrt{2|B|}}\bigl(\Pi_x+i\,\mathrm{sign}(B)\Pi_y\bigr),
\end{align}
one verifies $[a,a^\dagger]=1$ and obtains
\begin{equation}
\Pi_x^2+\Pi_y^2=2|B|\left(a^\dagger a+\frac12\right).
\end{equation}
Thus the transverse spectrum is quantized into Landau levels labeled by $n=a^\dagger a=0,1,2,\dots$. The perpendicular Hamiltonian becomes
\begin{equation}
H_\perp=v_F\gamma_0(\gamma_1\Pi_x+\gamma_2\Pi_y)
+v_F\gamma_0 |B|\left(a^\dagger a+\frac12\right).
\end{equation}

The LLL corresponds to the oscillator ground state $a|0\rangle=0$. Proof of this is in appendix \ref{Hamiltonianlattice}.  Consequently, only the quadratic zero-point contribution remains, resulting in a constant shift of {$\frac{|B|}{2}\gamma_0$}.

We define {$\theta=p_z-E_3t$}, and the effective Hamiltonian within the LLL becomes
\begin{equation}\label{hth}
H_{\mathrm{LLL}}(\theta)
=v_F\gamma_0\gamma_3\sin\theta
+v_F\gamma_0\left(m+1-\cos\theta+\frac{|B|}{2}\right).
\end{equation}
In explicit matrix form,
\begin{equation}
H_{\mathrm{LLL}}=v_F
\begin{pmatrix}
\sin\theta & m+1-\cos\theta+\frac{|B|}{2} \\
m+1-\cos\theta+\frac{|B|}{2} & -\sin\theta
\end{pmatrix}.
\end{equation}
The energy spectrum is
\begin{equation}\label{egvalue}
E_\pm(\theta)=\pm v_F\sqrt{\sin^2\theta
+\left(m+1-\cos\theta+\frac{|B|}{2}\right)^2}.
\end{equation}
{From this point onward, every occurrence of $B$ inside the scalar combinations $m+B/2$, $2+2m+B$ and $C(\theta)$ is to be understood as $|B|$; the signed field is retained only in $\mathrm{sign}(B)$ and in $\mathbf E\!\cdot\!\mathbf B$.}

These results demonstrate that, in the strong-field limit, the original three-dimensional problem reduces to an effective one-dimensional two-band system aligned with the magnetic field direction. The magnetic field constrains transverse motion and produces the positive-energy threshold $v_F|m+|B|/2|$ at $\theta=0$; {the band-to-band gap is twice this value}. Simultaneously, the electric field causes $\theta$ to evolve linearly with time, {$\theta=p_z-E_3t$}, resulting in Bloch oscillations throughout the Brillouin zone. Consequently, the effective LLL Hamiltonian encapsulates the key low-energy physics that governs transport, quantum oscillations, and non-equilibrium dynamics in Dirac and Weyl systems subjected to parallel electric and magnetic fields.

\section{Keldysh Formalism}\label{2}
\vspace{0.3cm}
In this section, we present the Keldysh formalism ~\cite{Keldysh1964DiagramTF} for Dirac fermions on a lattice~\cite{BOHRA2026116276}, adapting the continuum approach to a tight-binding description suitable for crystalline solids. The analysis employs the upper-triangle representation as described in Refs.~\cite{Arseev2015,Banerjee2022}, and generally follows Refs.~\cite{kamenev2005manybodytheorynonequilibriumsystems,Kamenev2023}, with the addition of lattice regularization. Subsequently, the strong magnetic field limit is examined within the lattice framework.

Beginning with the retarded Green function for a non-interacting lattice Hamiltonian, specifically for tight-binding Dirac fermions, the remaining components of the Keldysh propagator $\hat{G}$ can be constructed, and the inverse Keldysh operator $\hat{Q}$ is subsequently defined~\cite{Keldysh1964DiagramTF}:

\begin{equation}\label{eq:lattice_keldysh_def}
\begin{aligned}
\hat{G} &= \begin{pmatrix}
G^{R} & G^{<} \\
0 & G^{A}
\end{pmatrix} = \hat{Q}^{-1}, \quad
\hat{Q} = \begin{pmatrix}
Q^{R} & Q^{<} \\
0 & Q^{A}
\end{pmatrix},
\end{aligned}
\end{equation}
with the normalization condition
\begin{equation}\label{eq:lattice_norm}
 \begin{pmatrix} 1 &  \\  & 1 \end{pmatrix} =
\begin{pmatrix} Q^{R} & Q^{<} \\ & Q^{A} \end{pmatrix}
\begin{pmatrix} G^{R} & G^{<} \\ & G^{A} \end{pmatrix},
\end{equation}
and the components related by
\begin{equation}\label{eq:lattice_components}
G^{R(A)} =  Q^{R(A)-1},\quad 
G^{<} = -G^{R} Q^{<} G^{A}.
\end{equation}
Such definition, suggested in Ref.~\cite{kamenev2005manybodytheorynonequilibriumsystems}, is convenient for the path-integral representation of the Keldysh formalism on a lattice.

The retarded and advanced components are derived from the lattice Dirac propagator. For a simple cubic tight-binding model with two orbitals per site, representing the two chiralities, the propagator takes the following form:
\begin{equation}\label{eq:lattice_q}
Q^{R(A)}(\mathbf{k},\omega) = \omega \pm i\epsilon - H_{\mathrm{lattice}}(\mathbf{k}),
\end{equation}
where $H_{\mathrm{lattice}}(\mathbf{k})$ is the lattice Hamiltonian. For a minimal model of a Dirac semimetal on a lattice, we use the discretization with $m=0$, {(the finite mass is included via the substitution $\sum_i(1-\cos (k_i a)) \to m a + \sum_i(1-\cos(k_i a))$)}:
\begin{equation}\label{eq:lattice_hamiltonian}
\begin{aligned}
H_{\mathrm{lattice}}(\mathbf{k}) =& v_F \sum_{i=1}^{3} \alpha_i \frac{\sin(k_i a)}{a} \\ 
&+ \gamma_0 \left(\frac{v_F}{a}\Bigl(m a+\sum_{i=1}^{3}\bigl(1-\cos(k_i a)\bigr)\Bigr) \right),
\end{aligned}
\end{equation}
Here, $a$ denotes the lattice spacing, $\vec{\alpha} = \gamma_0\vec{\gamma}$, and $\gamma_{\mu}$ represent the gamma matrices with Minkowski signature. The term proportional to $(1-\cos(k_i a))$ eliminates fermion doubling by assigning mass to the additional Dirac points at the Brillouin zone boundaries, while maintaining the low-energy behavior near $\mathbf{k}=0$. {The constant $m$ provides the physical Dirac mass $\Delta = \hbar v_F m$ of the low-energy fermions.} Taking the limit $a \rightarrow 0$ with $k_i a \ll 1$ yields the continuum Hamiltonian. The condition $\epsilon \rightarrow +0$ guarantees the correct analytic properties of the Green functions.

The lesser Keldysh-Green function is derived based on the principle that the lesser component of the free Keldysh operator serves solely as a regularization \cite{kamenev2005manybodytheorynonequilibriumsystems}.
\begin{equation}\label{eq:lattice_qless}
Q^{<}(\mathbf{k},\omega) = -2i\epsilon n(\omega).
\end{equation}
With the initial spin-independent isotropic distribution $n(\omega)$ the standard result emerges
\begin{equation}\label{eq:lattice_gless}
\begin{aligned}
G^{<}(\mathbf{k},\omega) &= (G^{A}(\mathbf{k},\omega) - G^{R}(\mathbf{k},\omega))n(\omega) \\
&\approx 2\pi i\Delta_{\epsilon}(\omega,E_{\mathbf{k}}) \\
&\quad \times \Biggl(\omega + v_F\sum_i \alpha_i \frac{\sin(k_i a)}{a} \\
&\quad + \gamma_0\Bigl(\frac{v_F}{a}\bigl(m a+{\textstyle\sum_i}(1-\cos(k_i a))\bigr)\Bigr)\Biggr)n(\omega),
\end{aligned}
\end{equation}
where $E_{\mathbf{k}}$ is the lattice energy spectrum
\begin{equation}\label{eq:lattice_spectrum}
E_{\mathbf{k}} = \sqrt{v_F^2\sum_i \frac{\sin^2(k_i a)}{a^2} + \Bigl(\frac{v_F}{a}\bigl(m a+{\textstyle\sum_i}(1-\cos(k_i a))\bigr)\Bigr)^2}.
\end{equation}
The equilibrium distribution function at finite temperature is given by the Fermi distribution
\begin{equation}\label{eq:lattice_fermi}
n(\omega) = (e^{\frac{\omega - \mu}{T}} + 1)^{-1}\to \Theta (\mu -\omega).
\end{equation}
In this paper, the limit $|\mu|\gg T$ is relevant, and the distribution is degenerate.

The broadened delta-function on the lattice is defined as
\begin{equation}\label{eq:lattice_delta}
\begin{aligned}
\Delta_{\epsilon}(\omega,E_{\mathbf{k}}) &= \frac{\delta_{\epsilon}(\omega - E_{\mathbf{k}}) - \delta_{\epsilon}(\omega + E_{\mathbf{k}})}{2E_{\mathbf{k}}},\\
\delta_{\epsilon}(x) &= \frac{\epsilon}{\pi(x^2 + \epsilon^2)},
\end{aligned}
\end{equation}
which in the limit $\epsilon \rightarrow +0$ gives $\delta_{\epsilon}(x)\rightarrow \delta(x)$ and $\Delta_{\epsilon}(\omega,E_{\mathbf{k}})\rightarrow \mathrm{sign}(\omega)\delta(\omega^2 - E_{\mathbf{k}}^2)$.

The approximate form of Eq.~\eqref{eq:lattice_gless} is employed, substituting the bare dissipation rate $\epsilon \rightarrow 0$ with the finite value obtained from the fermion self-energy, under the assumption that this value is significantly smaller than $\mu$. The finite level halfwidth, or dissipation rate $\epsilon$, is introduced via the imaginary component of the self-energy.

To calculate analytically the `lesser' component of a product of Keldysh-Green functions on the lattice, we employ the formula \cite{Arseev2015,Banerjee2021}
\begin{equation}\label{eq:lattice_product}
\begin{aligned}
\bigl(\prod_{i = 1}^{n}\hat{K}_i(\mathbf{k},\omega)\bigr)^{<} =& \sum_{l = 1}^{n}\bigl(\prod_{i = 1}^{l - 1}K_i^R(\mathbf{k},\omega)\bigr) K_l^<(\mathbf{k},\omega) \\
&\bigl(\prod_{j = l + 1}^{n}K_j^A(\mathbf{k},\omega)\bigr).
\end{aligned}
\end{equation}
In this expression the product $(\prod_{i = 1}^{l - 1}K_i^R)$ for $l<2$ is assumed to be equal to unity. For example, for $n = 2$ the formula reads $(\hat{K}_1\hat{K}_2)^{<} = K_1^R K_2^< + K_1^< K_2^A$.

When $\hat{G}$ or $\hat{Q}$ is written without arguments, it refers to the operator. In contrast, the Green function and its inverse correspond to the matrix elements in momentum-frequency space: $\hat{G}(\mathbf{k},\mathbf{k}',\omega,\omega') = \langle \mathbf{k},\omega|\hat{G}|\mathbf{k}',\omega'\rangle$ and $\hat{Q}(\mathbf{k},\mathbf{k}',\omega,\omega') = \langle \mathbf{k},\omega|\hat{Q}|\mathbf{k}',\omega'\rangle$. If the Green function is translation-invariant in space and time, we have $\hat{G}(\mathbf{k},\mathbf{k}',\omega,\omega') = \hat{G}'(\mathbf{k},\omega)\delta(\mathbf{k}-\mathbf{k}')\delta(\omega-\omega')$. We will omit the prime symbol and denote the Green function itself and $\hat{G}'$ by the same letter.

The lattice Wigner-Weyl calculus \cite{zubkov2023discrete} (see also ~\cite{chernodub2017scale,zhang2020influence,suleymanov2019wigner})  is  employed in the present work. It continues the ideas of the original Wigner - Weyl calculus designed for the reformulation of quantum mechanics in phase space \cite{Szmigielski2025, Weyl1927, Mayer1947, Wigner1932}.   Wigner transformation of the matrix elements of an operator $A$ is defined as

\begin{equation}\label{eq:lattice_wigner}
\begin{aligned}
A_W(\mathbf{x},t|\mathbf{p},\omega) =  &\int d^{D}\mathbf{k} d \eta e^{i(2\mathbf{x}\cdot\mathbf{k} - \eta t)} \prod_{i=1,2,3} \frac{1+e^{ik_i}}{2}\\
&\langle \mathbf{p} - \mathbf{k}, \omega - \eta/2 |A|\mathbf{p} + \mathbf{k}, \omega + \eta/2\rangle ,
\end{aligned}
\end{equation}
where the integral over  $\mathbf{k}$  is restricted to the first Brillouin zone $[-\pi/a,\pi/a]^3$.

The Wigner transform~\cite{Roux2019} of a product of operators on the lattice results in the lattice star (Moyal) product~\cite{Moyal_1949}:
\begin{equation}\label{eq:lattice_star}
\begin{aligned}
(AB)_W(\mathbf{x},t|\mathbf{k},\omega) =& A_W(\mathbf{x},t|\mathbf{k},\omega) \star B_W(\mathbf{x},t|\mathbf{k},\omega) \\
=& A_W(\mathbf{x},t|\mathbf{k},\omega) \\
&e^{-i(\overleftarrow{\partial}_{\mathbf{x}}\cdot\overrightarrow{\partial}_{\mathbf{k}} - \overleftarrow{\partial}_{\mathbf{k}}\cdot\overrightarrow{\partial}_{\mathbf{x}} - \overleftarrow{\partial}_t \overrightarrow{\partial}_\omega + \overleftarrow{\partial}_\omega \overrightarrow{\partial}_t)/2}\\ &B_W(\mathbf{x},t|\mathbf{k},\omega),
\end{aligned}
\end{equation}
where derivatives with respect to $\mathbf{k}$ are understood as derivatives on the Brillouin zone torus.

In the strong magnetic field regime and at sufficiently low temperatures, the system's dynamics is confined to the lowest Landau level (LLL). On a lattice, the magnetic field is incorporated through the Peierls substitution~\cite{Peierls1933}, which alters the hopping amplitudes. 

In the symmetric gauge $\vec{A}(\mathbf{r}) =- \frac{1}{2}\mathbf{r}\times \mathbf{B}$ on a lattice, the Feynman LLL propagator for a lattice Dirac fermion in magnetic field reads
\begin{equation}\label{eq:lattice_propagator}
\begin{aligned}
G_F(\mathbf{x},t;\mathbf{y},t') =& {e^{iq\int_{\mathbf{x}}^{\mathbf{y}} d\mathbf{z}\cdot \mathbf{A}(\mathbf{z})}} e^{-i\omega (t-t')} S_{\mathrm{lattice}}(\mathbf{x}-\mathbf{y},\omega)\\
=& {e^{-iq(\mathbf{x}-\mathbf{y})\cdot \mathbf{A}((\mathbf{x}+\mathbf{y})/2)}} e^{-i\omega (t-t')} \\
&S_{\mathrm{lattice}}(\mathbf{x}-\mathbf{y},\omega),
\end{aligned}
\end{equation}
In this context, the exponential denotes the lattice parallel transporter along the straight line connecting the end-points $\mathbf{x}$ and $\mathbf{y}$. The Fourier transform of $S_{\mathrm{lattice}}$ within the lowest Landau level (LLL) approximation is given by
\begin{equation}\label{eq:lattice_lll}
\begin{aligned}
&-iS_{\mathrm{LLL}}^{\mathrm{lattice}}(\mathbf{k},\omega) = 2e^{-\frac{\mathbf{k}_\perp^2}{|B|}} \\
& \times {\frac{\omega \gamma_0 + v_F \frac{\sin(k_3 a)}{a} \gamma_3 + \frac{v_F}{a}\bigl(m a + 1-\cos(k_3 a) + |B|a^2/2\bigr)}{\omega^2 - v_F^2\frac{\sin^2(k_3 a)}{a^2} - \left[\frac{v_F}{a}\bigl(m a+1-\cos(k_3 a)+|B|a^2/2\bigr)\right]^2}} \\
&O^{-},
\end{aligned}
\end{equation}
where $\mathbf{k}_\perp = (k_1, k_2)$ and the Dirac algebra matrix~\cite{Abramchuk_2026}
\begin{equation}\label{eq:lattice_projection}
O^{-} = \frac{1}{2}\bigl(1 - i\gamma_1\gamma_2\,\mathrm{sign}(B)\bigr),
\end{equation}
represents the projection operator acting on fermion states whose spins are aligned with the magnetic field. The Gaussian factor $e^{-\mathbf{k}_\perp^2/|B|}$ emerges from the lattice Landau level wavefunctions in the strong field limit, where the magnetic length $l_B = \sqrt{1/|B|}$ is much larger than the lattice spacing $a$, ensuring the continuum-like form is recovered. Note that $S_{\mathrm{LLL}}^{\mathrm{lattice}}$, being defined with $\bar{\psi} = \psi^\dagger \gamma^0$, differs by an overall factor $\gamma^0$ from the Green function $\hat{G}$ (defined with $\psi^\dagger$); the reduced Green functions used in all subsequent calculations correspond to the latter convention, cf. Appendix \ref{Green detail}.

The LLL fermions couple only to the components of the electric field parallel to the magnetic field because of the projection operator $O^{-}$
\begin{equation}\label{eq:lattice_coupling}
O^{-}\gamma_{\mu}O^{-} = O^{-}\gamma_{\parallel,\mu}.
\end{equation}

Next, we present the Keldysh-Green functions within the lowest Landau level (LLL) approximation for electrons in a lattice Dirac semimetal, assuming no applied electric field:
\begin{equation}\label{eq:lattice_gauge}
\begin{aligned}
\hat{G}(\mathbf{x},t;\mathbf{y},t') &= {e^{{iq}\int_{\mathbf{x}}^{\mathbf{y}} d\mathbf{z}\cdot \mathbf{A}(\mathbf{z})}} \hat{G}^{(LLL)}(\mathbf{x}-\mathbf{y},t-t')\\
&= {e^{-iq(\mathbf{x}-\mathbf{y})\cdot \mathbf{A}((\mathbf{x}+\mathbf{y})/2)}} \hat{G}^{(LLL)}(\mathbf{x}-\mathbf{y},t-t').
\end{aligned}
\end{equation}
For the sake of brevity, we will omit the subscript (LLL) in the following discussion unless its inclusion is necessary to avoid confusion. The translationally invariant components of the Green functions will therefore be denoted by the same symbols as the complete functions. The Fourier transforms of these components, arranged in the $2\times 2$ Keldysh matrix, are given by:
\begin{equation}\label{eq:lattice_factorized}
\begin{aligned}
\hat{G}^{(LLL)}(\mathbf{k},\omega)& \equiv \hat{G}^{(LLL)}(k_1,k_2,k_3,\omega)\\
& \approx 2e^{-\frac{k_1^2+k_2^2}{| B|}} \hat{G}(k_3,\omega) O^{-},
\end{aligned}
\end{equation}
where now $\hat{G}(k_3,\omega)$ is the reduced $(1+1)$ dimensional lattice Green function along the magnetic field direction. We denote the transverse factor as
\begin{equation}\label{eq:lattice_transverse}
G_{\perp}(\mathbf{k}_\perp) = 2e^{-\frac{k_1^2+k_2^2}{| B|}}.
\end{equation}
In the strong magnetic field limit, the notation for the reduced $(1+1)$D lattice Keldysh-Green functions is used
\begin{equation}
\begin{aligned}
\omega_{\pm} &= \omega \pm i\epsilon,\quad \tilde{G}^{R(A)}(k_3,\omega) = \bigl(\omega_{\pm} - \mathcal{H}_{\mathrm{latt}}(k_3)\bigr)^{-1},
\end{aligned}
\end{equation}
where
\begin{align}\label{eq:lattice_reduced}
&\mathcal{H}_{\mathrm{latt}}(k_3) = v_F \alpha_3 \frac{\sin(k_3 a)}{a} + \gamma_0\Biggl(\frac{v_F}{a}\bigl(m a+1-\cos(k_3 a)\bigr) \nn\\
&\quad \quad+ v_F B {a}/2\Biggr),
\end{align}

The eigenvalues of $ \mathcal{H}_{\mathrm{latt}}(k_3)$ are given by 
\begin{align}
&\tilde{E}_{\mathrm{latt}}(k_3) \nn\\
&= \sqrt{\frac{\sin^2(k_3 a)}{v_F^{-2}a^2} + \Biggl(\frac{v_F}{a}\bigl(m a+1-\cos(k_3 a)\bigr) + v_F B a/2\Biggr)^2},
\end{align}

 we have:
\begin{align}
&\Delta_{\epsilon} = \Delta_{\epsilon}(\omega,\tilde{E}_{\mathrm{latt}}(k_3)),\nn\\
&\tilde{G}^{<}(k_3,\omega) = 2\pi i\bigl(\omega + \mathcal{H}_{\mathrm{latt}}(k_3)\bigr) \Delta_{\epsilon} n(\omega). 
\end{align}

Due to Eq.~\eqref{eq:lattice_gauge} we obtain the following expression for the Wigner transformation of the lattice Green function:
\begin{equation}\label{eq:lattice_wigner_g}
\hat{G}_W(\mathbf{x},t|\mathbf{k},\omega) = \hat{G}^{(LLL)}\bigl(\mathbf{k} - \mathbf{A}(\mathbf{x}), \omega - \phi(\mathbf{x})\bigr),
\end{equation}
where $\phi(\mathbf{x})$ is the scalar potential. This reflects the minimal coupling substitution on the phase space.

We denote the Keldysh functions for the interacting theory on the lattice, up to a specified order in perturbation theory, using bold symbols. Interactions with phonons and impurities preserve the general structure of the Green function, and
\begin{equation}\label{eq:lattice_interacting}
\begin{aligned}
\hat{\mathbf{G}}(\mathbf{x},t;\mathbf{y},t') = e^{{i}\int_{\mathbf{x}}^{\mathbf{y}} d\mathbf{z}\cdot \mathbf{A}(\mathbf{z})} &e^{-i \phi((\mathbf{x}+\mathbf{y})/2)(t-t')} \\&\hat{\mathbf{G}}^{(LLL)}(\mathbf{x}-\mathbf{y},t-t').
\end{aligned}
\end{equation}
The Fourier transformation of $\hat{\mathbf{G}}^{(LLL)}$ is given by
\begin{equation}\label{eq:lattice_interacting_fourier}
\hat{\mathbf{G}}^{(LLL)}(\mathbf{k},\omega) \approx G_\perp(\mathbf{k}_\perp) \hat{\mathbf{G}}_{k_3,\omega} O^{-},
\end{equation}
while
\begin{equation}\label{eq:lattice_dyson}
\hat{\mathbf{G}}^{-1}(k_3,\omega) = \hat{G}^{-1}(k_3,\omega) - \hat{\Sigma}(\omega),
\end{equation}
where $\hat{\Sigma}(\omega)$ is the self-energy on the lattice. Due to the lattice symmetries, $\hat{\Sigma}(\omega)$ contains only the unit matrix and $\gamma_0$ components.

By $\hat{\mathbf{G}}$ we denote the $2\times 2$ matrix in Keldysh space
\begin{equation}\label{eq:lattice_keldysh_matrix}
\hat{\mathbf{G}} = \begin{pmatrix}
\hat{\mathbf{G}}^R & \hat{\mathbf{G}}^< \\
0 & \hat{\mathbf{G}}^A
\end{pmatrix}.
\end{equation}
At least in the first order of perturbation theory, just like in the absence of interactions
\begin{equation}\label{eq:lattice_relation}
\tilde{\mathbf{G}}^<(k_3,\omega) = \bigl(\tilde{\mathbf{G}}^A(k_3,\omega) - \tilde{\mathbf{G}}^R(k_3,\omega)\bigr) n(\omega).
\end{equation}

Electron energy dissipation arising from interactions with phonons and lattice disorder leads to broadening of the lesser Green function. Specifically, for electrons subjected to an external magnetic field on the lowest Landau level of the lattice model,
\begin{equation}\label{eq:lattice_broadened}
\begin{aligned}
\mathbf{G}^{<}(k_3,\omega) &\approx 2\pi  i {\Delta_{\epsilon_B}\bigl(\omega,\tilde{E}_{\mathrm{latt}}(k_3)\bigr)} \\
&\times \Biggl(\omega + v_F \alpha_3 \frac{\sin(k_3 a)}{a} \\ &+ \gamma_0 \Bigl(\frac{v_F}{a}\bigl(m a+1-\cos(k_3 a)\bigr) + v_F B a/2\Bigr)\Biggr) n(\omega),
\end{aligned}
\end{equation}
where $\Delta_{\epsilon_B}$ is given by Eq.~\eqref{eq:lattice_delta} with  some finite level half-width $\epsilon = \epsilon_B$. 

Such energy level broadening leads to a significant qualitative change in the system. To illustrate this effect, consider the case of a degenerate Fermi distribution as described by Eq.~\eqref{eq:lattice_fermi}. In the absence of dissipation ($\epsilon_B \rightarrow 0$), the system consists of non-interacting lattice electrons, with all energy levels below $\mu$ occupied and the remaining levels vacant. When $\epsilon_B$ is finite, this occupation pattern is completely disrupted. The Fourier transform of the lesser Green function $G^{<}(k_3,\omega)$ with respect to $\omega$ describes the time evolution of a Bloch state with a given lattice momentum $k_3$. For $\epsilon_B = 0$, the evolution is determined by the phase factor $e^{-i\tilde{E}_{\mathrm{latt}}(k_3)t}$, indicating that the crystal momentum remains constant during time evolution. At finite $\epsilon_B$, the evolution includes a damping factor that vanishes as $t \rightarrow \infty$, with a decay time of $1/(2\epsilon_B)$. Consequently, the probability of finding the particle in the same Bloch state approaches zero for $t \gg 1/(2\epsilon_B)$. Therefore, the occupied states in the lattice Brillouin zone do not constitute truly stationary states when a finite dissipation rate is present. Instead, electron states possess a finite lifetime, implying that an electron initially in such a state will leave it with finite probability. However, an electron does not disappear entirely from the system; rather, it undergoes transition to another Bloch state, which was vacated by a different electron due to the same finite lifetime. As a result, an occupied state in the lattice may become vacant with a certain probability, allowing another electron to temporarily occupy it. The transition of an electron state, characterized by crystal momentum $k_3$, is accompanied by phonon emission or scattering on impurities, which removes the excess momentum and ensures momentum conservation modulo reciprocal lattice vectors.

The lattice formulation inherently incorporates ultraviolet regularization through the Brillouin zone, which is crucial for analyzing the chirality relaxation mechanism. The connection between left-handed and right-handed branches at the Brillouin zone boundaries, as described by the Nielsen-Ninomiya theorem, is explicitly represented in the lattice Hamiltonian of Eq.~\eqref{eq:lattice_hamiltonian}, where the branches converge at $k_3 = \pm \pi/a$. This structure establishes a natural cutoff $\Lambda \sim \pi/a$ and enables an explicit chirality-mixing mechanism, even in the $m=0$ limit, as discussed in another section.

With the lattice notation established, the linear response to an electric field can be calculated for a specified level width. The level width is determined separately in a subsequent section within the lattice model.

\section{Contribution to the self-energy due to scattering on impurities}

{
We take the scalar disorder to have zero mean and a white-noise correlator,
\begin{equation}
 \langle U(\mathbf{x})\rangle=0,\qquad
 \langle U(\mathbf{x})U(\mathbf{y})\rangle
 =u_0^2n_{\rm imp}\,\delta^{(3)}(\mathbf{x}-\mathbf{y}).
\end{equation}
The term linear in $U$ therefore vanishes after disorder averaging.  The leading nonzero (Born) contribution contains two impurity insertions and gives
\begin{equation}
 \hat\Sigma^{\rm imp}(x,y)=u_0^2n_{\rm imp}\,
 \delta^{(3)}(\mathbf{x}-\mathbf{y})\,\hat G(x,y).
\end{equation}
Equivalently, for a translationally invariant disorder average,
\begin{equation}
 \hat\Sigma^{\rm imp}(\omega)=u_0^2n_{\rm imp}
 \int_{\rm BZ}\frac{d^3p}{(2\pi)^3}\,\hat G(\omega,\mathbf p).
\end{equation}
On a lattice, $\delta^{(3)}(\mathbf{x}-\mathbf{y})$ is replaced by
$\delta_{\mathbf r_x,\mathbf r_y}/a^3$ and the integral is over the first
Brillouin zone.  Thus the Born self-energy is local in position space and is
proportional to the local (momentum-integrated) Green function.}

Under a strong magnetic field, Landau quantization significantly alters the transverse momentum integrals. The number of flux quanta passing through an area \(A\) is given by \(N_\phi = |B|A/(2\pi)\), and each Landau level contains precisely \(N_\phi\) degenerate states. This observation motivates the following replacement
\begin{equation}
\int \frac{dp_x dp_y}{(2\pi)^2} \rightarrow \frac{|B|}{2\pi} \sum_{n=0}^\infty,
\end{equation}
so that the full three-dimensional momentum integration becomes
\begin{equation}
\int \frac{d^3p}{(2\pi)^3} G(\omega,\mathbf{p}) \rightarrow \frac{|B|}{2\pi} \sum_{n=0}^\infty \int \frac{dp_z}{2\pi} G_n(\omega, p_z),
\end{equation}
where \(G_n(\omega, p_z)\) is the Green's function projected onto the \(n\)-th Landau level. Incorporating this into the self-energy expression yields
\begin{equation}
\hat{\Sigma}_{\text{imp}}(\omega) = u_0^2 n_{\text{imp}} \cdot \frac{|B|}{2\pi} \sum_{n=0}^\infty \int \frac{dp_z}{2\pi} G_n(\omega, p_z).
\end{equation}

In this study, we focus on the LLL regime, where the system is confined to the \(n = 0\) subspace. Upon projection onto the LLL, the self-energy simplifies to
\begin{equation}
\hat{\Sigma}_{\text{imp}}(\omega) = u_0^2 n_{\text{imp}} \cdot \frac{|B|}{2\pi} \int \frac{dp_z}{2\pi} G_0(\omega, p_z).
\end{equation}
The retarded, advanced, and lesser components follow by inserting the appropriate causal prescriptions:
\begin{align}
\hat{\Sigma}_{\text{imp}}^{(R)}(\omega) &= u_0^2 n_{\text{imp}} \cdot \frac{|B|}{2\pi} \int \frac{dp_z}{2\pi} \frac{1}{p_0 - H_{\text{LLL}} + i\epsilon},\\
\hat{\Sigma}_{\text{imp}}^{(A)}(\omega) &= u_0^2 n_{\text{imp}} \cdot \frac{|B|}{2\pi} \int \frac{dp_z}{2\pi} \frac{1}{p_0 - H_{\text{LLL}} - i\epsilon},\\
\hat{\Sigma}_{\text{imp}}^{(<)}(\omega) &=  {u_0^2 n_{\text{imp}} \cdot\frac{|B|}{2\pi}\int \frac{dp_z}{2\pi} 2\pi i\delta(p_0 - H_{\text{LLL}}) f(p_0)},
\end{align}
where $p_0=\omega$ stands for the initial energy.

\section{Impurity self-energy integral in the absence of electric field}

The impurity self-energy integral is now evaluated explicitly for the LLL regime. The lesser component of the impurity self-energy, as previously derived, is given by
\begin{equation}
\hat{\Sigma}^{(<)}_{\mathrm{imp}}(\omega) =  {u_0^2 n_{\mathrm{imp}} \cdot \frac{|B|}{2\pi}\int \frac{dp_z}{2\pi} \,2\pi i \, \delta(p_0 - H_{\text{LLL}}) f(p_0)},
\end{equation}
In this context, \(p_0\) corresponds to the frequency \(\omega\). The expression includes the Landau level degeneracy factor \(|B|/(2\pi)\) as well as the LLL-projected Green's function.

Given that \(H_{\text{LLL}}\) is a matrix Hamiltonian, the delta function is interpreted through its spectral decomposition. This approach expresses the matrix-valued function as a sum of scalar delta functions, each weighted by the corresponding projector onto an eigenstate, changing variables from \(p_z\) to \(\theta\) (with \(dp_z = d\theta\)), resulting in:

{
\begin{align}
 \delta(p_0-H_{\rm LLL})={}&\frac{E(\theta)+H_{\rm LLL}}{2E(\theta)}
 \delta\bigl(p_0-E(\theta)\bigr)\nn\\
 &+\frac{E(\theta)-H_{\rm LLL}}{2E(\theta)}
 \delta\bigl(p_0+E(\theta)\bigr).
\end{align}}

where $E(\theta)$ is the positive eigenvalue of \(H_{\text{LLL}}\). The process of getting it is in appendix \ref{dirac delta}. {In the remainder of this section we restrict explicitly to the conduction band, $\omega>0$ (and later $\mu>0$).} Substituting and factoring out the Fermi function \(f(\omega)\), we arrive at
\begin{align}
\hat{\Sigma}_{\text{imp}}^{(<)}(\omega) &= i f(\omega) \cdot \frac{u_0^2 n_{\mathrm{imp}} |B|}{2\pi}\nn\\& \times \int d\theta \frac{\omega + H_{\text{LLL}}(\theta)}{2E(\theta)} \delta(E(\theta) - \omega).
\end{align}

The remaining task is to evaluate the integral over longitudinal momentum, denoted as:
\begin{equation}
I(\omega) = \int d\theta \frac{\omega + H_{\text{LLL}}(\theta)}{2E(\theta)} \delta(E(\theta) - \omega).
\end{equation}

The equation \(E(\theta) = \omega\) is equivalent to
\begin{equation}
(2+2m+B)(1-\cos\theta) + \Bigl(m+\frac{B}{2}\Bigr)^2 = \frac{\omega^2}{v_F^2}.
\end{equation}
Defining the dimensionless parameter
\begin{equation}
K = \frac{\dfrac{\omega^2}{v_F^2} - \Bigl(m+\dfrac{B}{2}\Bigr)^2}{2+2m+B},
\end{equation}
this reduces to \(\cos\theta = 1-K\). Physical solutions require \(0 \le K \le 2\), ensuring the right-hand side lies within \([-1,1]\). For each period, there are two roots \(\theta = \pm \theta_0\) with \(\theta_0 = \arccos(1-K)\).

Using the standard identity for integrating a delta function of a function \cite{gelfand1964generalized},
\begin{equation}
\int d\theta \, \delta(E(\theta)-\omega) F(\theta) = \sum_{\theta_r} \frac{F(\theta_r)}{|E'(\theta_r)|}
\end{equation}
where the sum runs over the roots within one period. The derivative \(E'(\theta)\) is obtained from differentiating \(2E(\theta) E'(\theta) = (2+2m+B)\sin\theta\), giving
\begin{equation}
E'(\theta) = \frac{v_F^2(2+2m+B)\sin\theta}{2E(\theta)}.
\end{equation}
At the roots, \(E(\theta) = \omega\) and \(\sin\theta = \pm\sqrt{K(2-K)}\). Thus
\begin{equation}
\frac{1}{|E'(\theta_r)|} = \frac{2\omega}{v_F^2(2+2m+B)\sqrt{K(2-K)}}.
\end{equation}
With two roots $\pm \theta_0$ per period, the integral over one period becomes
\begin{align}
&\int_{\text{period}} d\theta \frac{\omega + H_{\text{LLL}}(\theta)}{2\omega} \delta(E(\theta)-\omega)\nn\\ =& \sum_{\theta_r = \pm \theta_0} \frac{\omega + H_{\text{LLL}}(\theta_r)}{2\omega} \cdot \frac{\omega}{v_F^2\left(1 + m + \frac{B}{2}\right)\sqrt{K(2-K)}} \nonumber\nn \\
=& \frac{1}{2v_F^2\left(1 + m + \frac{B}{2}\right)\sqrt{K(2-K)}} \sum_{\theta_r = \pm \theta_0} (\omega + H_{\text{LLL}}(\theta_r)).
\end{align}

Substituting this result back into the self-energy expression, we obtain
\begin{align}
\hat{\Sigma}_{\text{imp}}^{(<)}(\omega) &= i f(\omega) \cdot \frac{u_0^2 n_{\mathrm{imp}} |B|}{2\pi} \cdot \frac{2}{v_F^2(2+2m+B)\sqrt{K(2-K)}}\nn\\&\times \left[ \omega \mathbf{1} + \gamma_0 v_F \left( \frac{\dfrac{\omega^2}{v_F^2} + 2m + B + \bigl(m+\frac{B}{2}\bigr)^2}{2+2m+B} \right) \right]\nn\\&\times{\Theta(\omega-E_{\min})\Theta(E_{\max}-\omega)}.
\end{align}

{Here, with $a=1$, $E_{\min}=v_F\lvert m+|B|/2\rvert$ and
$E_{\max}=v_F\lvert m+2+|B|/2\rvert$.  The upper step function is required
because the lattice band has a finite upper edge.  The displayed expression is
the imaginary (spectral) Born contribution; the principal-value real part of
$\Sigma^R$ is absorbed into a renormalization of the band parameters.}

The expression represents the combined effects of Landau quantization, the electric field via the variable \(\theta\), and impurity scattering. 

Let us restore in this expression the value of lattice spacing, and assume that {$\mu a/v_F \ll 1$, $|B| a^2 \ll 1$, and $m a \ll 1$}. Moreover, we suppose that these small parameters are subject to the hierarchy {$\mu a/v_F \gg ma+|B|a^2$}. Then we arrive at $K = \frac{\dfrac{\omega^2a^2}{v_F^2} - \bigl(ma+\dfrac{Ba^2}{2}\bigr)^2}{2+2ma+Ba^2} \approx \frac{\omega^2 a^2}{2 v_F^2}$ and 
	\begin{equation}
		\hat{\Sigma}_{\text{imp}}^{(<)}(\omega) = 2 i \epsilon f(\omega){\Theta(\omega-E_{\min})\Theta(E_{\max}-\omega)}\label{Sigma}
	\end{equation}
with
\begin{equation}
	\epsilon = \frac{u_0^2 n_{\mathrm{imp}} |B|}{4\pi v_F} 
\end{equation}
{With $a$ restored, the two band edges are
	$E_{\min}=(v_F/a)|ma+|B|a^2/2|$ and
	$E_{\max}=(v_F/a)|ma+2+|B|a^2/2|$.}
{The edge factors may be replaced by unity only when the positive chemical potential lies well inside the LLL band.}

For sufficiently large values of magnetic field, such that $\mu/v_F \sim m + Ba$ we obtain Eq. (\ref{Sigma}), in which we should keep theta function, and with the essentially different expression for the dissipation rate (depending on  energy):
\begin{equation}
	{\epsilon(\omega) = \frac{u_0^2 n_{\mathrm{imp}} |B|}{4\pi v_F\,\mathcal D\sqrt{1-|M(\omega)|^2}},\qquad
\mathcal D=\frac{2+2ma+|B|a^2}{2}.}
\end{equation}	
{The expression without $\mathcal D$ is its low-energy
approximation $\mathcal D\simeq1$.}
{The equilibrium width above is evaluated to leading
Born/on-shell order. Appendix~\ref{app:kernel} uses this width as input and
solves the corresponding linearized Schwinger--Dyson equation for
$\hat\Sigma^{(1)}$ self-consistently, thereby resumming the impurity ladder and
satisfying the Ward identity at leading on-shell order.} { This approximation may be used for sufficiently weak disorder. }


{

\section{Field-induced correction to the self-energy: vertex correction and the chirality flip rate}
\label{vertex}

The impurity self-energy of Section IV is a functional of the Green function itself, $\hat{\Sigma} = u_0^2 n_{\text{imp}} \int \frac{d^3p}{(2\pi)^3} \hat{\mathbf{G}}$. Consequently, when the external electric field modifies the Green function, it modifies the self-energy as well. The complete linear response therefore obeys the closed equation
\begin{align}
\hat{G}^{(1)}_W =& -\frac{i e}{2} F_{\mu\nu}\, \hat{G}_W \star (\partial_{p_\mu}\hat{Q}_W) \star \hat{G}_W \star (\partial_{p_\nu}\hat{Q}_W) \star \hat{G}_W \nn\\
&+ \hat{G}_W \star \hat{\Sigma}^{(1)} \star \hat{G}_W ,
\label{eq:ladder}
\end{align}
where
\begin{equation}
\hat{\Sigma}^{(1)}(\omega) = u_0^2 n_{\text{imp}}\, \frac{|B|}{2\pi} \int \frac{dp_3}{2\pi}\, \hat{G}^{(1)}(p_3,\omega)
\label{eq:sigma1def}
\end{equation}
is the correction to the self-energy linear in $E_3$ (the transverse integration has already been performed as in Section IV). Iteration of Eq.~(\ref{eq:ladder}) generates the impurity ladder (the vertex correction). {In this section we solve the leading weak-disorder, on-shell form of the linearized problem exactly; the details of the calculations are collected in Appendices \ref{app:source} and \ref{app:kernel}.} The observables are computed in Section \ref{observables}; {the vertex correction introduces the response denominator $\epsilon_{5,{\rm resp}}$ of Eq.~(\ref{eq:eps5}), which must be distinguished from the microscopic single-particle width $\epsilon$.}

Equivalently, after the transverse reduction the Keldysh Green function obeys
\begin{align}
&\hat{G} \star (\hat{Q} - \hat{\Sigma}) = 1, \nn\\
&\hat{\Sigma}(\omega) = g \int \frac{dp_3}{2\pi}\, \hat{G}(\omega, p_3), \qquad g = u_0^2 n_{\rm imp}\, \frac{|B|}{2\pi},
\label{eq:scba}
\end{align}
where all quantities are the upper-triangular matrices in the Keldysh space. Expanding $\hat{G} = \hat{G}^{(0)} + \hat{G}^{(1)}$, $\hat{\Sigma} = \hat{\Sigma}^{(0)} + \hat{\Sigma}^{(1)}$ to the linear order in the electric field (the zeroth order being the equilibrium problem solved in Sections IV, V), we obtain
\begin{equation}
\hat{G}^{(1)} = \hat{G}_d + \hat{G}^{(0)}\, \hat{\Sigma}^{(1)}\, \hat{G}^{(0)},
\label{eq:linG}
\end{equation}
where $\hat{G}_d$ is the first (gradient) term of Eq.~(\ref{eq:ladder}). Since for the point-like impurities $\hat{\Sigma}^{(1)}$ does not depend on momentum, integration of Eq.~(\ref{eq:linG}) over $p_3$ produces a closed \emph{linear} equation for a function of frequency alone:
\begin{equation}
\hat{\Sigma}^{(1)} = \hat{\Sigma}_d + \hat{F}[\hat{\Sigma}^{(1)}], \qquad \hat{\Sigma}_d = g \int \frac{dp_3}{2\pi}\, \hat{G}_d ,
\label{eq:linSigma}
\end{equation}
where for the lesser component, which dominates the solution, the map $\hat{F}$ reduces to
\begin{equation}
\hat{F}[X] = g \int \frac{dp_3}{2\pi}\, \tilde{G}^R\, X\, \tilde{G}^A ,
\label{eq:selfcons}
\end{equation}
and the formal solution is $\hat{\Sigma}^{(1)} = (1 - \hat{F})^{-1}\, \hat{\Sigma}_d$, the Neumann expansion of which reproduces the ladder rung by rung.

The drive (the first iteration) is calculated in Appendix \ref{app:source}:
\begin{align}
\hat{\Sigma}_d^{<}(\omega) &= i E_3\, \frac{v_F^2 \sin\theta_\omega}{\omega}\, n'(\omega)\, \gamma_0\gamma_3 , \nn\\
\hat{\Sigma}_d^{R(A)} &= O(\epsilon E_3/\mu) ,
\label{eq:sigma1}
\end{align}
where $\theta_\omega > 0$ is the positive root of $E_+(\theta_\omega) = \omega$; the explicit closed form of the retarded and advanced components is given by Eq.~(\ref{eq:sigmaRA}). Three facts established in Appendix \ref{app:source} determine the further treatment: the drive is concentrated at the Fermi surface ($n'(\omega) = -\delta(\omega - \mu)$ at $T = 0$); it is of the \emph{zeroth} order in $u_0^2 n_{\rm imp}$, because the impurity factor in Eq.~(\ref{eq:sigma1def}) is cancelled by the $1/\epsilon$ singularity of the mixed product $\tilde{G}^R \ldots \tilde{G}^A$, so that the perturbation theory in $u_0^2$ breaks down and the ladder must be summed; and it is purely odd under the reflection $p_3 \to -p_3$, $X \to \gamma_0 X \gamma_0$, being proportional to the group velocity at the Fermi points.

To formulate the solution we recast the equilibrium self-energy of Section V in the compact form (using $(\omega^2/v_F^2 + 2m + B + (m+B/2)^2)/(2+2m+B) = C(\theta_\omega)$, where $C(\theta) = m + 1 - \cos\theta + B/2$, and $\sqrt{K(2-K)} = \sin\theta_\omega$)
\begin{equation}
\hat{\Sigma}^{(0)<}(\omega) = 2 i\, \epsilon(\omega)\, n(\omega)\, \bigl( 1 + |M(\omega)|\, \gamma_0 \bigr),
\label{eq:sigma0compact}
\end{equation}
with $\epsilon(\omega) = \frac{u_0^2 n_{\rm imp} |B|}{2\pi}\, \frac{\omega}{v_F^2 (2+2m+B) \sin\theta_\omega}$ and
\begin{equation}
|M(\omega)| = \frac{v_F\, C(\theta_\omega)}{\omega} = \bigl| \langle u_+(-\theta_\omega) | u_+(\theta_\omega) \rangle \bigr| ,
\label{eq:overlap}
\end{equation}
where the last equality identifies $|M|$ as the overlap of the eigenspinors at the two Fermi points, i.e. the amplitude of impurity backscattering, controlled by the chiral symmetry breaking term $C(\theta)$ of the effective Hamiltonian; the on-shell level width is $\epsilon_{\rm on} = \epsilon(\omega)(1 + |M|^2)$. \zd{Directly at the Fermi level, $\omega = \mu$, the overlap of Eq.~(\ref{eq:overlap}) can be expressed explicitly through the parameters of the model: using $\cos\theta_\mu = 1 - K$ we obtain $C(\theta_\mu) = m + B/2 + K$, so that
\begin{align}
|M(\mu)| &= \frac{v_F}{\mu}\left( m + \frac{B}{2} + \frac{\mu^2/v_F^2 - (m+B/2)^2}{2+2m+B} \right) \nn\\
&\approx \frac{v_F m}{\mu} + \frac{v_F B}{2\mu} + \frac{\mu}{2 v_F},
\label{eq:Mvalue}
\end{align}
where the expansion holds for $m,\, B,\, \mu/v_F \ll 1$ in the units $a=1$ of this section. The first term is the contribution of the Dirac mass, $\Delta/\mu$ with $\Delta = v_F m$; the second one is due to the magnetic-field-induced gap $v_F B/2$ of the lowest Landau level; the third one originates from the band curvature (the lattice term). Restoring the lattice spacing, $|M(\mu)| \approx \hbar v_F m/\mu + |e|B v_F a/(2\mu) + \mu a/(2\hbar v_F)$, cf. Eq.~(\ref{eq:Mexp}) of Section~\ref{applications}.} {In Appendix \ref{app:kernel} the singular on-shell part of the momentum integral defining $\hat{F}$ is evaluated and diagonalized exactly.  The resulting leading kernel is an algebraic map on the four-dimensional space of matrices spanned by $\{1, \gamma_0, \gamma_3, \gamma_0\gamma_3\}$, with the spectrum}
\begin{align}
&\hat{F}\bigl[\,1 + |M|\gamma_0\,\bigr] = 1 \cdot \bigl(1 + |M|\gamma_0\bigr), \nn\\
&\hat{F}\bigl[\gamma_0\gamma_3\bigr] = q\, \gamma_0\gamma_3, \qquad q = \frac{1 - |M|^2}{1 + |M|^2}, \nn\\
&\hat{F}\bigl[\gamma_3\bigr] = 0, \qquad \hat{F}\bigl[\,|M| - \gamma_0\,\bigr] = 0 .
\label{eq:spectrum}
\end{align}
The unit eigenvalue expresses the Ward identity (particle conservation): its eigenvector is precisely the matrix structure of $\hat{\Sigma}^{(0)<}$, and the corresponding zero mode of $(1 - \hat{F})$ is fixed by the conservation of the particle density and does not contribute to the observables (Appendix \ref{app:kernel}). {For $0<|M|<1$ the drive (\ref{eq:sigma1}) is orthogonal to this even zero mode and is an eigenvector of the leading on-shell kernel with eigenvalue $q$.  Keeping the lattice group-velocity factor, define $\mathcal{D}=(2+2m+B)/2$ for $a=1$ (or $\mathcal{D}=(2+2ma+Ba^2)/2$ with $a$ restored).  Since $|E_+'(\theta_\omega)|=v_F\mathcal{D}\sqrt{1-|M|^2}$, $\epsilon(\omega)=g/(2|E_+'|)$ and $\epsilon_{\rm on}=\epsilon(\omega)(1+|M|^2)$, the consistently projected drive is $\hat{\Sigma}_d^{<}=iE_3v_F\mathcal{D}\,n'(\omega)q\,\gamma_0\gamma_3$.  The Neumann series is therefore geometric:}
{
\begin{equation}
\hat{\Sigma}^{(1)<}(\omega) = \frac{\hat{\Sigma}_d^{<}}{1 - q} = i E_3 v_F\mathcal{D}\, n'(\omega)\, \frac{1 - |M|^2}{2|M|^2}\, \gamma_0\gamma_3 .
\label{eq:directsol}
\end{equation}
}
{In the low-energy normalization used in the response formulas below, $m,B\ll1$ and hence $\mathcal{D}=1+O(m,B)\simeq1$; Eq.~(\ref{eq:directsol}) then reduces to the expression without $\mathcal{D}$.} Together with $\hat{G}^{(1)} = \hat{G}_d + \hat{G}^{(0)R}\, \hat{\Sigma}^{(1)<}\, \hat{G}^{(0)A} + O(\epsilon E_3/\mu)$, {Eq.~(\ref{eq:directsol}) constitutes the leading weak-disorder, on-shell solution} of Eq.~(\ref{eq:scba}) to the linear order in $E_3$. It holds provided the chirality-flip probability exceeds the level width measured in units of the bandwidth, $2|M|^2/(1+|M|^2) \gg \epsilon\, a/(\hbar v_F)$, and the Fermi level stays away from the band edges. We finally note that the described scheme is a conserving approximation in the sense of Baym and Kadanoff: the self-energy of Eq.~(\ref{eq:scba}) is the functional derivative of the Born skeleton functional, $\hat{\Sigma} = \delta\Phi/\delta\hat{G}$, and it is the self-consistent treatment of $\hat{\Sigma}^{(1)}$ that guarantees the Ward identity. Away from the DC, zero-temperature limit Eq.~(\ref{eq:linSigma}) remains valid, $\hat{F}$ becoming a one-dimensional integral kernel in frequency; its numerical inversion provides a practical route to the finite-temperature and finite-frequency extensions mentioned in the Conclusion.

}

\section{Axial Charge Density and Electric Current in the Lowest Landau Level}
\label{observables}

Within the Wigner--Keldysh framework ~\cite{Banerjee2021a}, transport quantities can be expressed in terms of the lesser Green's function and the inverse propagator operator $\hat{Q}$. The electric current density is given by~\cite{Zhang2020}:
\begin{equation}
\begin{aligned}
J_k(x) = -\frac{i}{2} \int \frac{d^{D+1}p}{(2\pi)^{D+1}} 
\mathrm{tr} &\Bigl[
(\partial_{p_k}\hat{Q}^R_W)\hat{G}^<_W \\
& + \hat{G}^<_W(\partial_{p_k}\hat{Q}^A_W)
\Bigr],
\end{aligned}
\end{equation}
while the axial current is defined as
\begin{equation}
\begin{aligned}
J^\mu_5(x) = -\frac{i}{2} \int \frac{d^{D+1}p}{(2\pi)^{D+1}} 
\mathrm{tr} &\Bigl[
\gamma_5(\partial_{p_\mu}\hat{Q}^R_W)\hat{G}^<_W \\
&+ \gamma_5 \hat{G}^<_W(\partial_{p_\mu}\hat{Q}^A_W)
\Bigr].
\end{aligned}
\end{equation}
The axial charge density corresponds to the temporal component $\rho_5(x)=J^0_5(x)$.

To compute the response to an external electromagnetic field, the Green's function is expanded to first order in the field strength tensor $F_{\mu\nu}$~\cite{Onoda2006ev}:
\begin{equation}
\hat{G}_W^{(1)} = -\frac{i e}{2} F_{\mu\nu}
\, \hat{G}_W \star (\partial_{p_\mu}\hat{Q}_W)
\star \hat{G}_W \star (\partial_{p_\nu}\hat{Q}_W)
\star \hat{G}_W.
\end{equation}
Substituting this into the expression for $\rho_5$ yields a contribution linear in the electric field.

{ In this expression the variation of the impurity self-energy caused by the electric field is neglected: since $\hat{\Sigma}$ is a functional of $\hat{\mathbf{G}}$, the complete linear response contains the additional term $\hat{G}_W \star \hat{\Sigma}^{(1)} \star \hat{G}_W$, with $\hat{\Sigma}^{(1)}$ calculated in Section~\ref{vertex}. In the present section we first evaluate the contribution of the gradient term written above; the contribution of $\hat{\Sigma}^{(1)}$ is added at the end of the section.}

{\subsection{Gradient contribution}}

In the presence of a strong magnetic field, the dominant contribution arises from the Lowest Landau Level (LLL). In this regime, the Green's function factorizes as~\cite{Miransky2015, Gusynin1995}
\begin{equation}
\hat{G}_{\mathrm{LLL}}(p_\perp,p_0,p_3)
=
G_\perp(p_\perp)\,\tilde{G}(p_0,p_3)\,\mathcal{O}_-,
\end{equation}
where $\mathcal{O}_-$ is the LLL projection operator and
\begin{equation}
G_\perp(p_\perp)=2\exp\left(-\frac{p_\perp^2}{B}\right).
\end{equation}

The transverse momentum integration can be carried out independently, leading to an effective dimensional reduction~\cite{Hupfer2001, Miransky2015, Gusynin1996},
\begin{equation}
\int \frac{dp_1 dp_2}{(2\pi)^2} \;\longrightarrow\; \frac{B}{2\pi}.
\end{equation}
Thus, the problem reduces to a $(1+1)$-dimensional theory involving only $(p_0,p_3)$.


After integrating out the transverse degrees of freedom, the axial charge density becomes
\begin{align}
\rho_5(x)
=&
-\frac{E_3\, B}{4\pi}
\mathrm{Re}
\int \frac{dp_0\,dp_3}{(2\pi)^2}
\mathrm{tr}\bigg(\left[
\gamma_5 \mathcal{O}^- \, \mathcal{I}(p_0,p_3)
\right]^{<}\nn\\& \left( \partial_{p_0} \tilde{Q}^A \right)\bigg),
\end{align}
where $\mathcal{I}(p_0,p_3) = \boldsymbol{\tilde{G}}  \partial_{p_{[0}} \boldsymbol{\tilde{Q}} \boldsymbol{\tilde{G}}  \partial_{p_3]} \boldsymbol{\tilde{Q}}  \boldsymbol{\tilde{G}}$ and $\partial_{p_0} \tilde{Q}^A=1$.

A crucial simplification follows from the identity:
\begin{equation}\label{trickg}
\gamma_5 \mathcal{O}^- = -\mathrm{sign}(B)\,\gamma_0\gamma_3 \mathcal{O}^-,
\end{equation}
which ensures that only the longitudinal Dirac structure contributes.


The integrand reduces to terms involving derivatives of delta functions such as $\delta(p_0-H)$ and its derivatives, along with the derivative of the Hamiltonian,
\begin{equation}
\partial_{p_3}H = v_F \gamma_0\gamma_3 \cos\theta + v_F \gamma_0 \sin\theta.
\end{equation}

After performing integration by parts and combining all contributions, the expression simplifies to
\begin{equation}
I = \frac{1}{i\epsilon}
\int \frac{dp_3}{4\pi^2}
\mathrm{tr}\left[\gamma_5 \mathcal{O}_- (\partial_{p_3}H)\, n'(H)\right].
\end{equation}

At zero temperature, the derivative of the distribution function gives $n'(p_0)=-\delta(\mu-p_0)$, restricting the integral to the Fermi surface.

Using the projector properties and gamma-matrix algebra, one finds:
\begin{equation}\label{tr1}
\mathrm{tr}\left[\gamma_5 \mathcal{O}_- \partial_{p_3}H\right]
=
-2\,\mathrm{sign}(B)\, v_F \cos\theta.
\end{equation}

Using \eqref{tr1}, the general form of axial current density is:\begin{equation}\label{rho5general}
\rho_5 =  i \;
\frac{E_3|B|}{2}\;
I
\end{equation}
where:
\begin{equation}
I = \frac{\operatorname{sign}(B)\,\mu\cos\theta_F}{i\pi^2\epsilon\, v_F(2+2m+B)\sqrt{1-\cos^2\theta_F}},
\end{equation}
with
\begin{equation}
\cos\theta_F = 1 - \frac{\mu^2/v_F^2 - (m+B/2)^2}{2+2m+B}.
\end{equation}

The computation is in appendix \ref{integral detail}. Restoring the finite value of lattice spacing we come to $I = \frac{\operatorname{sign}(B)\,\mu\,a\,\cos\theta_F}{i\pi^2\epsilon\, v_F(2+2ma+Ba^2)\sqrt{1-\cos^2\theta_F}} $ with $\cos\theta_F = 1 - \frac{\mu^2a^2/v_F^2 - (ma+Ba^2/2)^2}{2+2ma+Ba^2} $.  For {$\mu a/v_F \gg ma+|B|a^2$} we obtain   
\begin{equation}
\rho_5(x)
=
\frac{1}{4\pi^2  \epsilon}\, \frac{1 - |M|^2}{1 + |M|^2}\, (\mathbf{E}\cdot\mathbf{B}).\label{rho5}
\end{equation}
with $\epsilon = \frac{u_0^2 n_{\mathrm{imp}} |B|}{4\pi v_F}$.
{Equation~(\ref{rho5}) is the low-energy, positive-doping
($\mu>0$), leading on-shell result.  Outside the hierarchy regime the constant
$\epsilon$ must be replaced consistently by
$\epsilon(\mu)=\epsilon/[\mathcal D\sqrt{1-|M|^2}]$, rather than kept at its
low-energy value.  The result also contains both lattice-band edge factors
$\Theta(\mu-E_{\min})\Theta(E_{\max}-\mu)$; a lower threshold alone is
insufficient.}

{ The factor $(1-|M|^2)/(1+|M|^2) = q$ in Eq.~(\ref{rho5}), with $|M|$ of Eq.~(\ref{eq:overlap}), collects the leading on-shell corrections to the singular integral: the level width entering it is $\epsilon_{\rm on} = \epsilon(\mu)(1+|M|^2)$ rather than $\epsilon$, which produces the denominator, while the numerator arises from the two kinematic factors $\sqrt{1-|M|^2}$ -- the axial-vertex matrix element $\langle u_+(\pm\theta_F)|\gamma_0\gamma_3|u_+(\pm\theta_F)\rangle = \pm v_F \sin\theta_F/\mu$ and the energy dependence of the width, {$\epsilon(\mu) = \epsilon/[\mathcal D\sqrt{1-|M|^2}]$} (Appendices \ref{app:source}, \ref{app:kernel}). In the hierarchy regime $|M|^2 \ll 1$ and $\mathcal D\simeq1$, this factor tends to unity, and the original expression is recovered.}


The electric current can be derived in a completely analogous manner. After reduction to the LLL sector, it takes the form
\begin{equation}
\begin{aligned}
J_k(x)
=
-\frac{ E_3\, B}{4\pi}\,\delta_{k3}
\mathrm{Re}
\int \frac{dp_0\,dp_3}{(2\pi)^2}
\mathrm{tr}&\Bigl(\left[
\mathcal{O}^- \, \mathcal{I}_k(p_0,p_3)
\right]^{<}\\
&\left( \partial_{p_3} \tilde{Q}^A \right)\Bigr).
\end{aligned}
\end{equation}

where $\mathcal{I}_k(p_0,p_3) =\boldsymbol{\tilde{G}} \partial_{p_{[0}} \boldsymbol{\tilde{Q}} \boldsymbol{\tilde{G}} \partial_{p_3]} \boldsymbol{\tilde{Q}} \boldsymbol{\tilde{G}}$ and \\
$\partial_{p_3} \tilde{Q}^A = -v_F \gamma_0 \gamma_3 \cos\theta - \gamma_0 v_F \sin\theta$.
Combine with \eqref{trickg}, we get:\begin{align}
J^k(x)=  &-\frac{v_F E_3 |B|\delta^{k3}}{4\pi} \operatorname{Re} \int \frac{dp_0 dp_3}{(2\pi)^2} \operatorname{tr} \Bigl(\bigl[\gamma_0\gamma_3O^- \, \boldsymbol{\tilde{G}} \partial_{p_{[0}} \boldsymbol{\tilde{Q}} \nn\\
& \boldsymbol{\tilde{G}} \partial_{p_3]} \boldsymbol{\tilde{Q}} \boldsymbol{\tilde{G}} \bigr]^{<} v_F\left(\gamma_3  \sin\theta - \cos\theta \right)\Bigr).
\end{align}
Similarly the details are in appendix \ref{integral detail}. We find only the longitudinal component ($k=3$) survives, reflecting the one-dimensional propagation of LLL states along the magnetic field direction.

A direct relation between the electric current and the axial charge density emerges,
define\begin{align}
    \rho_{x}= \bigl[&\alpha^3 O^- \, \boldsymbol{\tilde{G}}  \partial_{p_{[0}} \boldsymbol{\tilde{Q}}  \boldsymbol{\tilde{G}}  \partial_{p_3]} \boldsymbol{\tilde{Q}}  \boldsymbol{\tilde{G}} \bigr]^{<},
    \end{align}
and\begin{align}
    \rho_{5}(x) &= \frac{\operatorname{sign}(B) E_3 |B|}{4\pi} 
\;\mathrm{Re} \int \frac{d p_0 \, d p_3}{(2\pi )^2}
\, \mathrm{tr} [\rho_x],
\end{align}
and\begin{align}
    J^k(x) &= -v_F\operatorname{sign}(B)\delta^{k3}\frac{ E_3 |B|}{4\pi}\nn\\& \times\operatorname{Re} \int \frac{dp_0 dp_3}{(2\pi)^2}\text{tr}(\rho_x \left(\gamma_3 \sin\theta - \cos\theta \right)). 
\end{align}
{ After the low-energy, on-shell
projection used throughout this section, the valid relation is
\begin{equation}
 J^k=\delta^{k3}\operatorname{sign}(B)v_F\rho_5
 +O(\theta_F^2,ma,|B|a^2,\epsilon/\mu).
\end{equation}
Consequently the gradient (bare-bubble) current is
\begin{equation}
 J_3=E_3\frac{|B|v_F}{4\pi^2\epsilon}
 \frac{1-|M|^2}{1+|M|^2}\,
 \Theta(\mu-E_{\min})\Theta(E_{\max}-\mu),\qquad \mu>0.
\end{equation}
}

Restoring lattice spacing, and taking into account {$\mu a/v_F \ll 1$}, $Ba^2 \ll 1$ we obtain the following expression for the conductivity
\begin{align}
	\sigma_{zz} &= \;
	\frac{|B|v_F}{4 \pi^2 \epsilon}\,\frac{1 - |M|^2}{1 + |M|^2}\,\theta (|\mu| - v_F (m + |B| a/2))
	\label{eq:sigmagrad}
\end{align}
with  $\epsilon = \frac{u_0^2 n_{\mathrm{imp}} |B|}{4\pi v_F}$.

{ The exact on-shell factor $(1-|M|^2)/(1+|M|^2) = q$ has the same origin as in Eq.~(\ref{rho5}): the on-shell width $\epsilon_{\rm on} = \epsilon(\mu)(1+|M|^2)$ in the denominator and the two kinematic factors $\sqrt{1-|M|^2}$ (here from the velocity matrix element $\langle u_+(\pm\theta_F)|H'|u_+(\pm\theta_F)\rangle = \pm |E_+'|$ combined with the density of states, and from $\epsilon(\mu) = \epsilon/\sqrt{1-|M|^2}$) in the numerator. Thus the gradient contributions to both observables equal $q$ times the expressions with the bare constant $\epsilon$.}

The above results demonstrate that in the LLL regime the system effectively behaves as a collection of chiral one-dimensional modes. The appearance of the factor $\mathbf{E}\cdot\mathbf{B}$ in the axial charge density reflects the non-conservation of axial charge, i.e., the chiral anomaly. Furthermore, the proportionality between the electric current and the axial charge density shows that the current is carried by these chiral modes, propagating along the direction of the magnetic field.

{

\subsection{Vertex contribution and the complete result}

We now add the contribution of the second term of Eq.~(\ref{eq:linG}), inserting the complete $\tilde{G}^{(1)<} = [\tilde{G}_d]^{<} + \tilde{G}^{(0)R}\, \hat{\Sigma}^{(1)<}\, \tilde{G}^{(0)A}$ into the response formulas written above. Since the self-energy is momentum independent, $\partial_{p_0}\hat{Q}^{R(A)} = 1$ and $\partial_{p_3}\hat{Q}^{R(A)} = -H'$, and after the transverse reduction
\begin{align}
\rho_5 &= -i\,\frac{|B|}{2\pi}\int \frac{d\omega\, dp_3}{(2\pi)^2}\, {\rm tr}\bigl[\gamma_5 \mathcal{O}^-\, \tilde{G}^{(1)<}\bigr], \nn\\
J_3 &= +i\,\frac{|B|}{2\pi}\int \frac{d\omega\, dp_3}{(2\pi)^2}\, {\rm tr}\bigl[H'\, \tilde{G}^{(1)<}\bigr].
\label{eq:respred}
\end{align}
Both observables receive two contributions -- from the gradient term $\hat{G}_d$ (this is the calculation presented above in this section) and from the vertex term; the contributions of $\hat{\Sigma}^{(1)R(A)}$ are $O(\epsilon E_3/\mu)$ and lie in the null channel of the kernel. The required momentum integrals are: for the gradient term, from Appendix \ref{app:source} with the width $\epsilon_{\rm on}$,
\begin{equation}
\int \frac{dp_3}{2\pi}\, [\tilde{G}_d]^{<} = \frac{i E_3\, n'(\omega)}{2 \epsilon_{\rm on}}\, \frac{v_F \sin\theta_\omega}{\omega}\, \gamma_0\gamma_3 ,
\end{equation}
and for the vertex term, from the $\gamma_0\gamma_3$ eigenvalue of the kernel (\ref{eq:kernel}), (\ref{eq:spectrum}),
\begin{equation}
\int \frac{dp_3}{2\pi}\, \tilde{G}^{R}\, \gamma_0\gamma_3\, \tilde{G}^{A} = \frac{q}{g}\, \gamma_0\gamma_3 .
\end{equation}
The traces are evaluated with ${\rm tr}[\gamma_5\mathcal{O}^-\gamma_0\gamma_3] = -2\,{\rm sign}(B)$ for the axial density, and with the odd matrix elements $\langle u_+(\pm\theta_\omega)|H'|u_+(\pm\theta_\omega)\rangle = \pm|E_+'|$, $\langle u_+(\pm\theta_\omega)|\gamma_0\gamma_3|u_+(\pm\theta_\omega)\rangle = \pm v_F\sin\theta_\omega/\omega$ for the current; both vertices are odd and therefore match the odd structure of $\tilde{G}^{(1)<}$, so that the two Fermi points add. {Performing the $\omega$ integration with $n'(\omega)=-\delta(\omega-\mu)$, using $v_F\sin\theta_F/\mu=\sqrt{1-|M|^2}$, $\epsilon(\mu)=\epsilon/[\mathcal{D}\sqrt{1-|M|^2}]$ and the exact identity $\epsilon(\mu)|E_+'(\theta_F)|=v_F\epsilon$, and finally taking the low-energy normalization $\mathcal{D}\simeq1$,} one finds that the gradient terms contribute the factor $q$ and the vertex terms the factor {$q^2/(1-q)$}, in both observables alike:
\begin{align}
\rho_5 &= \frac{\mathbf{E}\cdot\mathbf{B}}{4\pi^2\epsilon}\Bigl[\, q + \frac{q^2}{1-q}\,\Bigr] = \frac{\mathbf{E} \cdot \mathbf{B}}{4\pi^2 \epsilon}\, \frac{1 - |M|^2}{2|M|^2}, \nn\\
\sigma_{zz} &= \frac{|B| v_F}{4\pi^2\epsilon}\Bigl[\, q + \frac{q^2}{1-q}\,\Bigr] = \frac{|B| v_F}{4\pi^2 \epsilon}\, \frac{1 - |M|^2}{2|M|^2},
\label{eq:exactobs}
\end{align}
where both expressions are implied to contain the factor $\theta\bigl(|\mu| - v_F (m + |B| a/2)\bigr)$ expressing the existence of the Fermi surface. {Within the leading on-shell approximation and the low-energy normalization $\mathcal{D}\simeq1$, Eq.~(\ref{eq:exactobs}) is the resummed result:} the first term of the bracket, $q$, is precisely the gradient contribution of the previous subsection, Eqs.~(\ref{rho5}), (\ref{eq:sigmagrad}), while the second term, $q^2/(1-q)$, is the contribution of the vertex correction. The power counting in $q$ is diagrammatic: the contribution with $n$ impurity rungs contains $n+1$ mixed $\tilde{G}^R\ldots\tilde{G}^A$ loops, and each loop traversal in the odd channel contributes one factor $q$ (one factor from the drive loop $\hat{\Sigma}_d \propto q$, one from the readout loop $\int \tilde{G}^R \gamma_0\gamma_3 \tilde{G}^A = (q/g)\gamma_0\gamma_3$, and $q$ per additional rung), so that the $n$-rung term equals {$q^{n+1}$} {and the vertex part sums to $\sum_{n\ge1} q^{n+1} = q^2/(1-q)$, the full series giving $\sum_{n\ge0} q^{n+1} = q/(1-q)$}. Thus the net effect of the vertex correction is the replacement, in the expressions with the bare constant $\epsilon$,
{
\begin{align}
 \epsilon_{5,{\rm resp}}&=\frac{2|M|^2}{1-|M|^2}\,\epsilon,
 &\tau_{5,{\rm resp}}&=\frac{1}{2\epsilon_{5,{\rm resp}}}.
 \label{eq:eps5}
\end{align}
This is the parameter inferred by rewriting Eq.~(\ref{eq:exactobs})
in the conventional anomaly-balance form.  }

Eq.~(\ref{eq:exactobs}) vanishes continuously at the gap edge $|M| \to 1$, where the two Fermi points merge and the band velocity vanishes. (This vanishing is a limiting statement of the approximation used: close to the gap edge, where the relaxation width ceases to be small compared to the distance of the Fermi level from the band edge, the Born approximation and the pinch expansion degrade, so that Eq.~(\ref{eq:exactobs}) applies for the Fermi level well outside the immediate vicinity of the gap.) Two further features of Eq.~(\ref{eq:exactobs}) deserve a comment. First, {within the same leading on-shell and low-energy approximation,} the relation $J_3 = {\rm sign}(B)\, v_F\, \rho_5$ holds including all ladder corrections: the resummation renormalizes the common response denominator, but not the proportionality between the current and the axial density, as required if the current is carried by the chiral modes. Second, {the algebraic identity $\rho_5=(\mathbf E\cdot\mathbf B)\tau_{5,{\rm resp}}/(2\pi^2)$ defines the response time $\tau_{5,{\rm resp}}$.}

In the continuum limit $a \to 0$ (at fixed $\mu$, $B$ and $m$) the lattice contributions to the overlap of Eq.~(\ref{eq:overlap}) vanish and $|M| \to v_F m/\mu$. For the massless fermions ($m = 0$) { $\epsilon_{5,{\rm resp}}$ vanishes}, and the chiral magnetic current becomes non-dissipative, in accordance with the exact chirality protection of the massless continuum theory; at finite mass the relaxation survives, being driven by the mass term alone. For $m = 0$ the finite chirality relaxation -- and hence the finite value of the conductivity in Eq.~(\ref{eq:exactobs}) -- is entirely the lattice effect encoded in the term $C(\theta)$ of the effective SSH Hamiltonian, in line with the Nielsen--Ninomiya mechanism discussed in the Introduction.

Let us also note that strictly at $T = 0$ the reduced one - dimensional channels are subject to Anderson localization at the length scale of the backscattering mean free path. The Drude - type expressions above refer to the samples shorter than this scale, or to the presence of a finite dephasing (caused, e.g., by a small finite temperature).

}
\section{Application: the semiclassical-to-quantum-limit crossover and the magnetoresistance reversal in ZrTe$_5$}
\label{applications}

The strong-field reduction used throughout this paper -- projection onto the lowest Landau level (LLL), leading to the effective 1D chain of Sec.~\ref{Hamiltonian} -- is itself only valid once the field is large enough that higher Landau levels are depopulated, $\ell_B \lesssim \sqrt{2}\,\ell_\mu$ with $\ell_\mu=\hbar v_F/\mu$. Solving for the field gives a lower bound
\begin{equation}
B_{\rm LLL}^{\rm min} = \frac{\mu^2}{2e\hbar v_F^2},
\label{eq:Bmin}
\end{equation}
which depends only on the chemical potential and the Fermi velocity along the field direction\zd{, and only weakly on the Dirac mass: for the massive dispersion the $n=1$ Landau level lies at $E_1=\sqrt{\Delta^2+2e\hbar v_F^2 B}$, so that the exact crossover condition $\mu=E_1$ gives $B_{\rm LLL}^{\rm min}=({\mu^2-\Delta^2})/({2e\hbar v_F^2})$, of which Eq.~(\ref{eq:Bmin}) is the small-mass limit. With the zero-field spectroscopic gap of ZrTe$_5$, $2\Delta\approx9.4$~meV~\cite{JiangZrTe5Mass2017}, the mass correction is below $0.3\%$ at $\mu\approx90$~meV.} Equation~(\ref{eq:Bmin}) marks a genuine crossover in the physics, not merely a technical convenience: for $B \lesssim B_{\rm LLL}^{\rm min}$ many Landau levels are populated and transport is in the semiclassical regime described by the heuristic dynamical-chiral-magnetic-effect picture -- an axial chemical potential $\mu_5$ builds up under $\mathbf{E}\cdot\mathbf{B}$ and relaxes diffusively, giving the familiar $\sigma_{\rm CME}=\sigma_0+a(T)B^2$ fit and the accompanying \emph{negative} magnetoresistance~\cite{CMEZrTe5}. For $B \gtrsim B_{\rm LLL}^{\rm min}$, only the LLL is occupied and the conductivity is instead given by the exact result of Eq.~(\ref{eq:exactobs}) of this paper, obtained directly from the Keldysh formalism without invoking $\mu_5$ as a separate equilibrium-like construct. Because $\sigma_{zz}\propto g(M)=(1-M^2)/(2M^2)$ is a monotonically decreasing function of $|M|$, and $|M|$ grows with $B$ once inside this regime (Eq.~\zd{(\ref{eq:Mexp})}), the present calculation predicts that the negative magnetoresistance of the semiclassical regime must reverse into positive magnetoresistance once $B$ exceeds $B_{\rm LLL}^{\rm min}$. The two calculations describe the \emph{same} chiral-magnetic transport coefficient in two different, non-overlapping field regimes, rather than two independent contributions to be added.

\zd{Throughout this section we use the explicit form of the backscattering amplitude of Eq.~(\ref{eq:overlap}) at the Fermi level. Expanding it in the small parameters $ma$, $\mu a/(\hbar v_F)$ and $a^2/\ell_B^2$, and restoring SI units, one finds
\begin{equation}
	|M| \approx \frac{\hbar v_F m}{\mu} + \frac{\mu a}{2 v_F \hbar} + \frac{v_F eB a}{2\mu} = \frac{\ell_\mu}{\ell_m} + \frac{a}{2\ell_\mu} + \frac{\ell_\mu a}{2 \ell_B^2},
	\label{eq:Mexp}
\end{equation}
where $\ell_m = 1/m$, $\ell_\mu = \hbar v_F/\mu$ and $\ell_B = \sqrt{\hbar/|eB|}$, so that the first term equals $\Delta/\mu$ with the Dirac mass $\Delta = \hbar v_F m$.}

\subsection{Quantitative check for ZrTe$_5$}

\begin{figure}[t]
\centering
\includegraphics[width=\columnwidth]{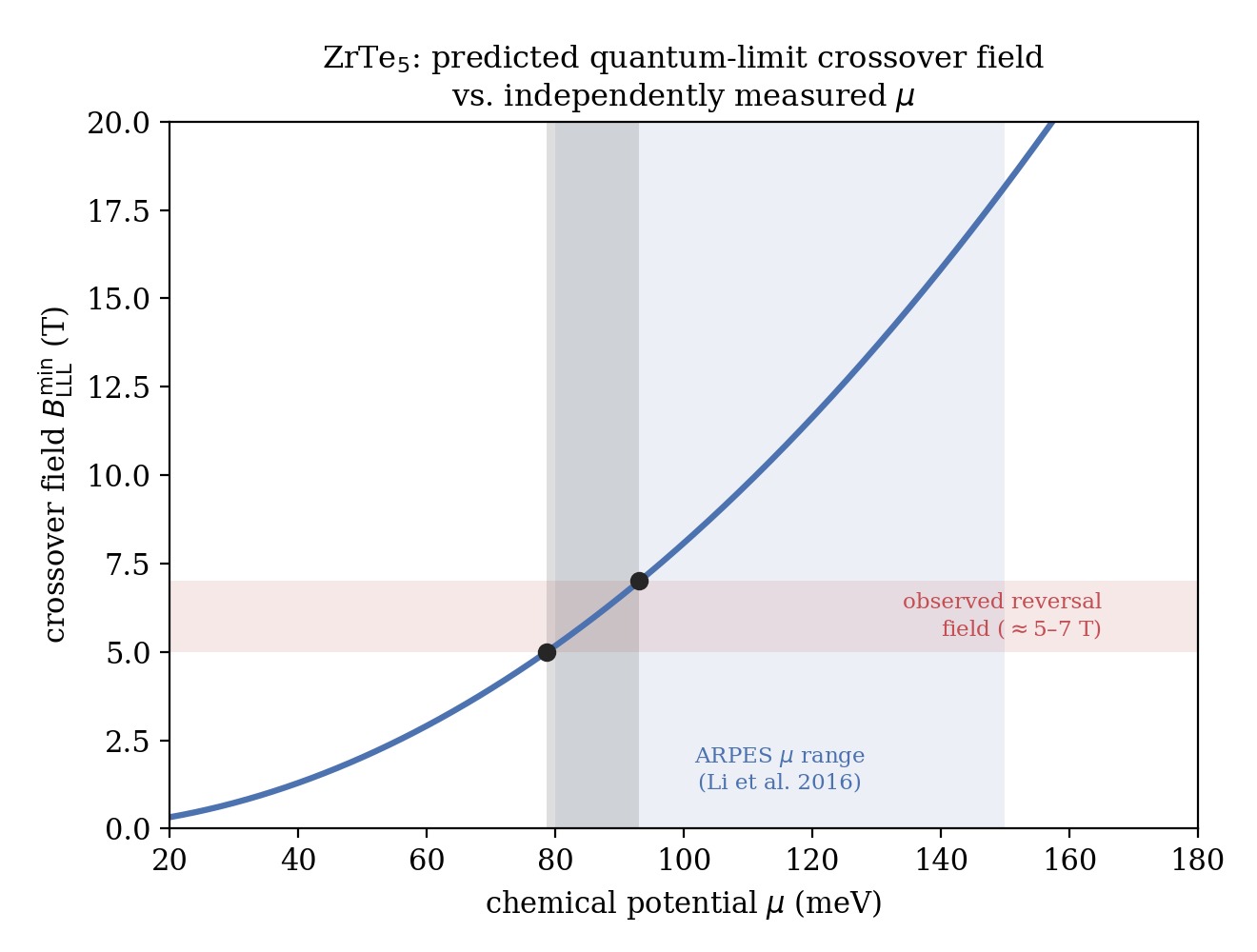}
\caption{Predicted quantum-limit crossover field $B_{\rm LLL}^{\rm min}(\mu)$ of Eq.~(\ref{eq:Bmin}) for ZrTe$_5$ (blue curve; $v_F=9.7\times10^5$~m/s, ARPES $a$-axis value~\cite{CMEZrTe5}), compared with the independently measured ARPES chemical potential range, $\mu=80$--$150$~meV (blue band), and with the field at which the raw low-temperature ZrTe$_5$ resistivity is observed to reverse from negative to positive magnetoresistance, $B\approx5$--$7$~T (red band)~\cite{CMEZrTe5}. The dark overlap marks the region where both independent constraints agree. \zd{As discussed in the text, in the anisotropic material the product of the transverse velocities $v_bv_c$ replaces $v_F^2$ in Eq.~(\ref{eq:Bmin}), so this comparison should be read as an order-of-magnitude consistency check.}}
\label{fig:validation}
\end{figure}

This gives a parameter-free way to check where the reversal should occur. Using the chain-direction ($a$-axis) Fermi velocity measured by ARPES, $v_F=6.4$~eV\AA$\simeq9.7\times10^5$~m/s -- the correct velocity for this geometry, since the CME measurement has $\mathbf{E}\parallel\mathbf{B}$ along the chain -- together with the ARPES estimate of the chemical potential, $\mu\sim100$~meV with a quoted sample-to-sample range of $80$--$150$~meV~\cite{CMEZrTe5}, Eq.~(\ref{eq:Bmin}) gives
\begin{equation}
B_{\rm LLL}^{\rm min} \approx 5\text{--}18~\text{T} \qquad (\mu = 80\text{--}150~\text{meV}).
\label{eq:BminZrTe5}
\end{equation}
The raw, low-temperature ($5$~K) magnetoresistance data for ZrTe$_5$ show negative magnetoresistance at low field, a minimum, and a reversal into positive magnetoresistance beginning at a field of approximately $5$--$7$~T~\cite{CMEZrTe5}. Inverting Eq.~(\ref{eq:Bmin}), a reversal in this range requires $\mu\approx79$--$93$~meV -- squarely inside the independently measured ARPES range quoted above. This is a nontrivial check: $\mu$ was not fit to the reversal field, it is an independently reported spectroscopic quantity, and the Dirac mass $\Delta=\hbar v_Fm$ \zd{enters Eq.~(\ref{eq:Bmin}) only at the negligible order $(\Delta/\mu)^2$}, so this comparison tests only the ($v_F,\mu$) sector of the theory. Figure~\ref{fig:validation} shows this comparison directly.

\zd{A caveat concerns the velocity entering Eq.~(\ref{eq:Bmin}). Within our isotropic model there is a single $v_F$, and it is the chain ($a$-axis) velocity that controls the transport quantities of the LLL regime, $M(B)$ and $\sigma_{zz}$, since the residual dispersion is along $\mathbf{B}$.  ARPES indicates that the out-of-plane ($b$-axis) velocity of ZrTe$_5$ is roughly an order of magnitude smaller than the in-plane ones~\cite{CMEZrTe5}. Therefore, the comparison of Fig.~\ref{fig:validation} should be read as an order-of-magnitude consistency check, whose precise position depends on the poorly characterized transverse velocities, rather than as a fully quantitative parameter-free agreement.}

\begin{figure}[t]
\centering
\includegraphics[width=\columnwidth]{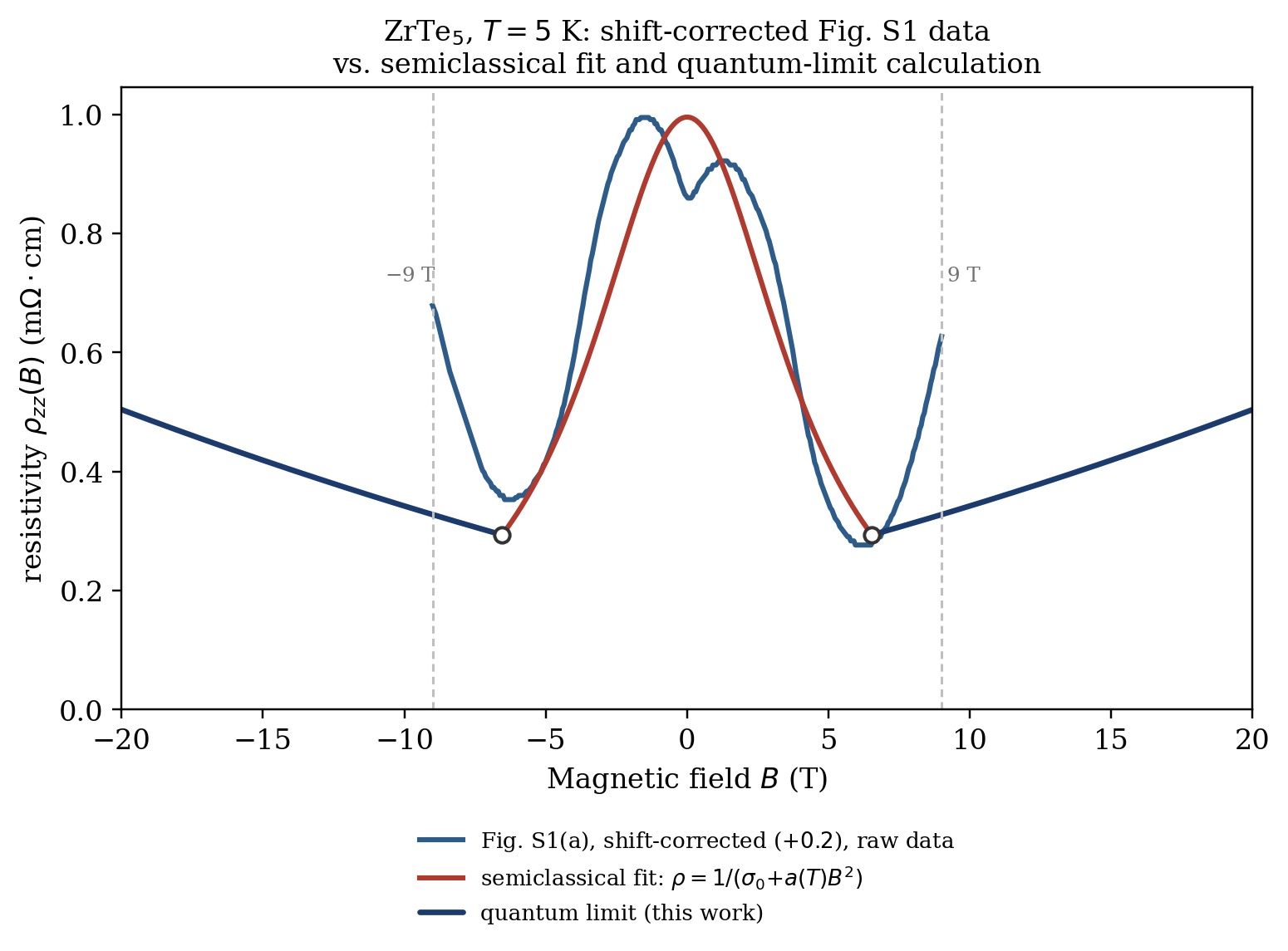}
\caption{ZrTe$_5$ at $T=5$~K: approximate shape of the measured resistivity (\zd{light blue}; read from Fig.~S1(a) of Ref.~\cite{CMEZrTe5}, not a digitization of the published curve\zd{; shifted upward by $0.2$~m$\Omega\,$cm to compensate the constant offset applied to the $5$~K curve of Fig.~S1(a) for clarity of presentation}), the semiclassical branch $\sigma_{\rm CME}=\sigma_0+a(T)B^2$ (red, $|B|<B_{\rm LLL}^{\rm min}$, coefficient fixed as described in the text), and the quantum-limit calculation of this paper (blue, $|B|>B_{\rm LLL}^{\rm min}$, Eq.~(\ref{eq:exactobs})), for $\mu=90$~meV ($B_{\rm LLL}^{\rm min}\approx6.5$~T)\zd{, $a=4.0$~\AA, and $\Delta\approx4.7$~meV (see text)}. The blue branch is anchored to the red one at $B_{\rm LLL}^{\rm min}$, not at $B=0$, since $M(B)$ and hence $\sigma_{zz}(B)$ are only defined once $B\ge B_{\rm LLL}^{\rm min}$ (see text). The measured data recovers faster, between $B_{\rm LLL}^{\rm min}$ and the $9$~T maximum field explored, than the asymptotic quantum-limit calculation; as discussed in the text, this region is expected to be a transition regime in which higher Landau levels are still being depopulated one by one, distinct from the pure-LLL asymptotics the present calculation describes, which should be approached only at higher fields.}
\label{fig:5Kcomparison}
\end{figure}

The semiclassical branch shown alongside our calculation in Fig.~\ref{fig:5Kcomparison} below is fixed independently, not fit to the data shown. Reference~\cite{CMEZrTe5} identifies the coefficient of its empirical fit $\sigma_{\rm CME}=\sigma_0+a(T)B^2$ with the dynamical chiral-magnetic conductivity of Fukushima, Kharzeev and Warringa,
\begin{equation}
a(T) = \frac{e^2}{\pi\hbar}\,\frac{3}{8}\,\frac{e^2}{\hbar c}\,\frac{v^3}{\pi^3}\,\frac{\tau_V}{T^2+\mu^2/\pi^2}, \qquad \tau_V^{-1}=\frac{\zd{\Delta_V}}{\hbar},
\label{eq:aT}
\end{equation}
which Ref.~\cite{CMEZrTe5} evaluates with the order-of-magnitude band parameters $\zd{\Delta_V}\simeq50$~meV, $\mu\sim100$~meV, $v\simeq c/300$ to reproduce the correct order of magnitude of $a(T)$, without claiming quantitative precision. \zd{We denote the energy scale entering Eq.~(\ref{eq:aT}) by $\Delta_V$ in order to distinguish it from the true Dirac mass $\Delta=\hbar v_F m$ of the fermions used throughout the previous sections. $\Delta_V$ is the phenomenological chirality-relaxation scale with which Ref.~\cite{CMEZrTe5} estimates the chirality-flip time $\tau_V$; it is identified there with the band gap only at the level of an order-of-magnitude estimate. The spectroscopically measured Dirac mass of ZrTe$_5$ is an order of magnitude smaller, $2\Delta\approx9.4$~meV~\cite{JiangZrTe5Mass2017}, and it is this $\Delta$, not $\Delta_V$, that enters the quantum-limit observables through Eq.~(\ref{eq:Mexp}).} We instead fix $a/\sigma_0$ directly from the two quantitative statements Ref.~\cite{CMEZrTe5} makes at $T=20$~K: the CME contribution equals the zero-field conductivity at $B=4$~T, and exceeds it by $\sim400\%$ at $B=9$~T. These give, respectively, $a/\sigma_0=1/16=0.0625$~T$^{-2}$ and $a/\sigma_0=4/81=0.0494$~T$^{-2}$; we use their average, $a/\sigma_0\approx0.055$~T$^{-2}$.

Figure~\ref{fig:5Kcomparison} compares the $T=5$~K data, not $T=20$~K, to the calculation, so we check whether the same coefficient still applies. By Eq.~(\ref{eq:aT}), $a(T)$ depends on temperature only through $T^2+\mu^2/\pi^2$ in the denominator. With $\mu\sim100$~meV, $\mu/\pi\approx32$~meV, while $k_BT\approx1.7$~meV at $20$~K and $0.43$~meV at $5$~K -- both negligible next to $\mu/\pi$. Consequently
\begin{equation}
\frac{a(5\,{\rm K})}{a(20\,{\rm K})} = \frac{(k_B\cdot20\,{\rm K})^2+(\mu/\pi)^2}{(k_B\cdot5\,{\rm K})^2+(\mu/\pi)^2} \approx 1.003,
\end{equation}
i.e.\ the two coefficients agree to better than $1\%$ within Eq.~(\ref{eq:aT}), and the $T=20$~K value of $a/\sigma_0$ can be used at $T=5$~K without adjustment.

\zd{The quantum-limit branch of Fig.~\ref{fig:5Kcomparison} requires, besides $\mu$ and $v_F$, the lattice spacing and the Dirac mass entering $M(B)$ of Eq.~(\ref{eq:Mexp}). It is drawn with the $a$-axis lattice constant $a\approx4.0$~\AA\ and with $\Delta\approx4.7$~meV, i.e.\ the zero-field spectroscopic gap $2\Delta\approx9.4$~meV of ZrTe$_5$ measured by Landau-level (magneto-infrared) spectroscopy~\cite{JiangZrTe5Mass2017}, consistent with the small-gap massive-Dirac picture of the magnetotransport in this material~\cite{ZhengZrTe5Transition2017}. This directly measured gap -- rather than the order-of-magnitude value $\Delta_V\simeq50$~meV used in Ref.~\cite{CMEZrTe5} to estimate $\tau_V$ in Eq.~(\ref{eq:aT}) -- is the appropriate mass here, because $\Delta$ enters the quantum-limit observables only through the backscattering amplitude of Eq.~(\ref{eq:Mexp}). The choice matters quantitatively. For $\mu=90$~meV the three terms of Eq.~(\ref{eq:Mexp}) are $\Delta/\mu\approx0.052$, $a/(2\ell_\mu)\approx0.028$ and $\ell_\mu a/(2\ell_B^2)\approx0.0022\,B/{\rm T}$, so that $|M|$ grows from $\approx0.094$ at $B_{\rm LLL}^{\rm min}\approx6.5$~T to $\approx0.12$ at $20$~T, and the resistivity $\rho_{zz}\propto 2|M|^2/(1-|M|^2)$ rises by a factor of $\approx1.7$ over this range -- this is the blue branch shown. Had the value $\Delta_V\simeq50$~meV been used for the Dirac mass instead, the field-independent term $\Delta/\mu\approx0.56$ would dominate $M(B)$ and the branch would be almost flat (a rise of only $\approx15\%$ up to $20$~T), while for $\Delta=0$ the resistivity would nearly triple over the same range. }

It is worth mentioning that  Ref.~\cite{CMEZrTe5} subtracts a misalignment-induced background from the raw magnetoresistance data before presenting Fig.~2 in~\cite{CMEZrTe5}, on the grounds that this contribution is due to an imperfect alignment between the magnetic field and the current. We are not fully convinced that this background subtraction procedure is appropriate, and \zd{in the comparisons drawn here we therefore used the original raw data of Fig.~S1 of Ref.}~\cite{CMEZrTe5}. One can see that our results qualitatively explain the growth of the resistivity at large values of magnetic field. However, we do not observe quantitative coincidence.  

\subsection{Other Dirac semimetals}

\zd{Both Cd$_3$As$_2$ and Na$_3$Bi differ from ZrTe$_5$ in one important respect: their Dirac points are protected by a crystalline rotational symmetry ($C_4$ and $C_3$, respectively), and this symmetry survives when the magnetic field is applied along the node-separation axis -- the geometry to which our model applies directly (in the actual experiments the field is not necessarily aligned with this axis, in which case the effective parameters interpolate between the crystal axes). The Dirac mass therefore vanishes, $\Delta=0$, and the backscattering amplitude of Eq.~(\ref{eq:Mexp}) is exhausted by its two lattice terms, $|M|\approx a/(2\ell_\mu)+\ell_\mu a/(2\ell_B^2)$, where $a$ is the lattice period along the node axis and $\ell_\mu=\hbar v_z/\mu$ is built from the velocity $v_z$ along that axis. We stress that for Cd$_3$As$_2$ the expansion leading to Eq.~(\ref{eq:Mexp}) is thereby used at the edge of its formal domain of validity ($\ell_\mu\sim a$, see below), so the estimates given here are qualitative.

For Cd$_3$As$_2$ the transverse (in-plane) velocity is large, $v_\perp\approx1.3\times10^6$~m/s~\cite{NeupaneCd3As2_2014}, so the crossover field of Eq.~(\ref{eq:Bmin}) (with $v_F^2\to v_\perp^2$, in accord with the discussion of the transverse velocities above) is $B_{\rm LLL}^{\rm min}\approx4$--$18$~T for $\mu=100$--$200$~meV -- reachable in the laboratory. The transport beyond the crossover is, however, controlled by the small node-axis velocity $v_z\sim10^5$~m/s and by the antifluorite sub-cell spacing $a\approx6.3$~\AA~\cite{Ali2014Cd3As2Crystal}: for $\mu\sim150$~meV one finds $\ell_\mu\approx4.4$~\AA, hence $a/(2\ell_\mu)\approx0.7$, while the field-dependent term is only $\ell_\mu a/(2\ell_B^2)\approx2\times10^{-4}\,B/{\rm T}$. The amplitude $|M|$ is thus large but almost field-independent: the model predicts a sizeable overall suppression of the quantum-limit conductivity, $(1-|M|^2)/(2|M|^2)\approx0.5$, but an essentially flat $\rho_{zz}(B)$, with the logarithmic slope ${\rm d}\ln\rho_{zz}/{\rm d}B\sim10^{-3}$~T$^{-1}$. Even taken at face value, our formulas therefore predict no observable magnetoresistance reversal in Cd$_3$As$_2$ at laboratory fields, consistent with the absence of any reported reversal~\cite{LiCd3As2NegativeMR2016}.

For Na$_3$Bi the situation is intermediate. The samples in which the chiral-anomaly-induced negative magnetoresistance was reported have a small chemical potential, $\mu\approx30$~meV~\cite{XiongNa3BiChiralAnomaly2015}; with the in-plane velocities $v_\perp\approx(2.5$--$3.7)\times10^5$~m/s this gives $B_{\rm LLL}^{\rm min}\approx5$--$11$~T. Along the node ($c$) axis, $v_z\approx10^5$~m/s and $c\approx9.7$~\AA\ give $\ell_\mu\approx24$~\AA, so that $\ell_\mu/c\approx2.5$ and the expansion of Eq.~(\ref{eq:Mexp}) is marginally under control. Then $|M|$ just above the crossover is $\approx0.2$, dominated by the field-independent term $c/(2\ell_\mu)\approx0.20$, with the field-dependent term $\approx1.8\times10^{-3}\,B/{\rm T}$; the predicted positive magnetoresistance beyond the crossover grows slowly, ${\rm d}\ln\rho_{zz}/{\rm d}B\approx2\%$ per tesla. Such a shallow reversal, setting in at $5$--$11$~T with a percent-level slope, would be difficult to separate from the misalignment and current-jetting backgrounds that complicate longitudinal-magnetoresistance measurements; the absence of a clearly reported reversal in Na$_3$Bi~\cite{XiongNa3BiChiralAnomaly2015} is compatible with this expectation. In both materials the pronounced crossover seen in ZrTe$_5$ is thus predicted to be absent or strongly muted -- in Cd$_3$As$_2$ because the large field-independent lattice term dominates $|M|$, in Na$_3$Bi because all contributions to $|M|$ remain small and the field dependence beyond the crossover is weak.}
\zd{\subsection{Related high-field observations in Weyl semimetals and in the ultra-quantum limit}

The reversal of the longitudinal magnetoresistance at strong magnetic field, predicted here for the Dirac semimetals, has direct experimental counterparts in the Weyl semimetals, where the two chiralities are separated in momentum space. In TaAs the longitudinal resistivity remains small and nearly field-independent up to $\sim 50$~T and then increases by two orders of magnitude; this behavior was attributed to the gap opening in the chiral lowest Landau levels once the magnetic field couples the Weyl fermions of opposite chirality, i.e. to the annihilation of the Weyl nodes in the extreme quantum limit \cite{RamshawTaAs2018}. In TaP the magnetic tunneling between the W1 Weyl pockets gaps their lowest Landau bands and produces a sharp sign reversal of the Hall resistivity at the field determined by the momentum-space separation of the nodes \cite{ZhangTaP2017}. These observations realize the same mechanism that controls the backscattering amplitude $|M|$ of the present paper: the field-induced mixing of the two chiralities gaps the chiral mode and destroys the anomaly-protected conduction. The difference is quantitative: for well-separated Weyl nodes the mixing turns on sharply when $\ell_B^{-1}$ becomes comparable to the node separation, while in the Dirac case considered here the two chiralities reside at the same point of the Brillouin zone, and $|M|$ grows smoothly with $B$, cf. Eq.~(\ref{eq:Mexp}). In ZrTe$_5$ itself, pulsed-field experiments deep in the ultra-quantum limit reveal a further sharp increase of the resistivity above $\sim 25$~T, interpreted as an interaction-induced dynamical generation of the Dirac mass \cite{LiuZrTe5UQL2016}, as well as a metal-insulator transition attributed to the magnetic freeze-out of the carriers on ionized impurities \cite{GourgoutZrTe5Freezeout2022}. Both mechanisms enhance the chirality mixing and act in the same direction as the single-particle mechanism discussed here; they are complementary to it rather than alternative, and their relative weight should depend on the sample quality and the carrier density. Let us finally note that the sign of the longitudinal magnetoresistance in the quantum limit is known to depend on the type of the scattering potential: for screened charged impurities positive longitudinal magnetoresistance in the quantum limit was obtained in Ref.~\cite{GoswamiPixleyDasSarma2015}, while the present paper treats the complementary case of short-range impurities, for which the positive magnetoresistance arises from the field dependence of the chirality-flip rate.}

\section{Conclusion}

In the present work we study negative magnetoresistance of Dirac semimetals. We consider the typical tight - binding model of Dirac semimetal with the idealized rectangular crystal lattice. At strong enough magnetic field the dynamics is reduced to that of the one - dimensional lattice model. The latter is a kind of the SSH model. 

 We use Keldysh technique of non - perturbative quantum field theory in order to calculate electric conductivity. We are interested in the behavior of the conductivity at small temperatures, where the effects of scattering on impurities dominates over interactions due to exchange by phonons. Coulomb interactions may still remain relevant, but we do not consider their effect in the present paper. The quasiparticle self-energy arising from elastic impurity scattering is explicitly evaluated. Subsequently, the Keldysh-Green's function formalism is employed to derive analytical expressions for both axial charge density and electric conductivity.

We demonstrated that {within the leading LLL, weak-disorder, low-energy approximation} the transport formulations for the $(3+1)$-dimensional Dirac semimetal manifest the functional equivalence observed in the corresponding $(1+1)$-dimensional configuration, which can be found in \cite{BOHRA2026116276}. The physical mechanism underlying this phenomenon stems from the effective Hamiltonian in the lowest Landau level approximation. We assume that the magnetic field is still sufficiently small to provide the smallness of the magnetic flux through the lattice cell compared to the elementary magnetic flux, that is magnetic length is much larger than the lattice spacing $\ell_B = \sqrt{\frac{\hbar}{|eB|}} \gg a$. Chemical potential (i.e. difference between Fermi energy and the level of Dirac point) is also assumed to be small compared to the lattice cutoff $\sim \frac{\hbar}{a}v_F$, i.e. length scale $\ell_\mu = \frac{\hbar v_F}{\mu} \gg a$. Strong magnetic field limit means the hierarchy of lengths $a \ll \ell_B \le \sqrt{2} \,\ell_\mu$. {The corresponding field in tesla is material- and density-dependent and cannot be fixed universally. However, we expect that for certain Dirac semimetals this regime is achieved  at around  $5$ T.}
	
 In this case we are able to consider dynamics in the plane orthogonal to the direction of magnetic field using effective low energy continuum field theory - because this dynamics is reduced to the only Landau level - the LLL, and magnetic length is much larger than the lattice spacing. We cannot consider such simplification in direction of magnetic field because there the covariant momentum {$\theta = p_3-|e|E_3t$} contains a term depending on time. Therefore, we cannot replace $\frac{1}{a}{\rm sin} \, (\theta a) $ by $\theta$ in general case for any $t$. The dynamics contains the domain of sufficiently large $\theta$ relevant for the chirality relaxation. When this mechanics is taken into account properly, the remaining calculation of the observed quantities ($\rho_5$ and $J$) is reduced to that of close to the Fermi surface, where we already may, in principle, apply the effective low energy continuum theory, as done in  \cite{Abramchuk_2026}.

The main result of \cite{Abramchuk_2026} (obtained using the effective continuum field theory) is that the electric conductivity is equal (up to a dimensional constant) to the response of axial charge density to  electric field. This result is reproduced in our work for sufficiently small lattice spacing $a$.  
	
{However, in \cite{Abramchuk_2026} a mistake was present in calculation of electric conductivity - the contribution of response of self energy to electric field was disregarded. In the present paper we demonstrate that this contribution modifies the value of conductivity presented in \cite{Abramchuk_2026} by factor $
\frac{1 - |M|^2}{2|M|^2}$, where 
\begin{equation}
	|M| \approx \frac{\hbar v_F m}{\mu} + \frac{\mu a}{2 v_F \hbar} + \frac{v_F eB a}{2\mu} = \frac{\ell_\mu}{\ell_m} + \frac{a}{2\ell_\mu} + \frac{\ell_\mu a}{2 \ell_B^2}.
	\label{eq:MhierC}
\end{equation}
with $\ell_m = 1/m$. \zd{(This expression was already introduced and used in Sec.~\ref{applications}, Eq.~(\ref{eq:Mexp}).)} For the realistic systems the second term in this sum is negligible while  
\begin{equation}
(a/\ell_B)^2 \ll a/\ell_\mu \label{cond}
\end{equation}
 holds for the majority of realistic systems at the values of magnetic field that are not much larger than $\sim 10$ T. Then only the first term in Eq. (\ref{eq:MhierC}) survives: $M = \frac{\Delta}{\mu}$. }
 
   The	 obtained results for the axial charge density $\rho_5$ \eqref{rho5} can be written in the SI units as follows
\begin{align}
    \rho_5 = \frac{e^2 (E_3B)}{4\pi^2 \hbar \epsilon}\frac{1 - |M|^2}{2|M|^2},
\end{align}
In the similar way for the electric conductivity we obtain in the SI units
\begin{align}
	\sigma_{zz} = \frac{|e|^3 |B| v_F}{4\pi^2 \hbar \epsilon} \frac{1 - |M|^2}{2|M|^2}
\end{align}
with
\begin{equation}
	\epsilon = \frac{u_0^2 n_{\mathrm{imp}} |eB|}{4\pi v_F \hbar^2} \label{disrate}
\end{equation}
{These SI expressions are the fixed-$\mu$, positive-doping,
low-energy response formulas and inherit the two band-edge step functions.
Here $\epsilon$ is the low-energy single-particle half-width.}
{At fixed $\mu$, and while $M$ is nearly field independent, the explicit
factor $|B|$ in $\sigma_{zz}$ cancels the Born width
$\epsilon\propto|B|$, producing a plateau within the model. } However, when magnetic fields are extra large  the last term of Eq. (\ref{eq:MhierC}) becomes of the same order as the first one, which results in decreasing value of conductivity (as a function of magnetic field). Such a dependence of conductivity on magnetic field is observed experimentally at small temperatures in $ZrTe_5$ (see \cite{CMEZrTe5}, FIG. S1 of ArXiV version).

 Notice that the above expressions disregard tunneling phenomenon that occurs when Fermi energy is within the gap, and cannot be calculated using the applied methodology. The corresponding probabilities contain factor that exponentially suppresses the effect due to nonzero gap $\sim \hbar v_F m + |eB|av_F/2$.

In order to investigate magnetoresistance for smaller magnetic fields (supposedly, smaller than about $  5  T$) we should take into account the higher Landau levels (HLL). In the framework of effective continuum field theory this has been done in \cite{ABRAMCHUK2026113374}.  Also this  may shed light on the other properties of Dirac semimetals. The interesting question is role of inter - electron interactions. Coulomb interactions may be strong and unlike exchange by phonons remain relevant even at very small temperatures. Taking them into account requires numerical simulations. Interactions with phonons come into the game at nonzero temperature. There are also some other effects related to finite temperature that worth to be considered. 

Besides, our approach may be  extended to the other systems, in particular, to quark matter  \cite{PhysRevD.52.4747,KHARZEEV2006260}, where the lattice regularization is a relevant way of study. The proposed approach may allow us to investigate directly the real time dynamics of quark matter in the the presence of  external fields. In particular, we expect that extension of the proposed methodology to investigation of non - dissipative transport effects may be relevant. Applications of the given study to description of  heavy ion collisions are expected it\cite{doi:10.1142/S0217751X09047570}. We postpone these and other  questions to future work.

\appendix

\section{Eigenstates of the transverse Hamiltonian $H_\bot$}

\label{Hamiltonianlattice}

Starting from the lattice Hamiltonian with Peierls substitution {$k_i \to \Pi_i = k_i + A_i$ for an electron of charge $q=-1$}
under symmetric gauge: $A_x=-\frac{B}{2}\,y, A_y=\frac{B}{2}\,x, A_z=0.$:
\begin{align}
    H_\perp& = v_F(\alpha_1 \sin\Pi_x + \alpha_2 \sin\Pi_y)
    \nn\\&+ v_F \gamma_0\bigl[(1-\cos\Pi_x)+(1-\cos\Pi_y)\bigr], 
\end{align}
where $\Pi_x = k_x -\frac{B}{2}\,y$ and $\Pi_y = k_y+\frac{B}{2}\,x$. In this appendix, we replace the lattice constant with unity. The transverse complex coordinate is written as:
\begin{align}
   & f=x+iy,\qquad \bar f=x-iy,\nn\\&
\partial_f=\tfrac12(\partial_x-i\partial_y),\qquad
\partial_{\bar f}=\tfrac12(\partial_x+i\partial_y).
\end{align}

The canonical commutation relation between $\Pi_x$ and $\Pi_y$ follows directly from
the gauge structure:
\begin{equation}
    [\Pi_x,\, \Pi_y] = {-iB}.
\end{equation}
This is algebraically identical to a harmonic oscillator with $[x, p]=i$.
Define ladder operators
\begin{equation}
    a = \frac{\Pi_x - i\Pi_y}{\sqrt{2B}},
    \qquad
    a^\dagger = \frac{\Pi_x + i\Pi_y}{\sqrt{2B}},
    \qquad [a,\, a^\dagger] = 1.
\end{equation}
Landau levels are labeled by $n = a^\dagger a = 0,1,2,\ldots$
The Lowest Landau Level (LLL) is the ground state defined by
\begin{equation}
    a\,|\psi_0\rangle = 0.
\end{equation}

Prove of it, start from the square of the linear part:\begin{align}
    H_{lin}^2 &= v_F^2\left(\gamma_0\gamma_1\Pi_x + \gamma_0\gamma_2\Pi_y\right)\left(\gamma_0\gamma_1\Pi_x + \gamma_0\gamma_2\Pi_y\right)\nn\\&= v_F^2\left[\Pi_x^2 + \Pi_y^2 - \gamma_1\gamma_2[\Pi_x,\Pi_y]\right]\nn\\&= {v_F^2\left[\Pi_x^2 + \Pi_y^2 + iB\gamma_1\gamma_2\right]}.
\end{align}
the eigenvalue of $i\gamma_1\gamma_2$ is $\pm 1$, so:\begin{align}
    H_{lin}^2\ket{\Psi_n} = {v_F^2\left[2B\left(a^\dagger a+\frac{1}{2}\right) + Bs\right]\ket{\Psi_n}},
\end{align}
where $s=\pm 1$. Notice it will have the same form of harmonic oscillator, so it will imply: $ a\,|\psi_0\rangle = 0.$

{Thus the eigenvalue of $H_{lin}^2$ is}\begin{align}
    {\lambda^2 = v_F^2 B(2n+1+s)}.
\end{align}
there are four eigenstates, for $s=1$, it implies:\begin{align}
    \chi_{+,+1} = \begin{pmatrix}1\\0\\0\\0\end{pmatrix}, \qquad 
\chi_{-,+1} = \begin{pmatrix}0\\0\\1\\0\end{pmatrix},
\end{align}
for $s=-1$:
\begin{align}
    \chi_{+,-1} = \begin{pmatrix}0\\1\\0\\0\end{pmatrix}, \qquad 
\chi_{-,-1} = \begin{pmatrix}0\\0\\0\\1\end{pmatrix},
\end{align}

For $H_\perp$, we have $H_\perp\ket{\psi_n}=E_\perp\ket{\psi_n}$:\begin{equation}
  a\,\psi_{n}=\sqrt{n}\,\psi_{n-1},\qquad
  a^{\dagger}\psi_{n}=\sqrt{n+1}\,\psi_{n+1}.
  \label{eq:aaction}
\end{equation}
Recall we have:\begin{align}
    \Pi_x^2+\Pi_y^2=\left[-\nabla^2+\frac{B^2}{4}(x^2+y^2)\right]{+}B\,L_z,
\end{align}
where \begin{align}
    L_z=-i(x\partial_y-y\partial_x) \qquad \nabla^2=\partial_x^2+\partial_y^2,
\end{align}
also we have the commute relation:\begin{align}
    [H_{\rm osc},L_z]=\Big[-\nabla^2+\tfrac{B^2}{4}r^2,\,L_z\Big]=0.
\end{align}
where $r=\sqrt{x^2+y^2}$, define:\begin{align}
&a_x=\frac{\sqrt B}{2}x+\frac{1}{\sqrt B}\partial_x,\qquad a_y=\frac{\sqrt B}{2}y+\frac{1}{\sqrt B}\partial_y,\nn\\&
    a_+=\frac{1}{\sqrt2}(a_x-ia_y),\qquad a_-=\frac{1}{\sqrt2}(a_x+ia_y).
\end{align}
Hence:\begin{align}
&\Pi_x^2+\Pi_y^2=H_{\rm osc}{+}BL_z\nn\\=&B((a_+^\dagger a_+))+(a_-^\dagger a_-)+1){+}B((a_+^\dagger a_+)-(a_-^\dagger a_-))\nn\\=&2B({(a_+^\dagger a_+)}+\tfrac12).
\end{align}
it is easy to find ${a_+}=ia$, {therefore,  $a_+^\dagger a_+=a^\dagger a$ (i.e. $a$ and $a_+$ are isomorphic).}
We will also use the guiding-center pair, mutually commuting with $a,a^\dagger$:
\begin{equation}
{b=\sqrt{\frac{2}{B}}\,\partial_{\bar f}+\sqrt{\frac{B}{8}}\,f,\qquad
b^\dagger=-\sqrt{\frac{2}{B}}\,\partial_f+\sqrt{\frac{B}{8}}\,\bar f},
\label{eq:bdef}
\end{equation}
\begin{align}
    &[b,b^\dagger]=1,\quad
[a,b]=[a,b^\dagger]=[a^\dagger,b]=[a^\dagger,b^\dagger]=0,\nn\\&
[H_\perp,b]=[H_\perp,b^\dagger]=0 .
\end{align}
It is easy to verify: $b={a_-}$.  The degeneracy of Landau levels refers to the angular momentum, which is also influenced by the magnetic fields.

The two ladder operators generate the complete orbital basis
\begin{equation}
\phi_{n,m}(f,\bar f)=\frac{(a^\dagger)^n}{\sqrt{n!}}\,\frac{(b^\dagger)^m}{\sqrt{m!}}\,
\phi_{0,0}(f,\bar f),\qquad n,m\ge0,
\label{eq:phinm}
\end{equation}
with the lowest-weight state annihilated by both $a$ and $b$:
\begin{equation}
a\,\phi_{0,0}=0,\qquad b\,\phi_{0,0}=0 .
\label{eq:annih}
\end{equation}
Hence
\begin{equation}
\phi_{0,0}(f,\bar f)=\left(\frac{B}{2\pi}\right)^{1/2}e^{-\frac{B}{4}|f|^2},
\end{equation}
The degeneracy tower at $n=0$ is
\begin{equation}\label{eq:phi0m}
\phi_{0,m}(f,\bar f)={N_m\,\bar f^{\,m}\,e^{-\frac{B}{4}|f|^2}},
\end{equation}
where \begin{align}
N_m=\sqrt{\frac{1}{\pi\,m!}\left(\frac{B}{2}\right)^{m+1}},
\end{align}
and the higher levels follow from \eqref{eq:phinm} in associated-Laguerre form
\begin{align}
   &\phi_{n,m}(f,\bar f)=e^{-\frac{B}{4}|f|^2}\times\nn\\&
\begin{cases}
{(-i)^n \sqrt{\frac{B}{2\pi} \frac{n!}{m!}} \left(\sqrt{\frac{B}{2}} \bar f\right)^{m-n} L_n^{m-n}\left(\frac{B}{2}|f|^2\right)}
 & m\ge n,\\[6pt]
{(-1)^mi^n \sqrt{\frac{B}{2\pi} \frac{m!}{n!}} \left(\sqrt{\frac{B}{2}} f\right)^{n-m} L_m^{n-m}\left(\frac{B}{2}|f|^2\right)} 
 & m<n.
\end{cases}
\label{eq:laguerre} 
\end{align}
In momentum space, we define the transverse momentum and its complex
combination
\begin{align}
   & \mathbf k_\perp=(k_x,k_y),\qquad k_\perp^2=k_x^2+k_y^2,\nn\\&
g=k_x+i k_y,\qquad \bar g=k_x-i k_y,
\end{align}
and the two-dimensional Fourier transformation:
\begin{equation}
\tilde\phi(\mathbf k_\perp)=\int D^2 r\;e^{-i\mathbf k_\perp\cdot\mathbf r}\,
\phi(f,\bar f),\qquad \mathbf k_\perp\cdot\mathbf r=k_x x+k_y y .
\end{equation}
in the LLL case, the Gaussian transforms into a Gaussian:
\begin{align}
    \tilde\phi_{0,0}(\mathbf k_\perp)
&=\left(\frac{B}{2\pi}\right)^{1/2}\!\int D^2 r\;e^{-i\mathbf k_\perp\cdot\mathbf r}
\,e^{-\frac{B}{4}|f|^2}
\nn\\&=\left(\frac{B}{2\pi}\right)^{1/2}\frac{4\pi}{B}\,e^{-k_\perp^2/B}
\nn\\&=\left(\frac{8\pi}{B}\right)^{1/2}e^{-k_\perp^2/B},
\label{eq:FT00}
\end{align}
hence:\begin{align}\label{eigenstate0}
    \tilde\phi_{0,m}(\mathbf k_\perp)
&={N_m\!\int d^2 r\;e^{-i\mathbf k_\perp\cdot\mathbf r}\,\bar f^{\,m}\,e^{-\frac{B}{4}|f|^2}}
\nn\\&={N_m\,\frac{4\pi}{B}\left(\frac{-2i}{B}\right)^{\!m}\!\bar g^{\,m}\,e^{-k_\perp^2/B}},
\end{align}
the general forms of the eigenfunctions are:
\begin{align}\label{eigenfunc}
    &\tilde{\phi}_{n,m}(\mathbf{k}_\perp) = e^{-\frac{k_\perp^2}{B}}\times\nn\\& \begin{cases}
{(-1)^n(-i)^{m}\! \sqrt{\frac{8\pi}{B} \frac{n!}{m!}} \left(\sqrt{\frac{2}{B}} \bar g\right)^{m-n} L_n^{m-n}\left(\frac{2}{B}k_\perp^2\right)} & m \geq n \\
{(i)^{m} \sqrt{\frac{8\pi}{B} \frac{m!}{n!}} \left(\sqrt{\frac{2}{B}} g\right)^{n-m} L_m^{n-m}\left(\frac{2}{B}k_\perp^2\right)} & m < n.
\end{cases}
\end{align}
It is convenient to introduce the spin raising/lowering combinations in
Dirac space,
\begin{equation}
  \alpha^{\pm}=\tfrac{1}{2}\big(\alpha_{1}\pm i\alpha_{2}\big),
  \label{eq:alphapm}
\end{equation}
where $\alpha_i=\gamma_0\gamma_i$, so that,
\begin{equation}
  H_{lin}
  = v_{F}\sqrt{2B}\,\big(\alpha^{+}a+\alpha^{-}a^{\dagger}\big).
  \label{eq:Hlin_ladder}
\end{equation}
With the
convention that $\chi_{+}$ is annihilated by $\alpha^{-}$ and raised by
$\alpha^{+}$ (and conversely for $\chi_{-}$),
\begin{align}
    &\alpha^{-}\chi_{+,+1}=\chi_{-,-1},\quad
\alpha^{+}\chi_{-,-1}=\chi_{+,+1},\nn\\&
\alpha^{-}\chi_{-,+1}=\chi_{+,-1},\quad
\alpha^{+}\chi_{+,-1}=\chi_{-,+1},
\label{eq:alphaaction}
\end{align}
evaluating the matrix of $H_{lin}$ in the ordered basis
\begin{equation}
\text{Pair I: }(\chi_{+,+1},\chi_{-,-1}),\qquad
\text{Pair II: }(\chi_{-,+1},\chi_{+,-1}),
\end{equation}directly:
\begin{align}
H_{\rm lin}\bigl(\phi_{N,m}\chi_{+,+1}\bigr)
&=v_F\sqrt{2B}\,(a\,\phi_{N,m})(\alpha^-\chi_{+,+1})
\nn\\&=v_F\sqrt{2NB}\,\phi_{N-1,m}\chi_{-,-1},\nn\\
H_{\rm lin}\bigl(\phi_{N-1,m}\chi_{-,-1}\bigr)
&=v_F\sqrt{2B}\,(a^\dagger\phi_{N-1,m})(\alpha^+\chi_{-,-1})
\nn\\&=v_F\sqrt{2NB}\,\phi_{N,m}\chi_{+,+1},
\end{align}
and the matirx form of $H_{lin}$ is:
\begin{align}
    H_{\mathrm{lin}}\big|_{N}
  = v_{F}\sqrt{2NB}
  \begin{pmatrix} 0 &0 &0 & 1 \\ 0 &0 &1 & 0\\0 &1 &0 &0 \\1&0 &0 &0  \end{pmatrix}.
  \label{eq:2x2}
\end{align}

The matrix \eqref{eq:2x2} is diagonalised immediately. Its eigenvalues
are
\begin{equation}
  E_{N}^{(\pm)} = \pm\,v_{F}\sqrt{2NB},
  \label{eq:Evalues}
\end{equation}
hence:
\begin{align}
 & \Psi^{\mathrm{I},(\lambda)}_{N,m}
=\frac{1}{\sqrt2}\Bigl(\phi_{N,m}\,\chi_{+,+1}
\pm\phi_{N-1,m}\,\chi_{-,-1}\Bigr),
\nn\\& E^{(\lambda)}_N=\pm v_F\sqrt{2NB}.  
\end{align}
symmetrically,
\begin{align}
 & \Psi^{\mathrm{II},(\pm)}_{N,m}
=\frac{1}{\sqrt2}\Bigl(\phi_{N,m}\,\chi_{-,+1}
\pm\phi_{N-1,m}\,\chi_{+,-1}\Bigr),
\nn\\& E^{(\lambda)}_N=\pm v_F\sqrt{2NB}.
\end{align}
{The higher-level formulas above refer to the linear Dirac part $H_{\rm lin}$ only; the quadratic Wilson term changes the higher-level eigenvectors and energies and is not needed for the LLL calculation. For the LLL, $n=0$ requires $s=-1$:}
\begin{equation}
\Psi^{(1)}_{0,m}=\phi_{0,m}\,\chi_{+,-1},\quad
\Psi^{(2)}_{0,m}=\phi_{0,m}\,\chi_{-,-1},\quad E_0=0,
\label{eq:LLLstates}
\end{equation}
both in the {$s=-1$ sector} selected by
$P_{\rm LLL}=\tfrac12\bigl(1-i\gamma_1\gamma_2\,\mathrm{sign}(B)\bigr)$ (which keeps
{$s=-1$ for $B>0$}).

Using $H_\parallel$ of the main text with {$\theta=p_z-E_3t$}, we obtain the effective Hamiltonian \eqref{hth}.
\begin{align}\label{eq:HLLLmat}
    \widehat H_{\rm LLL}(\theta)&=v_F
\begin{pmatrix}
\,m+1-\cos\theta+\tfrac{B}{2} & \sin\theta\,\\[4pt]
\,\sin\theta & -\bigl(m+1-\cos\theta+\tfrac{B}{2}\bigr)\,
\end{pmatrix}
\nn\\&\equiv v_F\begin{pmatrix}C&S\\ S&-C\end{pmatrix},
\end{align}
with $S=\sin\theta$, $C=m+1-\cos\theta+\tfrac{B}{2}$. Its eigenvalues are:
\begin{align}
    E_\pm(\theta)&=\pm v_F\sqrt{S^2+C^2}
\nn\\&=\pm v_F\sqrt{(2+2m+B)(1-\cos\theta)+\bigl(m+\tfrac{B}{2}\bigr)^2},
\end{align}
The complete LLL eigenstate is the product of the degenerate orbital factor
\eqref{eq:phi0m} and this longitudinal spinor:
\begin{equation}
\Psi^{(m,\pm)}_{\rm LLL}(f,\bar f;\theta)
=N_m\,f^{\,m}\,e^{-\frac{B}{4}|f|^2}
\;\;\eta_\pm(\theta),
\end{equation}
equivalently, in explicit 4-component form,
\begin{align}\label{eq:etaexplicit}
\eta_\pm(\theta)=\frac{1}{\mathcal N_\pm}
\begin{pmatrix} 0\\[2pt] S\\[2pt] 0\\[2pt] \pm\sqrt{S^2+C^2}-C\end{pmatrix},
\end{align}
where
\begin{align} 
\mathcal{N}_\pm=\sqrt{S^2+\bigl(\pm\sqrt{S^2+C^2}-C\bigr)^2}\,
\end{align}
in momentum space:\begin{align}
\tilde\phi_{0,m}(\mathbf k_\perp)
={\widetilde N_m\,(k_x-i k_y)^{m}\,e^{-k_\perp^2/B}},    
\end{align}
where
\begin{align}
\widetilde N_m=N_m\,\frac{4\pi}{B}\left(\frac{-2i}{B}\right)^{\!m}.
\end{align}
Hence we have:\begin{equation}
\tilde\Psi^{(m,\pm)}_{\rm LLL}(\mathbf k_\perp;\theta)
={\widetilde N_m\,(k_x-ik_y)^{m}\,e^{-k_\perp^2/B}\;\eta_\pm(\theta)}.
\label{eq:FTfull}
\end{equation}

\section{Green function in the LLL approximation}\label{Green detail}

We have the lattice Hamiltonian:
\begin{align}
H(\Pi)=v_F\sum_{i=1}^{3}\gamma_{0}\gamma_{i}\sin\Pi_{i}
+v_F\gamma_{0}\Bigl(m+\sum_{i=1}^{3}(1-\cos \Pi_{i})\Bigr),
\end{align}
Here we already replaced the lattice constant 'a' with unity and 
 {$k_i \rightarrow \Pi_i = k_i + |e|A_i$ for an electron} with the symmetric gauge:

\begin{equation}
A_x=-\frac{B}{2}y,\qquad A_y=\frac{B}{2}x,\qquad {A_z=-E_3t,\qquad A_0=0}.
\end{equation}

hence:

\begin{equation}
\Pi_x={k_x-\frac{B}{2}y},\qquad 
\Pi_y={k_y+\frac{B}{2}x},\qquad 
\Pi_z={k_z-E_3t}.
\end{equation}

We arrive at the effective  Hamiltonian within the LLL
\begin{equation}
H_{\mathrm{LLL}}(\theta)
=v_F\gamma_0\gamma_3\sin\theta
+v_F\gamma_0\left(m+1-\cos\theta+\frac{B}{2}\right).
\end{equation}

For example, for the  retarded Green function we have representation :
\begin{equation}\label{greenwithstate}
G^R(\omega) = \sum_n \frac{|u_n\rangle\langle u_n|}{\omega + i\epsilon - E_n}.
\end{equation}
The eigenfunctions \eqref{eigenfunc} in momentum space are expressed through $\phi_{n,m}=\braket{k|u_{n,m}}$, where $\bra{k}$ is the momentum basis in space. With this we get:
\begin{equation}
\hat{G}_{\mathrm{LLL}}(p_\perp,p_0,p_3)
=
G_\perp(p_\perp)\,\tilde{G}(p_0,p_3)\,\mathcal{O}_-,
\end{equation}
 i.e.  \eqref{eq:lattice_factorized} of the main text.

We start from the Retarded Green's function in the transverse plane within the lowest Landau level (LLL) approximation. In the symmetric gauge $\mathbf{A}(\mathbf{x}_\perp) = \frac{B}{2}(-x_2, x_1, 0)$ with $B > 0$, the degenerate LLL wavefunctions are given by \eqref{eq:phi0m}:
\begin{equation}
\phi_{0,m}(\mathbf{x}_\perp) = {\sqrt{\frac{1}{\pi\,m!}\left(\frac{B}{2}\right)^{m+1}} (x_1 - i x_2)^m e^{-\frac{B}{4}\mathbf{x}_\perp^2}},
\label{eq:app_wavefunction}
\end{equation}
where $m \ge 0$ is the angular momentum quantum number labeling the gauge degeneracy. Introducing another field $\mathbf{A}(\mathbf{y}_\perp) = \frac{B}{2}(-y_2, y_1, 0)$ the complex coordinates $l = x_1 + i x_2$ and $w = y_1 + i y_2$, the spectral summation over all degenerate states for the transverse propagator reads:
\begin{align}\label{eq:app_sum_def}
G^R(\mathbf{x}_\perp,\mathbf{y}_\perp)
=\sum_{s=\pm}\sum_{m\ge0}
\frac{\Psi^{(m,s)}_{\rm LLL}(\mathbf{x}_\perp;\theta)\,
\Psi^{(m,s)\dagger}_{\rm LLL}(\mathbf{y}_\perp;\theta)}
{\omega+i\epsilon-E_s(\theta)},
\end{align}
where we used the replacement:\begin{align}
    \sum_n |u_n\rangle\langle u_n|
\;\longrightarrow\;
\sum_{s=\pm}\sum_{m\ge0}\int\frac{dk_3}{2\pi}\,
\big|\tilde\Psi^{(m,s)}_{\rm LLL}\big\rangle\!\big\langle\tilde\Psi^{(m,s)}_{\rm LLL}\big|.
\end{align}
Hence:\begin{align}
    G^R=\left(\sum_{m\ge0}\phi_{0,m}(\mathbf{x}_\perp)\phi_{0,m}^{*}(\mathbf{y}_\perp)\right)
\sum_{s=\pm}\frac{\eta_s(\theta)\eta_s^\dagger(\theta)}{\omega+i\epsilon-E_s(\theta)}.
\end{align}
Substituting Eq.~\eqref{eq:app_wavefunction} into Eq.~\eqref{eq:app_sum_def}, we factor out the common Gaussian envelope and obtain:
\begin{align}\label{eq:app_sum_substituted}
    &\sum_m\Big[\tilde\phi_{0,m}(\mathbf{x}_\perp)\tilde\phi_{0,m}^{*}(\mathbf{y}_\perp)\Big] 
    \nn\\=&{\frac{B}{2\pi} e^{-\frac{B}{4}(\mathbf{x}_\perp^2 + \mathbf{y}_\perp^2)} \sum_{m=0}^{\infty} \frac{1}{m!} \left( \frac{B}{2} l^* w \right)^m}.
\end{align}

Employing the standard Taylor series for the exponential function, $\sum_{m=0}^{\infty} \frac{\xi^m}{m!} = e^\xi$, the summation simplifies yields a single complex exponential term:
\begin{align}
&\sum_m\Big[\tilde\phi_{0,m}(\mathbf{x}_\perp)\tilde\phi_{0,m}^{*}(\mathbf{y}_\perp)\Big] 
\nn\\=& {\frac{B}{2\pi} \exp \left[ -\frac{B}{4}(\mathbf{x}_\perp^2 + \mathbf{y}_\perp^2) + \frac{B}{2} l^* w \right]}.
\label{eq:app_sum_exponential}
\end{align}
To decouple the relative coordinates from the center-of-mass coordinates, we expand the complex product in the exponent:
\begin{equation}
{l^* w = (x_1 - i x_2)(y_1 + i y_2) = \mathbf{x}_\perp \cdot \mathbf{y}_\perp + i (\mathbf{x}_\perp \times \mathbf{y}_\perp)_z},
\label{eq:app_complex_product}
\end{equation}
where $(\mathbf{x}_\perp \times \mathbf{y}_\perp)_z = x_1 y_2 - x_2 y_1$. Inserting Eq.~\eqref{eq:app_complex_product} back into Eq.~\eqref{eq:app_sum_exponential} gives:
\begin{align}
&\sum_m\Big[\tilde\phi_{0,m}(\mathbf{x}_\perp)\tilde\phi_{0,m}^{*}(\mathbf{y}_\perp)\Big] 
    \nn\\=& {\frac{B}{2\pi} \exp \left[ -\frac{B}{4}(\mathbf{x}_\perp^2 + \mathbf{y}_\perp^2 - 2\mathbf{x}_\perp \cdot \mathbf{y}_\perp)  +i \frac{B}{2}(x_1 y_2 - x_2 y_1) \right]}.
\label{eq:app_unseparated}
\end{align}
The real part of the exponent forms a perfect square in terms of the relative coordinate $\mathbf{r}_\perp = \mathbf{x}_\perp - \mathbf{y}_\perp$, since $\mathbf{x}_\perp^2 + \mathbf{y}_\perp^2 - 2\mathbf{x}_\perp \cdot \mathbf{y}_\perp = (\mathbf{x}_\perp - \mathbf{y}_\perp)^2 = \mathbf{r}_\perp^2$. Consequently, the coordinate-space propagator factorizes exactly into:
\begin{align}
&\sum_m\Big[\tilde\phi_{0,m}(\mathbf{x}_\perp)\tilde\phi_{0,m}^{*}(\mathbf{y}_\perp)\Big] 
\nn\\=&{\frac{B}{2\pi} \exp \left( -\frac{B}{4}\mathbf{r}_\perp^2 \right) \exp \left[ -i \int_{\mathbf{y}_\perp}^{\mathbf{x}_\perp} \mathbf{A}(\mathbf{z}_\perp) \cdot d\mathbf{z}_\perp \right]},
\label{eq:app_factorized}
\end{align}
where the second factor is precisely the non-integrable Peierls phase factor (or the gauge link) linking $\mathbf{y}_\perp$ to $\mathbf{x}_\perp$ along a straight line path in the symmetric gauge.

According to the definition of the gauge-covariant Wigner transformation used in \eqref{eq:lattice_propagator} of section \ref{2}, the gauge link must be stripped away to restore translational invariance before moving to momentum space. The remaining gauge-invariant core depends solely on the relative coordinate:
\begin{equation}
S_{\text{lattice}}(\mathbf{r}_\perp) = \frac{B}{2\pi} \exp \left( -\frac{B}{4}\mathbf{r}_\perp^2 \right).
\label{eq:app_core}
\end{equation}
Finally, we perform a two-dimensional Fourier transform with respect to the relative coordinate $\mathbf{r}_\perp$ to obtain the propagator in terms of the covariant kinetic momentum $\mathbf{k}_\perp$:
\begin{align}
G_\perp(\mathbf{k}_\perp) &= \int d^2 r_\perp e^{-i \mathbf{k}_\perp \cdot \mathbf{r}_\perp} S_{\text{lattice}}(\mathbf{r}_\perp) \nn\\&= \frac{B}{2\pi} \int d^2 r_\perp e^{-i \mathbf{k}_\perp \cdot \mathbf{r}_\perp} e^{-\frac{B}{4}\mathbf{r}_\perp^2}.
\label{eq:app_fourier_integral}
\end{align}
Evaluating this standard Gaussian integral yields:
\begin{equation}
G_\perp(\mathbf{k}_\perp) = \frac{B}{2\pi} \left( \frac{4\pi}{B} e^{-\frac{\mathbf{k}_\perp^2}{B}} \right) = 2 e^{-\frac{\mathbf{k}_\perp^2}{B}}.
\label{eq:app_final_result}
\end{equation}
Eq.~\eqref{eq:app_final_result} is in line with Eq.~(51) and Eq.~(B32) of the main text. 

The spectral representation is equivalent to the operator inverse formula. To see this explicitly, act with $(\omega + i\epsilon - H)$ on $G^R(\omega)$:
\begin{align}
    \left(\omega + i\epsilon - H\right)G^R(\omega)& = \sum_n \frac{(\omega + i\epsilon - E_n)}{\omega + i\epsilon - E_n}|u_n\rangle\langle u_n|\nn\\& = \sum_n |u_n\rangle\langle u_n| = \mathbf{1},
\end{align}
which confirms that
\begin{equation}
    G^R(\omega) = \frac{1}{\omega + i\epsilon - H},
\end{equation}
and that the eigenstate expansion and the operator inverse are two equivalent representations of the same object, we substituting the Hamiltonian as $H_{LLL}$:
\begin{equation}
G^R(k,\omega) = \frac{1}{\omega+i\epsilon - H_{LLL}(k_3)},
\end{equation}
{multiply numerator and denominator by $\omega+i\epsilon+H_{LLL}$:}

\begin{equation}
{(\omega+i\epsilon-H_{LLL})(\omega+i\epsilon+H_{LLL}) = (\omega+i\epsilon)^2 - H_{LLL}^2}
\end{equation}

Using the Clifford algebra $\{\gamma_\mu,\gamma_\nu\} = 2g_{\mu\nu}$:

\begin{align}
  H_{LLL}^2& = v_F^2\sin^2 k_3 
+ v_F^2\left( m+1-\cos k_3
+ \frac{|B|}{2}\right)^2\nn\\& \equiv \tilde{E}^2_{latt}(k_3),  
\end{align}
therefore:
\begin{equation}
G^R(k_3,\omega) = {\frac{\omega+i\epsilon + H_{LLL}(k_3)}
{(\omega+i\epsilon)^2 - \tilde{E}^2_{latt}(k_3)}}.
\end{equation}
Combining the transverse Gaussian factor, the LLL spinor 
projector, and the longitudinal propagator:
\begin{align}
&-iS^{LLL}_{lattice}(k,\omega) = 2e^{-k_\perp^2/|B|}\gamma_0
\nn\\&\cdot\frac{\omega\gamma_0 + v_F\sin k_3 \gamma_3 
+ \left( v_F(m+1-\cos k_3) 
+ \frac{v_F|B|}{2}\right)}
{\omega^2 - \tilde{E}^2_{latt}(k_3)} \cdot O^-,
\end{align}
where:
\begin{equation}
\tilde{E}_{latt}(k_3) = v_F\sqrt{\sin^2 k_3 
+ \left( m+1-\cos k_3 
+ \frac{|B|}{2}\right)^2}
\end{equation}

This is the detail of section \ref{2}. The Gaussian 
factor $2e^{-k_\perp^2/|B|}$ comes from the LLL wavefunction, 
the $O^-$ projects onto the correct spinor subspace, and 
the longitudinal propagator captures the 1D SSH-like dynamics 
along the magnetic field direction. In the similar way we obtain the corresponding expressions for the advanced and lesser Green functions used in the main text.

\section{Axial  Current}\label{integral detail}
In this appendix we present calculation of axial charge density.
Recall the form of axial charge density:
\begin{align}
\rho_5(x) = \frac{\operatorname{sign}(B) E_3 |B|}{4\pi} \operatorname{Re} \int \frac{dp_0 dp_3}{(2\pi)^2} \operatorname{tr} \bigl[&\alpha^3 O^- \, \boldsymbol{\tilde{G}}  \partial_{p_{[0}} \boldsymbol{\tilde{Q}} \nn\\
& \boldsymbol{\tilde{G}}  \partial_{p_3]} \boldsymbol{\tilde{Q}}  \boldsymbol{\tilde{G}} \bigr]^{<}.
\end{align}
where we already use the trick $\gamma_5 O^- = -\operatorname{sign}(B) \alpha^3 O^-$. Using the property of `lesser' component of a product of Keldysh-Green functions \eqref{eq:lattice_product}, we have:
\begin{align}
     &\Big(
\tilde{G}\partial_{p_0} \tilde{Q} \, \tilde{G}\partial_{p_3} \tilde{Q} \tilde{G}
- \tilde{G}\partial_{p_3} \tilde{Q} \, \tilde{G} \partial_{p_0} \tilde{Q} \tilde{G}
\Big)^<\nn\\=&\Big(
-\partial_{p_0} \tilde{G} \, \partial_{p_3} \tilde{Q} \tilde{G}
+ \partial_{p_3} \tilde{G} \, \partial_{p_0} \tilde{Q} \tilde{G}
\Big)^< \nn\\=&-\partial_{p_0} \tilde{G}^R \, \partial_{p_3} \tilde{Q}^R \tilde{G}^<
-\partial_{p_0} \tilde{G}^R \, \partial_{p_3} \tilde{Q}^< \tilde{G}^A\nn\\
&-\partial_{p_0} \tilde{G}^< \, \partial_{p_3} \tilde{Q}^A \tilde{G}^A
+ \partial_{p_3} \tilde{G}^R \, \partial_{p_0} \tilde{Q}^R \tilde{G}^<\nn\\
&+ \partial_{p_3} \tilde{G}^R \, \partial_{p_0} \tilde{Q}^< \tilde{G}^A
+ \partial_{p_3} \tilde{G}^< \, \partial_{p_0} \tilde{Q}^A \tilde{G}^A.
\end{align}
Substituting this result to the axial charge density, and use the form of Keldysh Green-function, we get:
\begin{align}
   \rho_5 &=  i \;
\frac{ E_3|B|}{2}\;
\mathrm{Re}
\int \frac{d p_3 \, d p_0}{(2\pi )^2}
\nn\\
& \times \mathrm{tr}\Big[
\gamma_0 O^-\Big(
\partial_{p_0}\delta(p_0 - H)\,
n(p_0)\, (\partial_{p_3} H)\,
\frac{1}{p_0 - i\epsilon - H}
\nn\\
&- \delta(p_0 - H)\, n'(p_0)\,
(\partial_{p_3} H)\,
\frac{1}{p_0 - i\epsilon - H}
\nn\\
&+ \partial_{p_3}\delta(p_0 - H)\,
n(p_0)\,
\frac{1}{p_0 - i\epsilon - H},
\Big)
\Big]. 
\end{align}
Where $n(p_0)$ is the distribution function for fermion. There are three integrals to compute:
\begin{align}
    &I_1=\int \frac{dp_3dp_0}{4\pi^2} \operatorname{tr} \left[ \gamma_5 O^- \partial_{p_0}(\delta(p_0 - H)) n(p_0) (\partial_{p_3} H) D^{-1}\right],
   \nn\\&I_2=\int \frac{dp_3dp_0}{4\pi^2} \operatorname{tr} \left[ -\gamma_5 O^- \delta(p_0 - H) n'(p_0) (\partial_{p_3} H) D^{-1}\right],
   \nn\\&I_3 = \int \frac{dp_3dp_0}{4\pi^2} \operatorname{tr} \left[ \gamma_5 O^- \partial_{p_3}(\delta(p_0 - H)) n(p_0) D^{-1}\right].
\end{align}
where \begin{align}
    D^{-1} = \frac{1}{p_0 - H - i\epsilon},
\end{align}
for the integral $I_1$ we use integration by part and the property of delta function, we can get:\begin{align}
    I_1 = -\int \frac{dp_3}{4\pi^2} \operatorname{tr} \left[ \gamma_5 O^-(\partial_{p_3} H) \left( \frac{n(H)}{\epsilon^2} - \frac{n'(H)}{i\epsilon} \right) \right],
\end{align}
for the integral $I_2$ we use the property of delta function:\begin{align}
    I_2 = -\int \frac{dp_3}{4\pi^2}\operatorname{tr} \left[ \gamma_5 O^- (\partial_{p_3} H) n'(H)  \frac{1}{-i\epsilon}\right],
\end{align}
{For $I_3$ one must not use the scalar chain rule
$\partial_{p_3}\delta(p_0-H)=-H'\delta'(p_0-H)$ as a matrix identity,
because in the lattice model $[H,H']\ne0$.  The correct derivative follows
from the spectral resolution,
\begin{equation}
 \partial_{p_3}\delta(p_0-H)=\sum_{s=\pm}
 \left[P_s'\delta(p_0-E_s)-P_sE_s'\delta'(p_0-E_s)\right].
\end{equation}
In the leading weak-width intraband contribution used here,
$P_sH'P_s=E_s'P_s$; the $P_s'$ terms are interband and are regular rather
than pinch enhanced.  After projecting first and then integrating by parts,
the singular parts of $I_1$ and $I_3$ cancel as stated, whereas their omitted
interband remainder is of relative order $O(\epsilon/\mu)$.  Thus the result
used in the main text is the leading on-shell identity
\begin{equation}
 I = -\frac{1}{-i\epsilon}\int\frac{dp_3}{4\pi^2}
 \operatorname{tr}\!\left[\gamma_5O^-H'n'(H)\right]
 +O(\epsilon^0/\mu).
\end{equation}
It is not an exact matrix-chain-rule result.}
As $\rho_5=iE_3|B|I/2$, this determines the leading contribution.
As we only need the case in lower temperature, at zero temperature the distribution function is $n(p_0) = \Theta(\mu - p_0)$, so:
\begin{equation}
n'(p_0) = -\delta(\mu - p_0),
\end{equation}
and therefore $n'(H) = -\delta(\mu - H)$.
The matrix delta function restricts to the positive-energy eigenspace is:
\begin{equation}
\delta(\mu - H) = \delta(\mu - E_+(\theta))\,P_+(\theta),
\end{equation}
where the eigenvalues of $H$ are
\begin{equation}
E_{\pm}(\theta) = \pm v_F\sqrt{(2+2m+B)(1-\cos\theta) + \Bigl(m+\frac{B}{2}\Bigr)^2},
\end{equation}
and the energy projector can be shown as:\begin{align}
    H=E_{+}(\theta )P_{+}(\theta )+E_{-}(\theta )P_{-}(\theta ),
\end{align}
and $P_{\pm}(\theta )^2=P_{\pm}(\theta ),P_{+}(\theta )P_{-}(\theta )=P_{-}(\theta )P_{+}(\theta )=0$
\begin{equation}
P_+(\theta) = \frac{H(\theta) + \mu}{2\mu}.
\end{equation}
The trace with the projector reduces to
\begin{equation}
\operatorname{tr}\left[\gamma_5 O^- \partial_{p_3}H \cdot P_+\right] = -\operatorname{sign}(B)v_F\cos\theta.
\end{equation}
Thus:
\begin{equation}
\operatorname{tr}\left[\gamma_5 O^- \partial_{p_3}H \cdot n'(H)\right] = \operatorname{sign}(B)v_F\cos\theta \;\delta(\mu - E_+(\theta)).
\end{equation}
Using the eigenvalues \eqref{egvalue}, we have:\begin{align}
    I = \frac{\operatorname{sign}(B)v_F}{4\pi^2 i\epsilon} \int_{-\pi}^{\pi} d\theta\; \cos\theta \;\delta(\mu - E_+(\theta)).
\end{align}
The on-shell condition $E_+(\theta_F) = \mu$ gives:
\begin{equation}
(2+2m+B)(1-\cos\theta_F) + \Bigl(m+\frac{B}{2}\Bigr)^2 = \frac{\mu^2}{v_F^2},
\end{equation}
hence:
\begin{equation}
\cos\theta_F = 1 - \frac{\mu^2/v_F^2 - (m+B/2)^2}{2+2m+B}.
\end{equation}
The derivative of the dispersion is
\begin{equation}
E_+'(\theta) = \frac{v_F}{2\sqrt{f(\theta)}}(2+2m+B)\sin\theta, 
\end{equation}
where\begin{align}
    f(\theta) = (2+2m+B)(1-\cos\theta) + \Bigl(m+\frac{B}{2}\Bigr)^2.
\end{align}
At $\theta_F$, $\sqrt{f(\theta_F)} = \mu/v_F$, so
\begin{equation}
|E_+'(\theta_F)| = \frac{v_F^2(2+2m+B)|\sin\theta_F|}{2\mu}.
\end{equation}

There are two solutions in $[-\pi,\pi]$, $\theta_F$ and $-\theta_F$, both yielding the same $\cos\theta_F$ and $|\sin\theta_F| = \sqrt{1-\cos^2\theta_F}$. Therefore, using the property of delta function below:
\begin{align}
    \delta (g(x))=\sum _{i}\frac{\delta (x-x_{i})}{|g^{\prime }(x_{i})|},
\end{align}
the proof of it is in appendix \ref{dirac delta}, hence:
\begin{align}
    &\int_{-\pi}^{\pi} d\theta\; \cos\theta \;\delta(\mu - E_+(\theta)) = \frac{\cos\theta_F}{|E_+'(\theta_F)|} + \frac{\cos\theta_F}{|E_+'(-\theta_F)|}\nn\\& = \frac{4\mu\cos\theta_F}{v_F^2(2+2m+B)\sqrt{1-\cos^2\theta_F}}.
\end{align}
Finally, we get the form of axial charge density:
\begin{align}
    \rho_5= \;
\frac{E_3|B|}{2} \frac{\operatorname{sign}(B)\,\mu\cos\theta_F}{\pi^2\epsilon\, v_F(2+2m+B)\sqrt{1-\cos^2\theta_F}},
\end{align}
this is the general form of axial charge density we got.

\section{Proof of the Dirac Delta Identity}\label{dirac delta}

{
Let $H_{\rm LLL}^2=E(\theta)^2\mathbf 1$, with $E(\theta)>0$.  Its two
spectral projectors are
\begin{equation}
 P_\pm(\theta)=\frac{E(\theta)\pm H_{\rm LLL}(\theta)}{2E(\theta)},
 \qquad H_{\rm LLL}=E(P_+-P_-).
\end{equation}
Functional calculus for a Hermitian matrix then gives directly
\begin{align}
 \delta(p_0-H_{\rm LLL})={}&P_+(\theta)\delta\bigl(p_0-E(\theta)\bigr)
 +P_-(\theta)\delta\bigl(p_0+E(\theta)\bigr)\nn\\
={}&\frac{E(\theta)+H_{\rm LLL}}{2E(\theta)}
 \delta\bigl(p_0-E(\theta)\bigr)\nn\\
&+\frac{E(\theta)-H_{\rm LLL}}{2E(\theta)}
 \delta\bigl(p_0+E(\theta)\bigr).
\label{eq:matrixdelta_correct}
\end{align}
This identity can also be checked on an arbitrary eigenvector: it reduces to
$\delta(p_0-E)$ in the positive-energy subspace and to $\delta(p_0+E)$ in the
negative-energy subspace.  Notice in particular the minus sign multiplying
$H_{\rm LLL}$ in the negative-energy projector.  }

\section{Calculation of the field-induced self-energy}
\label{app:source}

{

In this appendix we calculate the drive $\hat{\Sigma}_d$ of Eq.~(\ref{eq:linSigma}) and the retarded and advanced components of the field-induced self-energy. We work in the LLL-reduced $(1+1)$D theory and set $a = 1$. We use the identities $\partial_{p_0}\tilde{G}^{R(A)} = -(\tilde{G}^{R(A)})^2$, $\partial_{p_3}\tilde{G}^{R(A)} = \tilde{G}^{R(A)} H' \tilde{G}^{R(A)}$ and $\tilde{G}^{<} = 2i\epsilon\, \tilde{G}^{R}\tilde{G}^{A}\, n(\omega)$, where $H' \equiv \partial_{p_3} H_{\mathrm{LLL}}$. The six terms of the lesser component of the antisymmetrized product entering Eq.~(\ref{eq:ladder}) (see Appendix~\ref{integral detail}) then collapse into two structures:
\begin{align}
\bigl[\hat{G}_d\bigr]^{<} =& -\frac{iE_3}{2}\Bigl\{ 2i\epsilon\, \bigl( \tilde{G}^{R}\tilde{G}^{A} H' \tilde{G}^{A} - \tilde{G}^{R} H' \tilde{G}^{R}\tilde{G}^{A} \bigr)\, n'(\omega) \nn\\
&+ \Bigl( (\tilde{G}^{R})^2 [H,H'] (\tilde{G}^{R})^2 \nn\\
&\quad - (\tilde{G}^{A})^2 [H,H'] (\tilde{G}^{A})^2 \Bigr)\, n(\omega) \Bigr\} .
\label{eq:g1less}
\end{align}
(Eq.~(\ref{eq:g1less}) holds up to terms carrying one extra power of $\epsilon$ in the numerator together with one mixed $\tilde{G}^R \ldots \tilde{G}^A$ product; such terms contribute to the observables at the relative order $\epsilon/\mu$ and are omitted throughout.) The contribution proportional to $n(\omega)$ contains only products of Green functions with poles on the same side of the real axis, and its intraband projection vanishes, since $\langle \pm | [H,H'] | \pm \rangle = 0$. It therefore remains finite in the limit $\epsilon \to 0$: only the interband terms survive, with denominators set by the band splitting $2E(\theta) \ge 2\mu$ rather than by $\epsilon$. After multiplication by $u_0^2 n_{\text{imp}} |B|/(2\pi) = 2\epsilon v_F$ this structure contributes to $\hat{\Sigma}_d$ at the order $O(\epsilon E_3/\mu)$ only. The same is true for the retarded and advanced components. In contrast, the contribution proportional to $n'(\omega)$ contains the mixed retarded-advanced products, singular in the limit $\epsilon \to 0$. Projecting it onto the eigenstates of $H_{\mathrm{LLL}}$ and keeping the singular intraband part, we obtain
\begin{align}
\int \frac{dp_3}{2\pi}\, \bigl[\hat{G}_d\bigr]^{<} &= \frac{iE_3\, n'(\omega)}{2\epsilon}\, \bigl[ P_+(\theta_\omega) - P_+(-\theta_\omega) \bigr] \nn\\
&= \frac{iE_3\, n'(\omega)}{2\epsilon}\, \frac{v_F \sin\theta_\omega}{\omega}\, \gamma_0\gamma_3 ,
\label{eq:g1int}
\end{align}
where $P_\pm(\theta)$ are the band projectors of Appendix~\ref{integral detail}. Substituting Eq.~(\ref{eq:g1int}) into Eq.~(\ref{eq:sigma1def}) and using $\epsilon = u_0^2 n_{\text{imp}} |B|/(4\pi v_F)$, we arrive at Eq.~(\ref{eq:sigma1}) of the main text.

The retarded and advanced components can be found in closed form at the first order in $g = u_0^2 n_{\rm imp} |B|/(2\pi)$. Owing to the triangular Keldysh structure they are built of retarded (advanced) functions only, and the two orderings of the antisymmetrized product combine into a commutator:
\begin{equation}
\bigl[\hat{G}_d\bigr]^{R(A)} = \frac{iE_3}{2}\, \bigl(\tilde{G}^{R(A)}\bigr)^2\, [H, H']\, \bigl(\tilde{G}^{R(A)}\bigr)^2 ,
\label{eq:g1ra}
\end{equation}
with
\begin{equation}
[H, H'] = -2 v_F^2 \Bigl( 1 - \bigl(1 + m + \tfrac{B}{2}\bigr) \cos\theta \Bigr)\, \gamma_3 ,
\label{eq:commutator}
\end{equation}
which is purely interband, $\langle u_\pm | [H,H'] | u_\pm \rangle = 0$. Consequently the four propagators in Eq.~(\ref{eq:g1ra}) reduce to the scalar factor $\bigl((\omega \pm i\epsilon)^2 - E^2(\theta)\bigr)^{-2}$, and the momentum integral is elementary in the variable $x = \cos\theta$: the denominator is linear in $x$, $(\omega \pm i\epsilon)^2 - E^2 = v_F^2 (2+2m+B)(x - x_0)$, where $x_0$ is the analytic continuation of $\cos\theta_\omega$, and
\begin{align}
\frac{1}{\pi}\int_{-1}^{1} \frac{\bigl[1 - (1+m+\tfrac{B}{2})x\bigr] dx}{(x - x_0)^2 \sqrt{1 - x^2}} &= \frac{x_0 - (1+m+\tfrac{B}{2})}{(x_0^2 - 1)^{3/2}} \nn\\
&= \frac{-\,C(\theta_\omega)}{(x_0^2 - 1)^{3/2}} ,
\end{align}
where we used $x_0 - 1 - m - B/2 = -[(\omega\pm i\epsilon)^2/v_F^2 + 2m + B + (m+B/2)^2]/(2+2m+B) = -C(\theta_\omega)$. With the branch $(x_0^2 - 1)^{3/2} = \pm i \sin^3\theta_\omega$ appropriate for the retarded (advanced) prescription we obtain, for $\omega$ above the gap,
\begin{align}
\hat{\Sigma}_d^{R}(\omega) &= -\hat{\Sigma}_d^{A}(\omega) = g E_3\, \frac{C(\theta_\omega)}{v_F^2 (2+2m+B)^2 \sin^3\theta_\omega}\, \gamma_3 \nn\\
&= \frac{2 \epsilon E_3 v_F\, |M(\omega)|}{(2+2m+B)^2\, \omega^2\, \bigl(1 - |M(\omega)|^2\bigr)^{3/2}}\, \gamma_3 ,
\label{eq:sigmaRA}
\end{align}
up to corrections of the next order in $\epsilon$, with $|M(\omega)|$ of Eq.~(\ref{eq:overlap}). Several properties of Eq.~(\ref{eq:sigmaRA}) are noteworthy. (a) It is explicitly of the first order in $g \propto \epsilon$; in the hierarchy regime its magnitude is $\epsilon E_3/(4\mu)$, which makes the estimate $O(\epsilon E_3/\mu)$ of Eq.~(\ref{eq:sigma1}) precise, including the coefficient. (b) Its Dirac structure is purely off-diagonal in the band basis, $\langle u_+(\theta)|\gamma_3|u_+(\theta)\rangle = 0$: Eq.~(\ref{eq:sigmaRA}) describes the field-induced interband (Zener) coherence -- the electric polarization of the band -- and does not modify the quasiparticle width at this order. (c) Below the gap $\theta_\omega$ is imaginary, $(x_0^2-1)^{3/2}$ is real, the expression becomes Hermitian and $\hat{\Sigma}_d^{R} = \hat{\Sigma}_d^{A}$: the retarded and advanced functions coincide on the part of the real axis free of spectral weight, as required by analyticity. (d) The spectral combination $(\hat{\Sigma}_d^{A} - \hat{\Sigma}_d^{R})\, n(\omega) = -2 \hat{\Sigma}_d^{R}\, n(\omega) = O(\epsilon E_3)$ is thereby negligible compared to the kinetic part of the lesser component, in accordance with the decomposition (\ref{eq:fdtsplit}) below. (e) Eq.~(\ref{eq:sigmaRA}) diverges at the band edges ($\sin\theta_\omega \to 0$), so the scheme requires the Fermi level to stay at a distance much larger than $\epsilon$ from the gap edge.

Three properties of Eq.~(\ref{eq:sigma1}) are worth noticing.

(i) The factor $u_0^2 n_{\text{imp}}$ has cancelled: $\hat{G}_d \propto 1/\epsilon \propto 1/(u_0^2 n_{\text{imp}})$, while the impurity line in Eq.~(\ref{eq:sigma1def}) contributes the factor $u_0^2 n_{\text{imp}}$. Therefore $\hat{\Sigma}_d$ is of the zeroth order in $u_0^2$, i.e. of the same order as the terms kept in the bare-bubble calculation. The same cancellation occurs in every subsequent rung of the ladder (one more factor $u_0^2 n_{\text{imp}}$ from the impurity line, one more factor $1/\epsilon$ from the on-shell product $\tilde{G}^{R}\tilde{G}^{A}$). Thus for the DC response the perturbation theory in $u_0^2$ breaks down, and the ladder must be resummed.

(ii) The physical meaning of Eq.~(\ref{eq:sigma1}) is transparent. Substituting the standard drift correction to the distribution function, $\delta n(\pm\theta_F) = \pm E_3 |E_+'(\theta_F)|\, n'(\omega)/(2\epsilon)$, into the equilibrium self-energy of Section IV reproduces Eq.~(\ref{eq:sigma1}) identically. The omitted term thus describes the scattering of electrons off the impurities from the actual (field-shifted) distribution rather than from the equilibrium one -- the ``in'' term of the collision integral. Furthermore, within the LLL subspace $\gamma_0\gamma_3\, \mathcal{O}^- = -\mathrm{sign}(B)\, \gamma_5\, \mathcal{O}^-$, so that the drive is an axial insertion localized at the Fermi surface: it is the dynamically generated analogue of the axial chemical potential $\mu_5$, which in the heuristic treatments is introduced by hand.

(iii) Substituting $\delta\hat{G}^{<} \approx \tilde{G}^{R}\, \hat{\Sigma}_d^{<}\, \tilde{G}^{A}$ back into the expression for the axial charge density, we obtain the second-iteration contribution
\begin{equation}
\frac{\delta\rho_5^{(2)}}{\rho_5^{(1)}} = 1 - O\Bigl( \theta_F^2,\; \bigl( v_F C(\theta_F)/\mu \bigr)^2 \Bigr) ,
\label{eq:ratio}
\end{equation}
where $C(\theta) = m + 1 - \cos\theta + B/2$, and $\rho_5^{(1)}$ is the bare-bubble result of Section \ref{observables}. (The subleading terms of this particular ratio depend on whether the bare width $\epsilon$ or the on-shell width $\epsilon_{\rm on}$ is used at the intermediate steps and are not needed in what follows; the exact ratio of the consecutive terms of the resummed series is the kernel eigenvalue $q$ of Appendix \ref{app:kernel}.) The second iteration reproduces the first one up to small corrections: the ladder is a geometric series with the ratio close to unity, and its sum is parametrically larger than its first term.

The smallness of the retarded and advanced components has a structural origin. Owing to the triangular structure of the Keldysh matrices, the retarded sector of Eq.~(\ref{eq:scba}) is autonomous: the equation $G^{R} \star (Q^{R} - \Sigma^{R}) = 1$ with $\Sigma^{R} = g \int G^{R}$ contains neither the distribution function nor the advanced functions. In the temporal gauge the electric field enters this closed sector only through the kinetic momentum {$\theta = p_3 - E_3 t$}, so that the equilibrium retarded function of the shifted argument solves the retarded equation up to regular gradient corrections. Consequently, all first-order retarded quantities are built of products of retarded functions alone; the poles of such products lie in one half-plane of the complex momentum, the momentum integrals contain no on-shell enhancement $1/\epsilon$, and $\hat{\Sigma}^{(1)R(A)} = O(\epsilon E_3/\mu)$, in agreement with the explicit calculation above.

At this point one could object that the relation $\Sigma^{<} = (\Sigma^{A} - \Sigma^{R})\, n(\omega)$ seems to force $\hat{\Sigma}^{(1)<}$ to be of the same order as $\hat{\Sigma}^{(1)R(A)}$. That relation is, however, the fluctuation--dissipation identity, valid in equilibrium only; the driven corrections are not constrained by it. Out of equilibrium the lesser component decomposes as
\begin{equation}
\hat{\Sigma}^{(1)<} = \bigl(\hat{\Sigma}^{(1)A} - \hat{\Sigma}^{(1)R}\bigr)\, n(\omega) \; + \; \hat{\Sigma}^{(1)}_{\rm kin},
\label{eq:fdtsplit}
\end{equation}
where the first (spectral) term is small together with $\hat{\Sigma}^{(1)R(A)}$, while $\hat{\Sigma}^{(1)}_{\rm kin}$ describes the field-induced deviation of the occupation from $n(\omega)$ and belongs to the sector that the fluctuation--dissipation identity does not govern. The static field modifies the spectral quantities only through the off-shell interband (Zener) mixing $\sim e E_3 a/\mu$, but it pumps the occupation continuously, the resulting shift being limited by the scattering alone, $\delta n \sim e E_3 |E_+'|\, n'(\omega)/(2\epsilon)$; hence the two sectors are of different orders in $1/\epsilon$, and the entire effect resides in $\hat{\Sigma}^{(1)}_{\rm kin}$. The mismatch is visible already in the supports of the two terms: at $T = 0$ the calculated lesser component $\propto n'(\omega) = -\delta(\omega - \mu)$ is concentrated at the Fermi level, whereas $(\hat{\Sigma}^{(1)A} - \hat{\Sigma}^{(1)R})\, n(\omega)$ is supported at all $\omega < \mu$.

}

\section{{Evaluation of the leading on-shell kernel and solution of the linearized equation}}
\label{app:kernel}

{

{For an auxiliary scalar width $\epsilon_G$, the momentum integral defining the map $\hat{F}$ of Eq.~(\ref{eq:selfcons}) can be reduced to elementary functions because $X=\hat{\Sigma}^{(1)<}(\omega)$ does not depend on $p_3$.  Writing $\tilde{G}^{R(A)}=(w_\pm+\mathcal{H}_{\rm latt})/(w_\pm^2-\tilde{E}^2)$ with $w_\pm=\omega\pm i\epsilon_G$, and noting that $\tilde{E}^2$ is linear in $x=\cos\theta$, the denominators factorize as $(w_+^2-\tilde{E}^2)(w_-^2-\tilde{E}^2)=v_F^4(2+2m+B)^2(x-x_+)(x-x_-)$, where $x_\pm$ are the analytic continuations of $\cos\theta_\omega$ to $\omega\pm i\epsilon_G$.  Partial fractions reduce the singular terms to $\frac{1}{\pi}\int_{-1}^{1}dx\,[(x-x_0)\sqrt{1-x^2}]^{-1}=-(x_0^2-1)^{-1/2}$, and the direct-current pinch singularity is contained in $(x_+-x_-)^{-1}=v_F^2(2+2m+B)/(-4i\omega\epsilon_G)$.  The physical self-consistent width is matrix-valued rather than scalar.  Nevertheless, in the weak-disorder pinch limit its leading $1/\epsilon_G$ contribution is obtained by replacing it with its on-shell projection.  The remaining matrix-width and off-shell contributions are regular, of relative order $\epsilon_Ga/(\hbar v_F)$.  Thus, in the limit $\epsilon_G\to0$, the leading on-shell kernel becomes the purely algebraic map}
\begin{align}
\hat{F}[X] = \frac{\epsilon(\omega)}{2\epsilon_G} \Bigl[\, X &+ |M|\, (X\gamma_0 + \gamma_0 X) \nn\\
&+ (1 - |M|^2)\, \gamma_0\gamma_3\, X\, \gamma_0\gamma_3 \nn\\
&+ |M|^2\, \gamma_0\, X\, \gamma_0 \,\Bigr]
\label{eq:kernel}
\end{align}
on the four-dimensional space of matrices spanned by $\{1, \gamma_0, \gamma_3, \gamma_0\gamma_3\}$, up to regular terms suppressed relative to Eq.~(\ref{eq:kernel}) by a factor of order $\epsilon_G a/(\hbar v_F)$. {The leading map (\ref{eq:kernel}) is diagonalized exactly; with the on-shell projected width $\epsilon_G = \epsilon_{\rm on} = \epsilon(\omega)(1 + |M|^2)$ its spectrum is given by Eq.~(\ref{eq:spectrum}) of the main text, and for $0<|M|<1$ this leading kernel has rank two.  The full finite-width kernel need not have rank two.}

We stress the role of the width used in $\tilde{G}^{R(A)}$: the exact level width is the matrix ${\rm Im}\,\hat{\Sigma}^{(0)R} = -\epsilon(\omega)(1 + |M|\gamma_0)$ following from Eq.~(\ref{eq:sigma0compact}), and $\epsilon_{\rm on} = \epsilon(\omega)(1+|M|^2)$ is its on-shell projection $\langle u_+|\ldots|u_+\rangle$. {Near either quasiparticle pole the singular part of the propagator therefore has the scalar denominator $\omega-E+i\epsilon_{\rm on}$, which justifies using $\epsilon_{\rm on}$ in Eq.~(\ref{eq:kernel}); it does not identify the complete matrix propagator with a scalar-width propagator at finite $\epsilon$.} If the bare scalar $\epsilon(\omega)$ were used instead (i.e., if the $\gamma_0$ component of the width were dropped), the even eigenvalue would equal $1 + |M|^2$, violating particle conservation at the same order $O(|M|^2)$ at which the deficit $1 - q$ is computed; the ratio of the odd and even eigenvalues, however, is insensitive to this choice, so that imposing the Ward identity $q_{\rm even} = 1$ fixes the normalization unambiguously. {At the level of the leading kernel the two null eigenvectors include $\gamma_3$, the channel in which the spectral components (\ref{eq:sigmaRA}) reside; its further propagation is therefore suppressed by $O(\epsilon_Ga/\hbar v_F)$ rather than being exactly zero in the full finite-width problem.}

{For $0<|M|<1$,} the unit eigenvalue makes $(1 - \hat{F})$ non-invertible on the full space of matrices, and one may worry that Eq.~(\ref{eq:linSigma}) has no solutions. The situation is governed by the Fredholm alternative: since the {leading map (\ref{eq:kernel})} is self-adjoint with respect to the scalar product $\langle X, Y \rangle = {\rm tr}(X^\dagger Y)$ (all its terms are symmetric combinations of left and right multiplications by Hermitian matrices), its left and right eigenvectors coincide, and the equation $(1 - \hat{F})[X] = \hat{\Sigma}_d$ is solvable if and only if the drive is orthogonal to the zero mode. This condition is satisfied identically, ${\rm tr}\bigl[(1 + |M|\gamma_0)\, \gamma_0\gamma_3\bigr] = 0$: the drive lives entirely in the odd channel, while the dangerous channel is empty. The general solution is then a one-parameter family,
\begin{equation}
\hat{\Sigma}^{(1)<} = \frac{\hat{\Sigma}_d^{<}}{1 - q} + c(\omega)\, \bigl(1 + |M|\gamma_0\bigr),
\label{eq:family}
\end{equation}
and the arbitrariness has a transparent physical meaning: the zero mode multiplies precisely the matrix structure of the equilibrium $\hat{\Sigma}^{(0)<}$, so that {the admixture $c(\omega)$ is a homogeneous, momentum-even perturbation of the energy-resolved occupation.  Because the disorder is elastic, the collision operator alone does not determine this function.  Fixing the total density supplies an integrated constraint on $c(\omega)$, while the physical adiabatic boundary condition selects the solution continuously connected to the finite-frequency response and gives $c(\omega)=0$.} The same conclusion follows from regularization: at a finite frequency $\Omega$ of the external field the even eigenvalue departs from unity by a term $\propto i\Omega/(2\epsilon_{\rm on})$, the operator $(1 - \hat{F})$ becomes invertible, and the even component of the solution equals the even component of the drive divided by that term -- i.e. zero for every $\Omega$, since the drive is odd; the DC limit is reached continuously with $c \equiv 0$. This structure is generic for conserving approximations: the zero mode is the $k \to 0$, $\Omega \to 0$ limit of the diffuson (at finite wavevector it would produce the diffusion pole $\sim 1/Dk^2$), and its presence in $(1 - \hat{F})$ is the fingerprint of the Ward identity rather than an inconsistency; equivalently, $\hat{F}[\hat{\Sigma}^{(0)<}] = \hat{\Sigma}^{(0)<}$ is nothing but the equilibrium equation (\ref{eq:scba}) itself. Finally, the observables are insensitive to the residual ambiguity even before it is fixed: for the zero-mode structure the intraband weights $\langle u_+(\pm\theta_F)| 1 + |M|\gamma_0 |u_+(\pm\theta_F)\rangle = 1 + |M|^2$ are even in $\theta$, while the vertices of both $\rho_5$ and $J_3$ are odd ($\propto \sin\theta_r$ and $E_+'(\theta_r)$, respectively), so the contribution of the zero mode to the axial density and to the electric current cancels identically between the two Fermi points.

That the drive contains no admixture of the even structure $1 + |M|\gamma_0$ is guaranteed by an exact symmetry: under the reflection $p_3 \to -p_3$ accompanied by $X \to \gamma_0 X \gamma_0$ the Hamiltonian is invariant, $\gamma_0 H(-\theta) \gamma_0 = H(\theta)$; the matrices $\{1, \gamma_0\}$ are even and $\{\gamma_0\gamma_3, \gamma_3\}$ are odd under this operation, while the drive, containing exactly one field vertex $\partial_{p_3}\hat{Q}$, is odd. On shell this is expressed by the group velocity factor: $P_+ H' P_+ = E_+'(\theta) P_+$ with $E_+'(\pm\theta_F) = \pm|E_+'|$, so that the two Fermi points enter with opposite signs, producing the difference $P_+(\theta_F) - P_+(-\theta_F) = (v_F \sin\theta_F/\omega)\, \gamma_0\gamma_3$ rather than the sum $P_+(\theta_F) + P_+(-\theta_F) = 1 + |M|\gamma_0$. Since $\hat{F}$ commutes with the reflection, the odd subspace is preserved along the entire Neumann series, and no even component is generated at any order. Moreover, since $\hat{\Sigma}_d \propto \gamma_0\gamma_3$ is an eigenvector of $\hat{F}$, every term of the Neumann series is computed explicitly,
\begin{equation}
\hat{F}^n[\hat{\Sigma}_d] = q^n\, \hat{\Sigma}_d , \qquad n = 0, 1, 2, \ldots,
\end{equation}
and {within the leading on-shell kernel} the series is geometric with the ratio $q$, leading to Eq.~(\ref{eq:directsol}) of the main text. The regular parts of the kernel, neglected in Eq.~(\ref{eq:kernel}), shift the eigenvalue $q$ by $O(\epsilon_G a/\hbar v_F)$, which is the origin of the validity condition quoted in the main text. {The derivation assumes $0<|M|<1$.  At $|M|=0$ one has $q=1$ and $\gamma_0\gamma_3$ is a second conserved channel; a nonzero odd drive then has no stationary homogeneous DC solution, corresponding to ballistic acceleration.  At $|M|=1$ the two Fermi points merge and the pinch expansion is invalid.  These exceptional limits are excluded, respectively, by the chirality-relaxation hierarchy and by the requirement that the Fermi energy remain away from the band edge.}

}

\bibliography{refs,citationsZ,new_references}

@article{chernodub2017scale,
  title={Scale magnetic effect in quantum electrodynamics and the Wigner-Weyl formalism},
  author={Chernodub, Maxim N and Zubkov, MA},
  journal={Physical Review D},
  volume={96},
  number={5},
  pages={056006},
  year={2017},
  publisher={APS}
}

@article{zhang2020influence,
	title={Influence of interactions on the anomalous quantum Hall effect},
	author={Zhang, CX and Zubkov, MA},
	journal={Journal of Physics A: Mathematical and Theoretical},
	volume={53},
	number={19},
	pages={195002},
	year={2020},
	publisher={IOP Publishing}
}

@article{zubkov2023effect,
	title={Effect of interactions on the topological expression for the chiral separation effect},
	author={Zubkov, M. A. and Abramchuk, R. A.},
	journal={Physical Review D},
	volume={107},
	number={9},
	pages={094021},
	year={2023},
	publisher={American Physical Society},
	doi={10.1103/PhysRevD.107.094021},
	url={https://journals.aps.org/prd/abstract/10.1103/PhysRevD.107.094021}
}

@article{zubkov2018momentum,
	title={Momentum space topology of QCD},
	author={Zubkov, MA},
	journal={Annals of Physics},
	volume={393},
	pages={264--287},
	year={2018},
	publisher={Elsevier}
}

@article{zubkov2012momentum,
	title={Momentum space topological invariants for the 4D relativistic vacua with mass gap},
	author={Zubkov, MA and Volovik, GE},
	journal={Nuclear Physics B},
	volume={860},
	number={2},
	pages={295--309},
	year={2012},
	publisher={Elsevier}
}

@article{volovik2017standard,
	title={Standard Model as the topological material},
	author={Volovik, GE and Zubkov, MA},
	journal={New Journal of Physics},
	volume={19},
	number={1},
	pages={015009},
	year={2017},
	publisher={IOP Publishing}
}

@article{volovik2013nambu,
	title={Nambu sum rule in the NJL Models: from superfluidity to top quark condensation},
	author={Volovik, Grigorii Efimovich and Zubkov, M.A.},
	journal={JETP letters},
	volume={97},
	number={6},
	pages={301--306},
	year={2013},
	publisher={Springer}
}

@article{bakker1999central,
	title={Central dominance and the confinement mechanism in gluodynamics},
	author={Bakker, BLG and Veselov, AI and Zubkov, MA},
	journal={Physics Letters B},
	volume={471},
	number={2-3},
	pages={214--219},
	year={1999},
	publisher={Elsevier}
}

@article{zhang2019hall,
	title={Hall conductivity as the topological invariant in the phase space in the presence of interactions and a nonuniform magnetic field},
	author={Zhang, CX and Zubkov, MA},
	journal={JETP letters},
	volume={110},
	number={7},
	pages={487--494},
	year={2019},
	publisher={Springer}
}

@article{katsnelson2013euler,
	title={Euler--Heisenberg effective action and magnetoelectric effect in multilayer graphene},
	author={Katsnelson, MI and Volovik, GE and Zubkov, MA},
	journal={Annals of Physics},
	volume={331},
	pages={160--187},
	year={2013},
	publisher={Elsevier}
}

@article{abramchuk2018anatomy,
	title={Anatomy of the chiral vortical effect},
	author={Abramchuk, Ruslan and Khaidukov, ZV and Zubkov, MA},
	journal={Physical Review D},
	volume={98},
	number={7},
	pages={076013},
	year={2018},
	publisher={APS}
}

@article{bakker2005standard,
	title={Standard Model with the additional Z6 symmetry on the lattice},
	author={Bakker, BLG and Veselov, AI and Zubkov, MA},
	journal={Physics Letters B},
	volume={620},
	number={3-4},
	pages={156--163},
	year={2005},
	publisher={Elsevier}
}

@article{zubkov2017topology,
  title={Topology of the momentum space, Wigner transformations, and a chiral anomaly in lattice models},
  author={Zubkov, Mikhail A and Khaidukov, Zakhar Viktorovich},
  journal={JETP Letters},
  volume={106},
  number={3},
  pages={172--178},
  year={2017},
  publisher={Springer}
}

@article{volovik2015scalar,
	title={Scalar excitation with Leggett frequency in He 3-B and the 125 GeV Higgs particle in top quark condensation models as pseudo-Goldstone bosons},
	author={Volovik, G.E. and Zubkov, M.A.},
	journal={Physical Review D},
	volume={92},
	number={5},
	pages={055004},
	year={2015},
	publisher={APS}
}

@article{selch2025non,
	title={Non-renormalization of the fractional quantum Hall conductivity by interactions},
	author={Selch, M. and Zubkov, M.A. and Pramanik, S. and Lewkowicz, M.},
	journal={Annals of Physics},
	pages={170202},
	year={2025},
	publisher={Elsevier}
}

@article{suleymanov2019wigner,
  title={Wigner--Weyl formalism and the propagator of Wilson fermions in the presence of varying external electromagnetic field},
  author={Suleymanov, M. and Zubkov, M.A.},
  journal={Nuclear Physics B},
  volume={938},
  pages={171--199},
  year={2019},
  publisher={Elsevier}
}

@article{JiangZrTe5Mass2017,
  title = {Landau-level spectroscopy of massive Dirac fermions in single-crystalline ${\mathrm{ZrTe}}_{5}$ thin flakes},
  author = {Jiang, Y. and Dun, Z. L. and Zhou, H. D. and Lu, Z. and Chen, K.-W. and Moon, S. and Besara, T. and Siegrist, T. M. and Baumbach, R. E. and Smirnov, D. and Jiang, Z.},
  journal = {Phys. Rev. B},
  volume = {96},
  issue = {4},
  pages = {041101(R)},
  numpages = {5},
  year = {2017},
  month = {Jul},
  publisher = {American Physical Society},
  doi = {10.1103/PhysRevB.96.041101},
  url = {https://link.aps.org/doi/10.1103/PhysRevB.96.041101}
}

@article{ZhengZrTe5Transition2017,
  author  = {Zheng, G. and Zhu, X. and Liu, Y. and Lu, J. and Ning, W. and Zhang, H. and Gao, W. and Han, Y. and Yang, J. and Du, H. and Yang, K. and Zhang, Y. and Tian, M.},
  title   = {Field-induced topological phase transition from a three-dimensional {W}eyl semimetal to a two-dimensional massive {D}irac metal in {ZrTe$_5$}},
  journal = {Phys. Rev. B},
  volume  = {96},
  pages   = {121401(R)},
  year    = {2017}
}

@article{NeupaneCd3As2_2014,
  author  = {Neupane, M. and Xu, S.-Y. and Sankar, R. and Alidoust, N. and Bian, G. and Liu, C. and Belopolski, I. and Chang, T.-R. and Jeng, H.-T. and Lin, H. and Bansil, A. and Chou, F. and Hasan, M. Z.},
  title   = {Observation of a three-dimensional topological {D}irac semimetal phase in high-mobility {Cd$_3$As$_2$}},
  journal = {Nature Communications},
  volume  = {5},
  pages   = {3786},
  year    = {2014}
}

@article{Ali2014Cd3As2Crystal,
  author  = {Ali, M. N. and Gibson, Q. and Jeon, S. and Zhou, B. B. and Yazdani, A. and Cava, R. J.},
  title   = {The Crystal and Electronic Structures of {Cd$_3$As$_2$}, the Three-Dimensional Electronic Analogue of Graphene},
  journal = {Inorganic Chemistry},
  volume  = {53},
  pages   = {4062--4067},
  year    = {2014}
}

@article{LiCd3As2NegativeMR2016,
  author  = {Li, H. and He, H. and Lu, H.-Z. and Zhang, H. and Liu, H. and Ma, R. and Fan, Z. and Shen, S.-Q. and Wang, J.},
  title   = {Negative magnetoresistance in {D}irac semimetal {Cd$_3$As$_2$}},
  journal = {Nature Communications},
  volume  = {7},
  pages   = {10301},
  year    = {2016}
}

@article{XiongNa3BiChiralAnomaly2015,
  author  = {Xiong, J. and Kushwaha, S. K. and Liang, T. and Krizan, J. W. and Hirschberger, M. and Wang, W. and Cava, R. J. and Ong, N. P.},
  title   = {Evidence for the chiral anomaly in the {D}irac semimetal {Na$_3$Bi}},
  journal = {Science},
  volume  = {350},
  pages   = {413--416},
  year    = {2015}
}

@article{RamshawTaAs2018,
  author  = {Ramshaw, B. J. and Modic, K. A. and Shekhter, A. and Zhang, Yi and Kim, Eun-Ah and Moll, P. J. W. and Bachmann, M. D. and Chan, M. K. and Betts, J. B. and Balakirev, F. and Migliori, A. and Ghimire, N. J. and Bauer, E. D. and Ronning, F. and McDonald, R. D.},
  title   = {Quantum limit transport and destruction of the {W}eyl nodes in {TaAs}},
  journal = {Nature Communications},
  volume  = {9},
  pages   = {2217},
  year    = {2018}
}

@article{ZhangTaP2017,
  author  = {Zhang, C.-L. and Xu, S.-Y. and Wang, C. M. and Lin, Z. and Du, Z. Z. and Guo, C. and Lee, C.-C. and Lu, H. and Feng, Y. and Huang, S.-M. and Chang, G. and Hsu, C.-H. and Liu, H. and Lin, H. and Li, L. and Zhang, C. and Zhang, J. and Xie, X.-C. and Neupert, T. and Hasan, M. Z. and Lu, H.-Z. and Wang, J. and Jia, S.},
  title   = {Magnetic-tunnelling-induced {W}eyl node annihilation in {TaP}},
  journal = {Nature Physics},
  volume  = {13},
  pages   = {979--986},
  year    = {2017}
}

@article{LiuZrTe5UQL2016,
  author  = {Liu, Y. and Yuan, X. and Zhang, C. and Jin, Z. and Narayan, A. and Luo, C. and Chen, Z. and Yang, L. and Zou, J. and Wu, X. and Sanvito, S. and Xia, Z. and Li, L. and Wang, Z. and Xiu, F.},
  title   = {Zeeman splitting and dynamical mass generation in {D}irac semimetal {ZrTe$_5$}},
  journal = {Nature Communications},
  volume  = {7},
  pages   = {12516},
  year    = {2016}
}

@article{GourgoutZrTe5Freezeout2022,
  author  = {Gourgout, A. and Leroux, M. and Smirr, J.-L. and Massoudzadegan, M. and Lobo, R. P. S. M. and Vignolles, D. and Proust, C. and Berger, H. and Li, Q. and Gu, G. and Homes, C. C. and Akrap, A. and Fauqu{\'e}, B.},
  title   = {Magnetic freeze-out and anomalous {H}all effect in {ZrTe$_5$}},
  journal = {npj Quantum Materials},
  volume  = {7},
  pages   = {71},
  year    = {2022}
}

@article{GoswamiPixleyDasSarma2015,
  author  = {Goswami, P. and Pixley, J. H. and Das Sarma, S.},
  title   = {Axial anomaly and longitudinal magnetoresistance of a generic three-dimensional metal},
  journal = {Phys. Rev. B},
  volume  = {92},
  pages   = {075205},
  year    = {2015}
}

@article{PhysRevLett.42.1698,
  title = {Solitons in Polyacetylene},
  author = {Su, W. P. and Schrieffer, J. R. and Heeger, A. J.},
  journal = {Phys. Rev. Lett.},
  volume = {42},
  issue = {25},
  pages = {1698--1701},
  numpages = {0},
  year = {1979},
  month = {Jun},
  publisher = {American Physical Society},
  doi = {10.1103/PhysRevLett.42.1698},
}

@article{Abramchuk_2026,
   title={Magnetoconductivity of Dirac semimetals and chiral magnetic effect from Keldysh technique},
   volume={38},
   ISSN={1361-648X},
   url={http://dx.doi.org/10.1088/1361-648X/ae350d},
   DOI={10.1088/1361-648x/ae350d},
   number={3},
   journal={Journal of Physics: Condensed Matter},
   publisher={IOP Publishing},
   author={Abramchuk, Ruslan A and Zubkov, M A},
   year={2026},
   month=Jan, pages={035601} }

@article{KHARZEEV2006260,
title = {Parity violation in hot QCD: Why it can happen, and how to look for it},
journal = {Physics Letters B},
volume = {633},
number = {2},
pages = {260-264},
year = {2006},
issn = {0370-2693},
doi = {https://doi.org/10.1016/j.physletb.2005.11.075},
url = {https://www.sciencedirect.com/science/article/pii/S0370269305017430},
author = {Dmitri Kharzeev}
}

@article{doi:10.1142/S0217751X09047570,
author = {SKOKOV, V. V. and ILLARIONOV, A. YU. and TONEEV, V. D.},
title = {ESTIMATE OF THE MAGNETIC FIELD STRENGTH IN HEAVY-ION COLLISIONS},
journal = {International Journal of Modern Physics A},
volume = {24},
number = {31},
pages = {5925-5932},
year = {2009},
doi = {10.1142/S0217751X09047570},

URL = {

        https://doi.org/10.1142/S0217751X09047570



},
eprint = {

        https://doi.org/10.1142/S0217751X09047570



}
}

@article{PhysRevD.52.4747,
  title = {Dynamical chiral symmetry breaking by a magnetic field in QED},
  author = {Gusynin, V. P. and Miransky, V. A. and Shovkovy, I. A.},
  journal = {Phys. Rev. D},
  volume = {52},
  issue = {8},
  pages = {4747--4751},
  numpages = {0},
  year = {1995},
  month = {Oct},
  publisher = {American Physical Society},
  doi = {10.1103/PhysRevD.52.4747},
  url = {https://link.aps.org/doi/10.1103/PhysRevD.52.4747}
}

@article{ABRAMCHUK2026113374,
title = {Negative magneto-resistance and Chiral Magnetic Effect in Dirac semimetals from Keldysh technique in Landau levels basis},
journal = {Journal of Physics and Chemistry of Solids},
volume = {210},
pages = {113374},
year = {2026},
issn = {0022-3697},
doi = {https://doi.org/10.1016/j.jpcs.2025.113374},
url = {https://www.sciencedirect.com/science/article/pii/S0022369725008273},
author = {Ruslan A. Abramchuk}
}

@article{BOHRA2026116276,
title = {Relation between chiral anomaly and electric transport in 1D Dirac semimetal},
journal = {Solid State Communications},
volume = {409},
pages = {116276},
year = {2026},
issn = {0038-1098},
doi = {https://doi.org/10.1016/j.ssc.2025.116276},
url = {https://www.sciencedirect.com/science/article/pii/S003810982500451X},
author = {Mustafa Bohra and Mikhail Zubkov}
}

@misc{kamenev2005manybodytheorynonequilibriumsystems,
      title={Many-body theory of non-equilibrium systems},
      author={Alex Kamenev},
      year={2005},
      eprint={cond-mat/0412296},
      archivePrefix={arXiv},
      primaryClass={cond-mat.dis-nn},
}

@Book{Kamenev2023,
  author    = {Kamenev, A.},
  year      = {2023},
  title     = {Field Theory of Non-Equilibrium Systems},
  isbn      = {9781108488259},
  publisher = {Cambridge University Press},
}

@article{CMEZrTe5,
    author = "Li, Qiang and Kharzeev, Dmitri E. and Zhang, Cheng and Huang, Yuan and Pletikosic, I. and Fedorov, A. V. and Zhong, R. D. and Schneeloch, J. A. and Gu, G. D. and Valla, T.",
    title = "{Chiral magnetic effect in ZrTe5}",
    eprint = "1412.6543",
    archivePrefix = "arXiv",
    primaryClass = "cond-mat.str-el",
    doi = "10.1038/nphys3648",
    journal = "Nature Phys.",
    volume = "12",
    pages = "550--554",
    year = "2016"
}

@article{Fukushima2008,
    author = "Fukushima, Kenji and Kharzeev, Dmitri E. and Warringa, Harmen J.",
    title = "{The Chiral Magnetic Effect}",
    eprint = "0808.3382",
    archivePrefix = "arXiv",
    primaryClass = "hep-ph",
    doi = "10.1103/PhysRevD.78.074033",
    journal = "Phys. Rev. D",
    volume = "78",
    pages = "074033",
    year = "2008"
}

@article{Onoda2006ev,
    author = "Onoda, Shigeki and Sugimoto, Naoyuki and Nagaosa, Naoto",
    title = "{Theoy of Non-equilibirum states driven by constant electromagnetic fields: Non-commutative quantum mechanics in the Keldysh formalism}",
    eprint = "cond-mat/0605363",
    archivePrefix = "arXiv",
    doi = "10.1143/PTP.116.61",
    journal = "Prog. Theor. Phys.",
    volume = "116",
    pages = "61",
    year = "2006"
}

@article{Arseev2015,
	author = {P. I. Arseev},
	title = {On the nonequilibrium diagram technique: derivation, some features and applications},
	publisher = {Physics-Uspekhi},
	year = {2015},
	journal = {Phys. Usp.},
	volume = {58},
	number = {12},
	pages = {1159-1205},
	url = {https://ufn.ru/en/articles/2015/12/b/},
	doi = {10.3367/UFNe.0185.201512b.1271}
}

@article{Gao2012,
    author = "Gao, Jian-Hua and Liang, Zuo-Tang and Pu, Shi and Wang, Qun and Wang, Xin-Nian",
    title = "{Chiral Anomaly and Local Polarization Effect from Quantum Kinetic Approach}",
    eprint = "1203.0725",
    archivePrefix = "arXiv",
    primaryClass = "hep-ph",
    reportNumber = "USTC-ICTS-12-02",
    doi = "10.1103/PhysRevLett.109.232301",
    journal = "Phys. Rev. Lett.",
    volume = "109",
    pages = "232301",
    year = "2012"
}

@Article{Kaushik2017,
  author    = {Kaushik, Sahal and Kharzeev, Dmitri E.},
  journal   = {Physical Review B},
  title     = {Quantum oscillations in the chiral magnetic conductivity},
  year      = {2017},
  issn      = {2469-9969},
  month     = June,
  number    = {23},
  pages     = {235136},
  volume    = {95},
  doi       = {10.1103/physrevb.95.235136},
  publisher = {American Physical Society (APS)},
}

@Article{Gorbar2014,
  author    = {Gorbar, E. V. and Miransky, V. A. and Shovkovy, I. A.},
  journal   = {Physical Review B},
  title     = {Chiral anomaly, dimensional reduction, and magnetoresistivity of Weyl and Dirac semimetals},
  year      = {2014},
  issn      = {1550-235X},
  month     = Feb,
  number    = {8},
  pages     = {085126},
  volume    = {89},
  doi       = {10.1103/physrevb.89.085126},
  publisher = {American Physical Society (APS)},
}

@Article{Lu2015,
  author    = {Lu, Hai-Zhou and Zhang, Song-Bo and Shen, Shun-Qing},
  journal   = {Physical Review B},
  title     = {High-field magnetoconductivity of topological semimetals with short-range potential},
  year      = {2015},
  issn      = {1550-235X},
  month     = July,
  number    = {4},
  pages     = {045203},
  volume    = {92},
  doi       = {10.1103/physrevb.92.045203},
  publisher = {American Physical Society (APS)},
}

@Article{Li2023,
  author    = {Li, Shuai and Lu, Hai-Zhou and Xie, X. C.},
  journal   = {Physical Review B},
  title     = {Impurity and dispersion effects on the linear magnetoresistance in the quantum limit},
  year      = {2023},
  issn      = {2469-9969},
  month     = June,
  number    = {23},
  pages     = {235202},
  volume    = {107},
  doi       = {10.1103/physrevb.107.235202},
  publisher = {American Physical Society (APS)},
}

@Article{Stephanov2012,
  author    = {Stephanov, M. A. and Yin, Y.},
  journal   = {Physical Review Letters},
  title     = {Chiral Kinetic Theory},
  year      = {2012},
  issn      = {1079-7114},
  month     = Oct,
  number    = {16},
  pages     = {162001},
  volume    = {109},
  doi       = {10.1103/physrevlett.109.162001},
  publisher = {American Physical Society (APS)},
}

@Article{Burkov2014,
  author    = {Burkov, A. A.},
  journal   = {Physical Review Letters},
  title     = {Chiral Anomaly and Diffusive Magnetotransport in Weyl Metals},
  year      = {2014},
  issn      = {1079-7114},
  month     = Dec,
  number    = {24},
  pages     = {247203},
  volume    = {113},
  doi       = {10.1103/physrevlett.113.247203},
  publisher = {American Physical Society (APS)},
}

@Article{Son2013,
  author    = {Son, D. T. and Spivak, B. Z.},
  journal   = {Physical Review B},
  title     = {Chiral anomaly and classical negative magnetoresistance of Weyl metals},
  year      = {2013},
  issn      = {1550-235X},
  month     = Sept,
  number    = {10},
  pages     = {104412},
  volume    = {88},
  doi       = {10.1103/physrevb.88.104412},
  publisher = {American Physical Society (APS)},
}

@book{gelfand1964generalized,
  title={Generalized Functions: Properties and operations, by I. M. Gelfand and G. E. Shilov, translated by E. Saletan},
  author={Gelfand, I.M. and Shilov, G.E.},
  isbn={9780122795015},
  lccn={63016960},
  series={Generalized Functions},
  url={https://books.google.co.il/books?id=oPdQAAAAMAAJ},
  year={1964},
  publisher={Academic Press}
}

@Article{Gorbar2016,
  author    = {Gorbar, E. V. and Shovkovy, I. A. and Vilchinskii, S. and Rudenok, I. and Boyarsky, A. and Ruchayskiy, O.},
  journal   = {Physical Review D},
  title     = {Anomalous Maxwell equations for inhomogeneous chiral plasma},
  year      = {2016},
  issn      = {2470-0029},
  month     = May,
  number    = {10},
  pages     = {105028},
  volume    = {93},
  doi       = {10.1103/physrevd.93.105028},
  publisher = {American Physical Society (APS)},
}

@Article{Lin2020,
  author    = {Lin, Shu and Yang, Lixin},
  journal   = {Physical Review D},
  title     = {Chiral kinetic theory from Landau level basis},
  year      = {2020},
  issn      = {2470-0029},
  month     = Feb,
  number    = {3},
  pages     = {034006},
  volume    = {101},
  doi       = {10.1103/physrevd.101.034006},
  publisher = {American Physical Society (APS)},
}

@Article{Hattori2017,
  author    = {Hattori, Koichi and Li, Shiyong and Satow, Daisuke and Yee, Ho-Ung},
  journal   = {Physical Review D},
  title     = {Longitudinal conductivity in strong magnetic field in perturbative QCD: Complete leading order},
  year      = {2017},
  issn      = {2470-0029},
  month     = Apr,
  number    = {7},
  pages     = {076008},
  volume    = {95},
  doi       = {10.1103/physrevd.95.076008},
  publisher = {American Physical Society (APS)},
}

@Article{Huang2018,
  author    = {Huang, Anping and Jiang, Yin and Shi, Shuzhe and Liao, Jinfeng and Zhuang, Pengfei},
  journal   = {Physics Letters B},
  title     = {Out-of-equilibrium chiral magnetic effect from chiral kinetic theory},
  year      = {2018},
  issn      = {0370-2693},
  month     = Feb,
  pages     = {177--183},
  volume    = {777},
  doi       = {10.1016/j.physletb.2017.12.025},
  publisher = {Elsevier BV},
}

@Article{Son2012,
  author    = {Son, Dam Thanh and Yamamoto, Naoki},
  journal   = {Physical Review Letters},
  title     = {Berry Curvature, Triangle Anomalies, and the Chiral Magnetic Effect in Fermi Liquids},
  year      = {2012},
  issn      = {1079-7114},
  month     = Nov,
  number    = {18},
  pages     = {181602},
  volume    = {109},
  doi       = {10.1103/physrevlett.109.181602},
  publisher = {American Physical Society (APS)},
}

@Article{Sekine2017,
  author    = {Sekine, Akihiko and Culcer, Dimitrie and MacDonald, Allan H.},
  journal   = {Physical Review B},
  title     = {Quantum kinetic theory of the chiral anomaly},
  year      = {2017},
  issn      = {2469-9969},
  month     = Dec,
  number    = {23},
  pages     = {235134},
  volume    = {96},
  doi       = {10.1103/physrevb.96.235134},
  publisher = {American Physical Society (APS)},
}

@Article{Sekine2021,
  author    = {Sekine, Akihiko and Nomura, Kentaro},
  journal   = {Journal of Applied Physics},
  title     = {Axion electrodynamics in topological materials},
  year      = {2021},
  issn      = {1089-7550},
  month     = Apr,
  number    = {14},
  volume    = {129},
  doi       = {10.1063/5.0038804},
  publisher = {AIP Publishing},
}

@Misc{Valgushev2015,
  author    = {Valgushev, S. N. and Puhr, M. and Buividovich, P. V.},
  title     = {Chiral Magnetic Effect in finite-size samples of parity-breaking Weyl semimetals},
  year      = {2015},
  copyright = {arXiv.org perpetual, non-exclusive license},
  doi       = {10.48550/ARXIV.1512.01405},
  publisher = {arXiv},
}

@Article{Buividovich2015,
  author    = {Buividovich, P. V. and Puhr, M. and Valgushev, S. N.},
  journal   = {Physical Review B},
  title     = {Chiral magnetic conductivity in an interacting lattice model of parity-breaking Weyl semimetal},
  year      = {2015},
  issn      = {1550-235X},
  month     = Nov,
  number    = {20},
  pages     = {205122},
  volume    = {92},
  doi       = {10.1103/physrevb.92.205122},
  publisher = {American Physical Society (APS)},
}

@Article{Buividovich2014,
  author    = {Buividovich, P. V.},
  journal   = {Physical Review D},
  title     = {Spontaneous chiral symmetry breaking and the chiral magnetic effect for interacting Dirac fermions with chiral imbalance},
  year      = {2014},
  issn      = {1550-2368},
  month     = Dec,
  number    = {12},
  pages     = {125025},
  volume    = {90},
  doi       = {10.1103/physrevd.90.125025},
  publisher = {American Physical Society (APS)},
}

@Article{Buividovich2014a,
  author    = {Buividovich, P.V.},
  journal   = {Nuclear Physics A},
  title     = {Anomalous transport with overlap fermions},
  year      = {2014},
  issn      = {0375-9474},
  month     = May,
  pages     = {218--253},
  volume    = {925},
  doi       = {10.1016/j.nuclphysa.2014.02.022},
  publisher = {Elsevier BV},
}

@Article{Zubkov2016,
  author    = {Zubkov, M.A.},
  journal   = {Annals of Physics},
  title     = {Wigner transformation, momentum space topology, and anomalous transport},
  year      = {2016},
  issn      = {0003-4916},
  month     = Oct,
  pages     = {298--324},
  volume    = {373},
  doi       = {10.1016/j.aop.2016.07.011},
  publisher = {Elsevier BV},
}

@Article{Zubkov2016a,
  author    = {Zubkov, M. A.},
  journal   = {Physical Review D},
  title     = {Absence of equilibrium chiral magnetic effect},
  year      = {2016},
  issn      = {2470-0029},
  month     = May,
  number    = {10},
  pages     = {105036},
  volume    = {93},
  doi       = {10.1103/physrevd.93.105036},
  publisher = {American Physical Society (APS)},
}

@Article{Vazifeh2013,
  author    = {Vazifeh, M. M. and Franz, M.},
  journal   = {Physical Review Letters},
  title     = {Electromagnetic Response of Weyl Semimetals},
  year      = {2013},
  issn      = {1079-7114},
  month     = July,
  number    = {2},
  pages     = {027201},
  volume    = {111},
  doi       = {10.1103/physrevlett.111.027201},
  publisher = {American Physical Society (APS)},
}

@Article{Yamamoto2015,
  author    = {Yamamoto, Naoki},
  journal   = {Physical Review D},
  title     = {Generalized Bloch theorem and chiral transport phenomena},
  year      = {2015},
  issn      = {1550-2368},
  month     = Oct,
  number    = {8},
  pages     = {085011},
  volume    = {92},
  doi       = {10.1103/physrevd.92.085011},
  publisher = {American Physical Society (APS)},
}

@Article{Banerjee2021,
  author    = {Banerjee, Chitradip and Lewkowicz, Meir and Zubkov, Mikhail A.},
  journal   = {Physics Letters B},
  title     = {Equilibrium chiral magnetic effect: Spatial inhomogeneity, finite temperature, interactions},
  year      = {2021},
  issn      = {0370-2693},
  month     = Aug,
  pages     = {136457},
  volume    = {819},
  doi       = {10.1016/j.physletb.2021.136457},
  publisher = {Elsevier BV},
}

@Article{Banerjee2022,
  author    = {Banerjee, Chitradip and Lewkowicz, Meir and Zubkov, Mikhail. A.},
  journal   = {Physical Review D},
  title     = {Chiral magnetic effect out of equilibrium},
  year      = {2022},
  issn      = {2470-0029},
  month     = Oct,
  number    = {7},
  pages     = {074508},
  volume    = {106},
  doi       = {10.1103/physrevd.106.074508},
  publisher = {American Physical Society (APS)},
}

@Article{Banerjee2021a,
  author    = {Banerjee, C. and Fialkovsky, I. V. and Lewkowicz, M. and Zhang, C. X. and Zubkov, M. A.},
  journal   = {Journal of Computational Electronics},
  title     = {Wigner-Weyl calculus in Keldysh technique},
  year      = {2021},
  issn      = {1572-8137},
  month     = Oct,
  number    = {6},
  pages     = {2255--2283},
  volume    = {20},
  doi       = {10.1007/s10825-021-01775-8},
  publisher = {Springer Science and Business Media LLC},
}

@Article{Hupfer2001,
  author    = {Hupfer, Thomas and Leschke, Hajo and Warzel, Simone},
  journal   = {Journal of Mathematical Physics},
  title     = {Upper bounds on the density of states of single Landau levels broadened by Gaussian random potentials},
  year      = {2001},
  issn      = {1089-7658},
  month     = Dec,
  number    = {12},
  pages     = {5626--5641},
  volume    = {42},
  doi       = {10.1063/1.1401138},
  publisher = {AIP Publishing},
}

@Article{Miransky2015,
  author    = {Miransky, Vladimir A. and Shovkovy, Igor A.},
  journal   = {Physics Reports},
  title     = {Quantum field theory in a magnetic field: From quantum chromodynamics to graphene and Dirac semimetals},
  year      = {2015},
  issn      = {0370-1573},
  month     = Apr,
  pages     = {1--209},
  volume    = {576},
  doi       = {10.1016/j.physrep.2015.02.003},
  publisher = {Elsevier BV},
}

@Article{Gusynin1995,
  author    = {Gusynin, V.P. and Miransky, V.A. and Shovkovy, I.A.},
  journal   = {Physics Letters B},
  title     = {Dimensional reduction and dynamical chiral symmetry breaking by a magnetic field in 3 + 1 dimensions},
  year      = {1995},
  issn      = {0370-2693},
  month     = Apr,
  number    = {4},
  pages     = {477--483},
  volume    = {349},
  doi       = {10.1016/0370-2693(95)00232-a},
  publisher = {Elsevier BV},
}

@Article{Gusynin1996,
  author    = {Gusynin, V.P. and Miransky, V.A. and Shovkovy, I.A.},
  journal   = {Nuclear Physics B},
  title     = {Dimensional reduction and catalysis of dynamical symmetry breaking by a magnetic field},
  year      = {1996},
  issn      = {0550-3213},
  month     = Mar,
  number    = {2-3},
  pages     = {249--290},
  volume    = {462},
  doi       = {10.1016/0550-3213(96)00021-1},
  publisher = {Elsevier BV},
}

@Article{Zhang2020,
  author    = {Zhang, C.X. and Zubkov, M.A.},
  journal   = {Physics Letters B},
  title     = {Feynman rules in terms of the Wigner transformed Green functions},
  year      = {2020},
  issn      = {0370-2693},
  month     = Mar,
  pages     = {135197},
  volume    = {802},
  doi       = {10.1016/j.physletb.2020.135197},
  publisher = {Elsevier BV},
}

@Misc{Roux2019,
  author    = {Roux, Filippus S. and Fabre, Nicolas},
  title     = {Wigner functional theory for quantum optics},
  year      = {2019},
  copyright = {arXiv.org perpetual, non-exclusive license},
  doi       = {10.48550/ARXIV.1901.07782},
  publisher = {arXiv},
}

@InBook{Szmigielski2025,
  author    = {Szmigielski, Jacek},
  pages     = {314--320},
  publisher = {Elsevier},
  title     = {Weyl-Wigner Quantization},
  year      = {2025},
  isbn      = {9780323957069},
  booktitle = {Encyclopedia of Mathematical Physics},
  doi       = {10.1016/b978-0-323-95703-8.00054-9},
}

@article{Moyal_1949, title={Quantum mechanics as a statistical theory}, volume={45}, DOI={10.1017/S0305004100000487}, number={1}, journal={Mathematical Proceedings of the Cambridge Philosophical Society}, author={Moyal, J. E.}, year={1949}, pages={99–124}}

@Article{Weyl1927,
  author    = {Weyl, H.},
  journal   = {Zeitschrift für Physik},
  title     = {Quantenmechanik und Gruppentheorie},
  year      = {1927},
  issn      = {1434-601X},
  month     = Nov,
  number    = {1-2},
  pages     = {1--46},
  volume    = {46},
  doi       = {10.1007/bf02055756},
  publisher = {Springer Science and Business Media LLC},
}

@Article{Mayer1947,
  author    = {Mayer, Joseph E. and Band, William},
  journal   = {The Journal of Chemical Physics},
  title     = {On the Quantum Correction for Thermodynamic Equilibrium},
  year      = {1947},
  issn      = {1089-7690},
  month     = Mar,
  number    = {3},
  pages     = {141--149},
  volume    = {15},
  doi       = {10.1063/1.1746445},
  publisher = {AIP Publishing},
}

@Article{Wigner1932,
  author    = {Wigner, E.},
  journal   = {Physical Review},
  title     = {On the Quantum Correction For Thermodynamic Equilibrium},
  year      = {1932},
  issn      = {0031-899X},
  month     = June,
  number    = {5},
  pages     = {749--759},
  volume    = {40},
  doi       = {10.1103/physrev.40.749},
  publisher = {American Physical Society (APS)},
}

@book{sakurai2017modern,
  title={Modern Quantum Mechanics},
  author={Sakurai, J.J. and Napolitano, J.},
  isbn={9781108422413},
  url={https://books.google.co.il/books?id=010yDwAAQBAJ},
  year={2017},
  publisher={Cambridge University Press}
}

@Book{Shankar1994,
  author    = {Shankar, R.},
  publisher = {Springer US},
  title     = {Principles of Quantum Mechanics},
  year      = {1994},
  isbn      = {9781475705768},
  doi       = {10.1007/978-1-4757-0576-8},
}

@Article{Su1980,
  author    = {Su, W. P. and Schrieffer, J. R. and Heeger, A. J.},
  journal   = {Physical Review B},
  title     = {Soliton excitations in polyacetylene},
  year      = {1980},
  issn      = {0163-1829},
  month     = Aug,
  number    = {4},
  pages     = {2099--2111},
  volume    = {22},
  doi       = {10.1103/physrevb.22.2099},
  publisher = {American Physical Society (APS)},
}

@book{bernevig2013topological,
  title={Topological insulators and topological superconductors},
  author={Bernevig, B Andrei},
  year={2013},
  publisher={Princeton university press}
}

@Article{Lima2026,
  author    = {Lima, L.S.},
  journal   = {Physica A: Statistical Mechanics and its Applications},
  title     = {Interplay of phase transition from purely real and line-gapped phase on quantum entanglement in the generalized Su–Schrieffer–Heeger model},
  year      = {2026},
  issn      = {0378-4371},
  month     = July,
  pages     = {131574},
  volume    = {694},
  doi       = {10.1016/j.physa.2026.131574},
  publisher = {Elsevier BV},
}

@Article{McCann2026,
  author    = {McCann, Edward},
  journal   = {Physical Review B},
  title     = {Non-Hermitian Su-Schrieffer-Heeger model with the energy levels of free parafermions},
  year      = {2026},
  issn      = {2469-9969},
  month     = Jan,
  number    = {4},
  volume    = {113},
  doi       = {10.1103/hdzd-94qm},
  publisher = {American Physical Society (APS)},
}

@article{Keldysh1964DiagramTF,
  author  = {Keldysh, L. V.},
  title   = {Diagram technique for nonequilibrium processes},
  journal = {Sov. Phys. JETP},
  volume  = {20},
  pages   = {1018--1026},
  year    = {1965},
  note    = {[Zh. Eksp. Teor. Fiz. 47, 1515 (1964)]; reprinted in: Selected Papers of Leonid V Keldysh (World Scientific, 2023), p. 47},
}

@Article{Peierls1933,
  author    = {Peierls, R.},
  journal   = {Zeitschrift für Physik},
  title     = {Zur Theorie des Diamagnetismus von Leitungselektronen},
  year      = {1933},
  issn      = {1434-601X},
  month     = Nov,
  number    = {11-12},
  pages     = {763--791},
  volume    = {80},
  doi       = {10.1007/bf01342591},
  publisher = {Springer Science and Business Media LLC},
}

@article{zubkov2023discrete,
  title={Discrete Wigner--Weyl calculus for the finite lattice},
  author={Zubkov, MA},
  journal={Journal of Physics A: Mathematical and Theoretical},
  volume={56},
  number={39},
  pages={395201},
  year={2023},
  publisher={IOP Publishing}
}

@article{volovik2017chiral,
  title={On the chiral magnetic effect in Weyl superfluid 3He-A},
  author={Volovik, Grigorii Efimovich},
  journal={JETP letters},
  volume={105},
  number={1},
  pages={34--37},
  year={2017},
  publisher={Springer}
}

\end{document}